Programmable self-assembly for scalable molecular electronic architectures

Seham Helmi* [1,2], Junjie Liu* [3,4], Keith G Andrews [5], Robert Schreiber [3], Jonathan Bath [1,3], Harry L Anderson [2], Andrew J Turberfield [1,3], Arzhang Ardavan [3]

1 The Kavli Institute for Nanoscience Discovery, University of Oxford, Oxford OX1 3QU, UK
2 Department of Chemistry, University of Oxford, Mansfield Road, Oxford OX1 3TA, UK
3 The Clarendon Laboratory, Department of Physics, University of Oxford, Oxford OX1 3PU, UK
4 School of Physical and Chemical Sciences, Queen Mary University of London, London E1 4NS, UK
5 Department of Chemistry, Durham University, Lower Mount Joy, South Rd, Durham, DH1 3LE, UK

*Abstract*

Molecular electronics and other technologies whose components comprise individual molecules have been pursued for half a century because the molecular scale represents the limit of miniaturisation of objects whose structure is tuneable for function[1–3]. Despite the promise, practical progress has been hindered by the lack of methodologies for directed assembly of arbitrary structures at the molecular scale[4–8]. DNA nanotechnology is an emerging framework that uses programmed synthetic oligomers to encode the design of self-assembling structures with atomic precision at the nanoscale[9,10].

Here, we leverage DNA-directed self-assembly to construct single-molecule electrical transport devices in high yield, by precisely positioning a metaloporphyrin[11] between two 60 nm gold nanoparticles. Following deposition of the assembled devices on $SiO_2$ substrates, we image and establish electrical contact via established nanofabrication techniques. Each step of the process has a high success rate and we demonstrate device yields dramatically better than is possible using conventional approaches. Our approach is inherently scalable and adaptable to devices incorporating multiple heterogenous functional molecular components, finally offering a realistic framework for the implementation of classical and quantum molecular technologies.

*Introduction*

Since Moore's observation in 1965[12] that circuit integration grows exponentially, the quest to reach the fundamental limit of miniaturization has focused on molecules—the smallest tuneable functional units. By the mid-1970s, Aviram and Ratner proposed the first molecular diode[13], highlighting the potential of molecular-scale devices to push classical electronics to their limits[1–3] and to exploit quantum effects for advanced functionalities[14–18]. However, practical progress has been hampered by the lack of scalable methods to integrate individual molecules into functional electrical devices[4–8].

Current methods for contacting single molecules, such as incorporation in electromigrated break junctions[4,5], provide valuable insights but are limited by low, stochastic yields (typically <1%). Scanning tunnelling microscopy (STM) provides higher reproducibility at the cost of complex instrumentation[6,7]. Intermediate approaches, such as STM break junctions and mechanically controlled break junctions[6], offer a balance between reliability and speed of data acquisition[6]. Recently, hydrogen-plasma-etched graphene junctions have been reported to yield single-molecule devices with impressive reliability and reproducibility[19]. None of these methods can conceivably support scalable assembly of devices incorporating multiple integrated molecular components[8].

Despite these limitations, proof-of-concept demonstrations have shown the promise of molecular electronics, including, as an example of particular relevance for quantum technologies, nuclear quantum spin manipulation and projective measurement in $TbPc_2$ junctions[20–22]. In this context porphyrin derivatives offer versatile functionalities as molecular quantum dots[23,24], wires[25,26], and quantum spin components[27–29]. However, half a century after their potential was recognized, integrated molecular devices remain elusive due to the mismatch between the ~10 nm resolution of top-down fabrication and the <0.1 nm precision required for molecular interfacing. Bridging this gap demands a paradigm shift in assembly, one that our DNA-based strategy delivers.

DNA self-assembly exploits the sequence specificity of synthetic oligonucleotides[9,30] to achieve programmable molecular construction. Advances in DNA synthesis now enable low-cost fabrication of custom sequences, providing a versatile toolkit for nanoscale engineering. The predictable hybridization of DNA strands forms stiff duplexes (~2 nm diameter) and branched junctions, with sticky-end overhangs which serve as modular connectors for assembling discrete and extended structures[31–35]. DNA origami[36] expands this concept by folding a long single-stranded scaffold using hundreds of staple strands to form well-defined 2D and 3D nanostructures which can extend from ~100 nm to micron-scale assemblies[36–41] . Each component of a DNA nanostructure can be identified through its base sequence and functionalized through site-specific modification[42]. Moreover, DNA nanotechnology enables the precise spatial positioning of functional components on lithographically patterned surfaces[43,44]. The positional control that it affords makes it an ideal tool for integrating and probing molecular-scale systems.

Here, we establish a DNA-assisted assembly strategy that enables scalable, reliable, and reproducible construction of molecular electronic devices. Our approach achieves high-yield integration of single-molecule electronic and quantum components while using DNA solely as a structural element to impart precise spatial control. Each step in the process is optimized for high yield, ensuring that the framework permits assembly of increasingly complex molecular devices reproducibly and at scale.

*A DNA-guided molecular electronic device architecture*

We validate the approach on the simplest molecular electronic device – a two-terminal molecular junction – where two metallic terminals bridge a single functional "target" molecule, enabling measurement of its electrical transport properties. In our design (Fig. 1a), a DNA nanostructure precisely controls the location and orientation of the target molecule. The target molecule is covalently conjugated to two distinct DNA adapters, which facilitate controlled placement within the DNA nanostructure through strand complementarity. DNA-functionalized gold nanoparticles (AuNPs, yellow) self-assemble onto the nanostructure, forming a nanogap bridged by the molecular component. We refer to this fully assembled structure as a "package".

The AuNPs serve as electrical contacts, and their tens-of-nanometre scale ensures compatibility with standard nanolithography techniques, allowing direct interfacing between the molecular junction and macroscopic laboratory equipment for electrical characterization and functional control.

To implement this approach, we focused on three key aspects to ensure precise, reproducible, and scalable fabrication of single-molecule devices. First, we covalently functionalized the target molecule with two distinct DNA oligomers, enabling precise positioning within the DNA nanostructure and eliminating stochastic placement issues. Second, we optimized the DNA nanostructure design, ensuring controlled spacing between the target molecule and AuNPs to enhance electronic coupling and device performance. Third, we developed methods for surface deposition and device interfacing, allowing DNA-assembled packages to be positioned on substrates, imaged, and contacted for electrical characterization.

*Covalent-functionalization of the target molecule with DNA adapters*

As the prototype molecular electronic component, we selected a single-metalloporphyrin quantum dot[11] (Fig. 1b), where the porphyrin core acts as the quantum dot island, and metalation with transition metal cations enables tuning of its magnetic and redox properties. For this study, we synthesized a diamagnetic Zn-porphyrin derivative modified for covalent attachment. The molecule was designed with two terminal alkyne groups, allowing site-specific conjugation to azide-modified DNA adapters via copper-catalysed alkyne-azide cycloaddition (CuAAC). Additionally, the SMe groups facilitate coordination with gold contacts positioned about 2.4 nm apart, while the internal alkyne backbone provides a π-conjugated charge transport pathway.

To ensure precise molecular positioning and orientation, we hetero-bi-functionalized this molecule using two distinct DNA adapters (A and B in Fig. 1c), allowing us to anchor the molecule at two tethering points within the DNA nanostructure. The adapters were labelled with Cy3 and Cy5 fluorophores at the ends opposing the reactive azide modifications, enabling fluorescence-based tracking of the conjugation reaction and subsequent incorporation into the DNA nanostructure.

To obtain the hetero-bi-functionalized porphyrin-DNA hybrid (A-porphyrin-B), we employed a DNA-templated conjugation approach[45]. A third, template, DNA strand, hybridizes to both adapters to localize their reactive azide groups around a spacer domain designed to accommodate the porphyrin molecule. This configuration increased the local concentration of reactive groups, selectively favouring hetero-adapter conjugation (Fig. 1c). An untemplated reaction also generated A-porphyrin-B but with significant side-products A-porphyrin, A-porphyrin-A, B-porphyrin and B-porphyrin-B; while the yield of the desired product was lower, it was easier to gel-purify owing to the absence of the overlapping band of the template strand.

The CuAAC reaction was performed in a water/DMF mixture (see Supplementary Information), and the reaction products were analysed using denaturing polyacrylamide gel electrophoresis (PAGE). Fluorescence scanning of the gel confirmed successful conjugation, with a yellow band in the merged gel image (Fig. 1d, lanes 2 and 3) indicating colocalization of Cy3 and Cy5 confirming the formation of the hetero-DNA-porphyrin product (A-porphyrin-B). The product was purified using high-performance liquid chromatography (HPLC) (SI Fig. 11) or PAGE (Fig. 1d, lane 3).

The UV-Visible (UV-Vis) spectrum of the purified Zn (A-porphyrin-B) product showed characteristic porphyrin absorption peaks, along with distinct peaks attributed to Cy3, Cy5 fluorophores and DNA (Fig. 1e), aligning well with control spectra obtained before the reaction (SI Fig. 2).

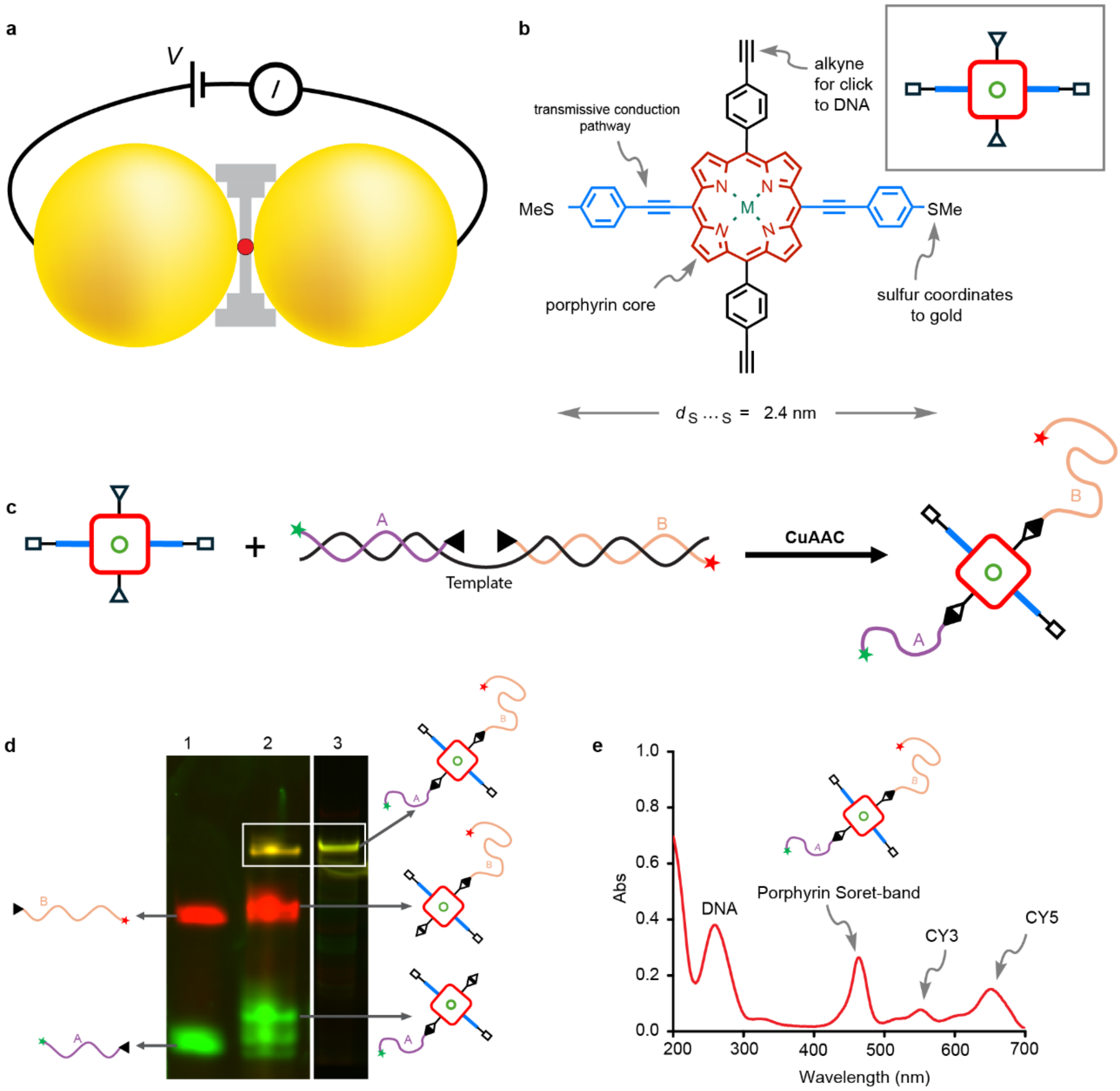

**Fig. 1:** (a) Schematic of the DNA-assembled molecular electronic device. A DNA nanostructure (grey) directs the self-assembly of a target molecule (red) precisely between two AuNP contacts (yellow), which are then interfaced with laboratory equipment (black) using nanofabrication methods. (b) Chemical structure of the metalloporphyrin molecular quantum dot, showing the porphyrin core (red) with coordinated transition metal cation (green) as the quantum dot. The inset illustrates the molecular structure schematically. Terminal alkynes (open triangles) enable site-specific DNA attachment via CuAAC click chemistry. SMe groups (open rectangles) facilitate gold coordination. (c) DNA-templated synthesis strategy for linking the porphyrin to two distinct DNA adapters (A and B), ensuring selective hetero-bi-functionalization. A template DNA strand hybridizes with the adapters, colocalizing azide groups (filled triangles) to facilitate selective hetero-conjugation. (d) Merged fluorescence image of a 20% denaturing PAGE gel showing successful CuAAc reaction products: lane 1, adapters A and B before the reaction (scans in SYBR® Gold, Cy3 and Cy5 channels are shown separately in SI); lane 2, templated reaction products; lane 3, HPLC-purified A-porphyrin-B product (yellow band). Cy3 and Cy5 fluorescence are green and red, respectively. (e) UV-Vis absorption spectrum of purified Zn A-porphyrin-B in TE buffer (10 mM Tris, 1 mM EDTA, pH 8), highlighting the characteristic porphyrin Soret peak along with distinct Cy3, Cy5, and DNA absorption.

*Design of the DNA nanostructure for molecular integration*

We designed a DNA nanostructure that precisely positions both the target molecule and AuNPs with a tunable spacing[46]. The design is based on a previously published five-layer DNA origami structure[47], assembled using the p7249 scaffold strand and measuring 45 × 36 × 10 nm (Fig. 2a, b). The central layer of parallel DNA helices forms a continuous scaffold sheet, while the top and bottom layers create bowl-shaped depressions that accommodate 60 nm AuNPs, ensuring stable dimer formation. To enable precise AuNP positioning, each depression displays a mix of 4- and 6-nucleotide single-stranded poly-A sequences which hybridize with thiolated 19-nucleotide poly-T sequences conjugated to the AuNP surface. This tuned interaction ensures optimal AuNP positioning while preventing entropic traps or multiple occupancies.

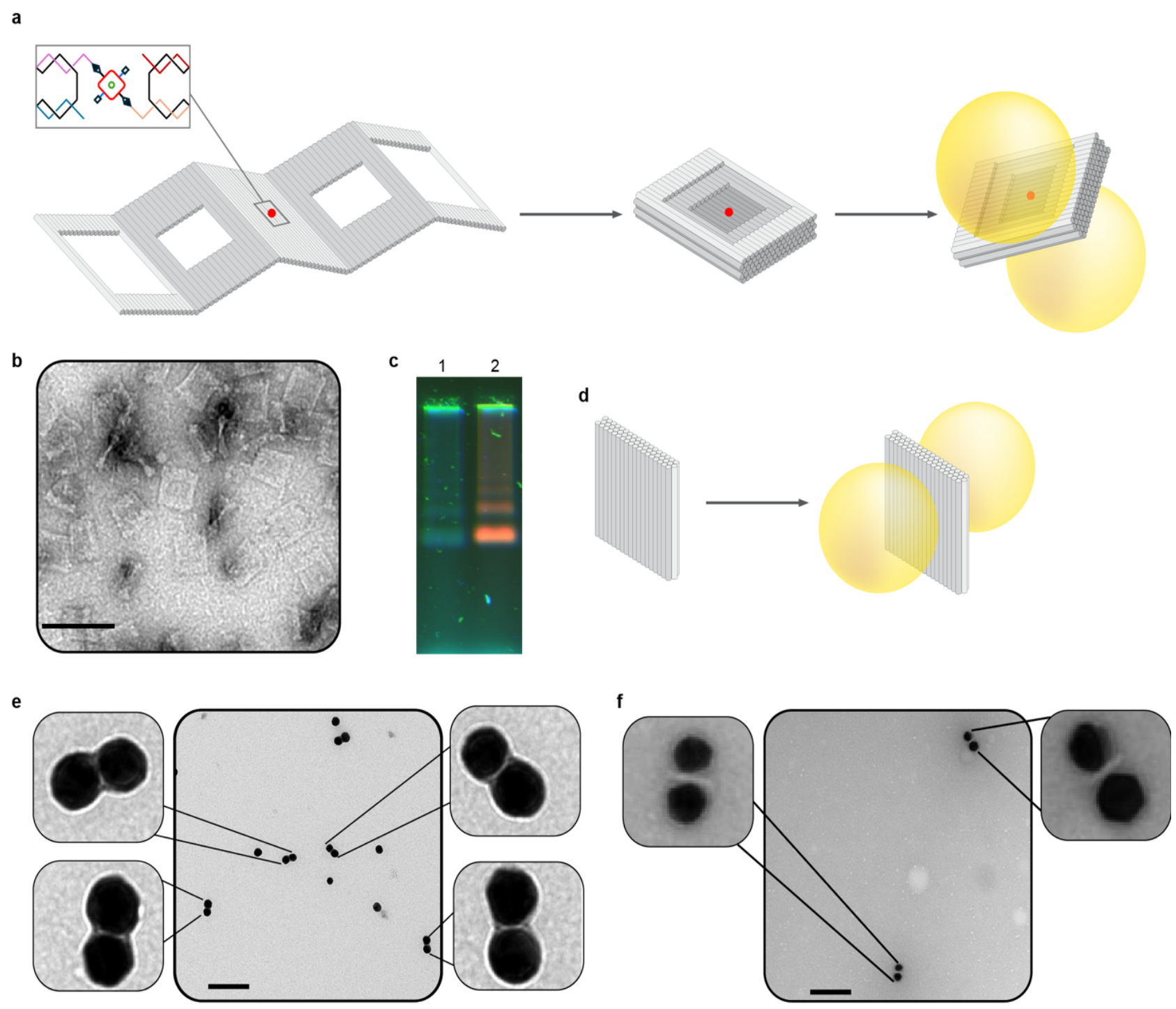

**Fig. 2:** Design and structural characterization of DNA-assembled nanogap architectures for molecular junction formation. (a) Schematic of the five-layer DNA origami nanostructure. Shallow depressions in both surfaces are designed to position 60 nm AuNPs on either side of a single, centrally positioned Zn-porphyrin “target” molecule. The two AuNPs are separated by a single DNA helix layer (2 nm), with the target molecule (red) positioned between them via hybridization of distinct DNA adapters to the DNA origami scaffold. (b) Transmission electron microscopy (TEM) image of the assembled five-layer DNA nanostructure, confirming structural integrity. (c) Merged fluorescence image of a 1.8% agarose gel showing the assembled five-layer structure without (lane 1) and with (lane 2) Zn A-porphyrin-B integration. (d) Alternative nanostructure with a central spacer comprising three layers of DNA helices and no Zn-porphyrin, creating a 6 nm nanogap, serving as a control for assessing the impact of the DNA structure on electronic transport. (e, f) TEM images of assembled structures with AuNPs assembled by (e) one-layer and (f) three-layer spacers, yielding nanogaps of 2 nm and 6 nm, respectively. TEM images show multiple correctly formed structures, highlighting high structural reproducibility. Scale bar: 100 nm.

This structural design serves two key functions. First, it ensures stable AuNP positioning within the depressions, directing them into well-defined locations within the DNA nanostructure. Second, it provides a central binding site for the Zn-porphyrin, ensuring that the molecule spans the nanogap between the AuNPs, enabling efficient electronic coupling. The nanoparticle separation is dictated by the thickness of the central single layer of DNA helices (2 nm), closely matching the 2.4 nm spacing between the Zn-porphyrin's gold coupling groups (Fig. 2a). This precise spacing ensures that the molecule effectively bridges the junction, forming a stable molecular electronic connection.

The Zn-porphyrin was incorporated in the structure through hybridization of the two covalently-linked DNA adapters (A and B) to complementary scaffold domains in the central layer of the structure. The DNA-functionalized target molecule was introduced with the staple strands during origami assembly, such that the porphyrin was integrated into the DNA origami structure during the annealing step (Fig. 2c). To promote incorporation of only one target molecule, the two adapters were designed with differential hybridization strengths, facilitating internal displacement of the weaker adaptor in case of competition between two molecules. This ensured that only one Zn-porphyrin molecule was integrated per package, in the correct orientation. We also assembled a control structure without the target molecule by replacing porphyrin-conjugated adapters with unmodified staple strands.

To investigate the effect of AuNP gap distance on electronic properties, we designed an additional control structure, assembled using the p7249 scaffold strand and featuring a 6 nm nanogap, achieved by stacking three layers of dsDNA instead of a single layer to form the spacer between AuNPs (Fig. 2d). AuNPs were bound by the same mechanism used in the single-layer design, ensuring reproducibility across both architectures.

Correctly assembled Zn-porphyrin-functionalized packages and control structures were purified using agarose gel electrophoresis and analysed via transmission electron microscopy (TEM) before proceeding to electrical measurements. TEM imaging confirmed our robust control over the design (Fig. 2b) allowing for precise tuning of nanoparticle separation (Fig. 2e and f) and highlighting the high reproducibility of the assembly process and the reliable formation of well-defined structures.

### *Device integration and electrical characterization*

The DNA self-assembly stage provides the critical link between molecular-scale components and traditional top-down nanofabrication, enabling reliable electrical interfacing. By depositing the assembled structures onto silicon wafers, the system becomes accessible for scanning electron microscopy (SEM) and atomic force microscopy (AFM) imaging, as well as for the fabrication of electrical connections via electron beam lithography (EBL).

We deposited nanostructures onto silicon wafers with a 100 nm $SiO_2$ insulating layer, pre-patterned with gold registration marks and micron-scale contact pads (Fig. 3a). Correctly assembled devices appear as AuNP dimers, readily distinguished by their scattering contrast in SEM and by their height in AFM.

Using SEM, we identified well-formed structures (e.g., the one highlighted in Fig. 3a) and determined their positions and orientations relative to the registration marks. The wafer was then spin-coated with a positive EBL resist (PMMA), chosen for its chemical compatibility with DNA nanostructures and suitability for low-temperature processing. The assembled structures remain immobilized on the $SiO_2$ surface, as confirmed by SEM imaging of sacrificial wafers post-PMMA deposition (SI).

Using the registration marks and previously established map of device positions, EBL exposure was used to define tracks connecting AuNP contacts to the pre-patterned wafer contact pads. A 120 nm gold layer was deposited over a 5 nm chromium adhesion layer to ensure Ohmic conduction and reliable electrical contact. Finally, the masked gold layer was lifted off, leaving precisely aligned gold contacts (Fig. 3b).

To evaluate the electronic properties of the metalloporphyrin target molecules, we analysed the current-voltage (I-V) characteristics of assembled single-molecule device packages. We collected data across multiple devices at a temperature of 2 K to assess reproducibility and reliability (Fig. 3c). Our primary metric for electrical device performance is the conductance, evaluated as $[\, I(+\delta V) - I(-\delta V)\,] / 2\delta V$ with $\delta V = 300$ mV.

We first studied control devices with no porphyrin and with 2 nm and 6 nm nanogap sizes. The 2 nm gaps exhibited a well-defined tunnelling transport behaviour (Fig. 3d upper panel), consistent with barrier-mediated electron transport at small separations, yielding conductances at small biases of the order of $10^{-7}$ $G_0$. The 6 nm gaps, by contrast, showed significantly lower conductance, typically below $10^{-8}$ $G_0$, and tunnelling-like behaviour was only observable in a small portion of the devices at high voltages (> 500 mV), consistent with the expected exponential dependence of tunnelling resistance on gap distance (SI). The scientific literature includes widely varying reports of the conductivity of DNA, depending on factors including sequence, structure and environmental conditions[48]. Our observations confirm that in our device environment, the DNA structure provides a highly electrically insulating substrate.

We then compared the I-V characteristics of 2 nm devices with and without Zn-porphyrin incorporation (Fig. 3d). Conduction of empty nanogaps followed the Simmons tunnelling dependence[49], with a fitted model (red trace) yielding a gap of 1.2 ± 0.02 nm and a tunnelling barrier of 3.2 ± 0.1 eV. Across a population (n = 79) of empty-gap devices, the conductance distribution was centred at $\log_{10}(G/G_0) \approx -6.8 \pm 0.6$, corresponding to a resistance of ~100 GΩ (Fig. 3e, blue). In this tunnelling model, variations in conductance reflect differences in gap width. Assuming the optical band gap of ~4 eV reported for double-stranded DNA[50] (comparable to the work functions of typical metals) the variation in conductance that we measure is consistent with sub-Angstrom variations of the gap widths.

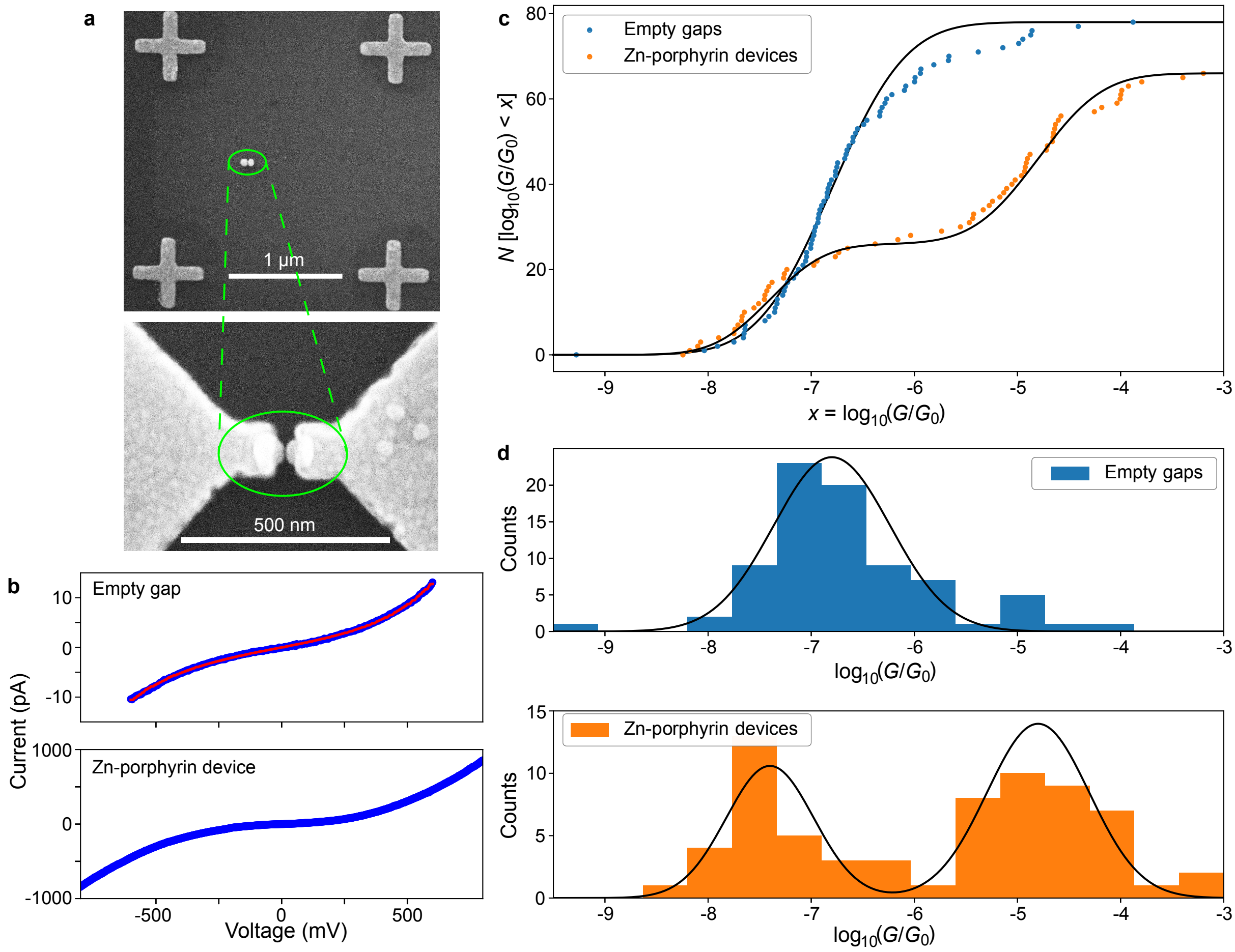

**Figure 3:** Fabrication and electrical characterization of DNA-assembled molecular junctions. (a, upper) SEM image of a DNA-assembled nanostructure deposited on a pre-patterned silicon wafer. Observation of a AuNP dimer indicates a successfully assembled device. (a, lower) Expanded SEM image of the same structure after electron-beam lithography (EBL) and metal deposition. (b) Representative current-voltage (I-V) characteristics for devices without (upper) and with (lower) a Zn-porphyrin molecule bridging the nanogap. The empty nanogap follows the Simmons tunnelling model[49] (red trace), while Zn-porphyrin incorporation enhances conductance, confirming molecular bridging (blue trace). (c) Cumulative distribution of device conductances measured at 2 K with a 300 mV bias voltage. (d) The same distributions as (c) plotted as histograms. Empty nanogaps exhibit high resistance (~100 GΩ), confirming that the DNA scaffold is electronically insulating. Zn-porphyrin integration leads to a bimodal conductance distribution, with a high-conductance peak around $10^{-5}$ $G_0$, consistent with prior STM studies of metalloporphyrin conduction[51]. The ~60% success rate of molecular bridging significantly surpasses electromigrated break-junction techniques, reflecting the degree of control and reproducibility inherent in our DNA-based assembly.

For Zn-porphyrin-containing devices (n = 67), conductance measurements revealed a bimodal distribution, indicating two distinct populations. A low-conductance population ($n_L$ = 27) at $\log_{10}(G/G_0) \approx -7.4 \pm 0.4$ was consistent with the empty nanogap control, while a high-conductance population ($n_H$ = 40) with $\log_{10}(G/G_0) \approx -4.8 \pm 0.5$ showed a significant enhancement in charge transport. This high conductance is consistent with STM break-junction studies on similar Zn-porphyrin derivatives, which reported conductance values of $\log_{10}(G/G_0) \approx -4.5$[51].

The bimodal conductance distribution suggests that Zn-porphyrin incorporation facilitates strong electronic coupling in a subset of devices, forming a stable conductive molecular junction. While about 27/67 ≈ 40% of devices remained in a low-conductance state, probably owing to variations in molecular alignment or weak coupling rather than failure to incorporate the porphyrin, the majority, 40/67 ≈ 60% exhibited a conductance several orders of magnitude higher, confirming robust molecular charge transport. These findings validate Zn-porphyrin as an effective electronic bridge in DNA-assembled molecular junctions, demonstrating the feasibility of using DNA self-assembly for scalable molecular electronics.

*Discussion and summary*

Our findings establish a high-yield, scalable realization of DNA-assembled molecular electronics, addressing a long-standing challenge in the field. The integration of Zn-porphyrin molecular junctions into DNA-templated architectures provides a reproducible approach for molecular-scale electronic devices, bridging the gap between molecular components and traditional top-down nanofabrication while maintaining an electronically insulating structural framework.

We were able to fabricate Zn-porphyrin-incorporated devices with a ~60% yield, far exceeding the 1% or lower success rates typical of electromigrated metal or graphene nanogaps. The platform demonstrates exceptional stability, with devices maintaining reproducible electrical characteristics up to ±0.5 V (corresponding to electric fields exceeding 250 MV/m) and remaining stable through multiple thermal cycles between room temperature and cryogenic conditions (~4 K).

Control structures with different nanogap sizes (2 nm and 6 nm) and no Zn-porphyrin exhibit insulating behaviour, confirming DNA as a suitable substrate for hosting molecular electronic componentry.

Beyond Zn-porphyrin junctions, the modularity of our approach enables broader applications, including tuneable electronic and spintronic properties through alternative metal centres. For example, replacing Zn with Cu is expected to yield a molecular quantum dot incorporating an electron spin-1/2 coupled with the molecular orbital providing itinerancy[27–29], a crucial ingredient for electrical detection of the spin state[20,52].

Our strategy of integrating functional molecular components by hetero-bi-functionalization with programmable DNA adapter sequences provides a natural extension to devices incorporating multiple molecular components with precise positioning and orientation. For example, incorporating a porphyrin ribbon supporting ballistic electronic transport[25] as an electrostatic gate connected to a third AuNP electrode will allow tuning of the quantum dot chemical potential, yielding a molecular single electron transistor, the basic building block for a wide range of potential molecular electronic technologies, both classical and quantum.

Based on our results, we can envisage constructing highly integrated three-dimensional networks incorporating large numbers of molecular components, and interfacing to the outside world via a smaller number of AuNP contacts. This vision will depend on processes for interconnecting (i.e., covalently bonding) components following the assembly of the DNA package, whether though insulating or electrically conductive bonds. A framework for addressing this challenge is offered by DNA-templated synthesis (DTS), pioneered by Liu[53], in which reactants are colocalized through covalently-attached DNA adapters. The specificity of DNA hybridisation is exploited to selectively enhance the effective concentration of reactants, thereby forming products which would arise in negligible yield without templating. Our procedure, described above, for hetero-bi-functionalisation of components is an instance of DTS, and it has been applied to a range of other objectives including, for example, forming sequence-controlled polymers [54,55] and materials discovery[56], demonstrating its versatility.

Our work marks the first indication that DNA-templated molecular assembly can be exploited to reliably fabricate single-molecule electronic devices at scale. We have established a key step in validating the 50-year-old hypothesis that molecular electronic components may be integrated into functional devices, laying the ground for functional nanoscale electronic and quantum technologies, and advancing molecular electronics towards practical implementation.

*References*

1. Joachim, C., Gimzewski, J. K. & Aviram, A. Electronics using hybrid-molecular and mono-molecular devices. *Nature* **408**, 541–548 (2000).
2. Ratner, M. A brief history of molecular electronics. *Nature Nanotechnology* **8**, 378–381 (2013).
3. Williams, R. S., Goswami, S. & Goswami, S. Potential and challenges of computing with molecular materials. *Nature Materials* **23**, 1475–1485 (2024).
4. Gehring, P., Thijssen, J. M. & van der Zant, H. S. J. Single-molecule quantum-transport phenomena in break junctions. *Nature Reviews Physics* **1**, 381–396 (2019).
5. Lörtscher, E. Wiring molecules into circuits. *Nature Nanotechnology* **8**, 381–384 (2013).
6. Pascual, J. I. *et al.* Seeing molecular orbitals. *Chem Phys Lett* **321**, 78–82 (2000).
7. Lu, X., Grobis, M., Khoo, K. H., Louie, S. G. & Crommie, M. F. Spatially Mapping the Spectral Density of a Single Molecule. *Phys Rev Lett* **90**, 4 (2003).
8. Joachim, C. & Ratner, M. A. Molecular electronics: Some views on transport junctions and beyond. *Proc Natl Acad Sci U S A* **102**, 8801–8808 (2005).
9. Seeman, N. C. & Sleiman, H. F. DNA nanotechnology. *Nature Reviews Materials* **3**, 1–23 (2017).
10. Dey, S. *et al.* DNA origami. *Nature Reviews Methods Primers* **1**, 13 (2021).
11. Jurow, M., Schuckman, A. E., Batteas, J. D. & Drain, C. M. Porphyrins as molecular electronic components of functional devices. *Coord Chem Rev* **254**, 2297–2310 (2010).
12. Moore, G. E. Cramming more components onto integrated circuits, Reprinted from Electronics, volume 38, number 8, April 19, 1965, pp.114 ff. *IEEE Solid-State Circuits Society Newsletter* **11**, 33–35 (2009).
13. Aviram, A. & Ratner, M. A. Molecular rectifiers. *Chem Phys Lett* **29**, 277–283 (1974).
14. Leuenberger, M. N. & Loss, D. Quantum computing in molecular magnets. *Nature* **410**, 789–793 (2001).

15. Tejada, J., Chudnovsky, E. M., Del Barco, E., Hernandez, J. M. & Spiller, T. P. Magnetic qubits as hardware for quantum computers. *Nanotechnology* **12**, 181 (2001).

16. Wasielewski, M. R. *et al.* Exploiting chemistry and molecular systems for quantum information science. *Nature Reviews Chemistry* **4**, 490–504 (2020).

17. Atzori, M. & Sessoli, R. The Second Quantum Revolution: Role and Challenges of Molecular Chemistry. *J Am Chem Soc* **141**, 11339–11352 (2019).

18. Gaita-Ariño, A., Luis, F., Hill, S. & Coronado, E. Molecular spins for quantum computation. *Nature Chemistry* **11**, 301–309 (2019).

19. Li, M. *et al.* Atomically precise construction of uniform single-molecule junctions for molecular electronics. *Nature Communications* **16**, 1–11 (2025).

20. Vincent, R., Klyatskaya, S., Ruben, M., Wernsdorfer, W. & Balestro, F. Electronic read-out of a single nuclear spin using a molecular spin transistor. *Nature* **488**, 357–360 (2012).

21. Thiele, S. *et al.* Electrically driven nuclear spin resonance in single-molecule magnets. *Science* **344**, 1135–1138 (2014).

22. Godfrin, C. *et al.* Operating Quantum States in Single Magnetic Molecules: Implementation of Grover's Quantum Algorithm. *Phys Rev Lett* **119**, 187702 (2017).

23. Chen, Z. *et al.* Connections to the Electrodes Control the Transport Mechanism in Single-Molecule Transistors. *Angewandte Chemie International Edition* **63**, e202401323 (2024).

24. Ornago, L. *et al.* Influence of Peripheral Alkyl Groups on Junction Configurations in Single-Molecule Electronics. *Journal of Physical Chemistry C* **128**, 1413–1422 (2024).

25. Deng, J. R., González, M. T., Zhu, H., Anderson, H. L. & Leary, E. Ballistic Conductance through Porphyrin Nanoribbons. *J Am Chem Soc* **146**, 3651–3659 (2024).

26. Chen, Z. *et al.* Phase-Coherent Charge Transport through a Porphyrin Nanoribbon. *J Am Chem Soc* **145**, 15265–15274 (2023).

27. Gimeno, I. *et al.* Broad-band spectroscopy of a vanadyl porphyrin: a model electronuclear spin qudit. *Chem Sci* **12**, 5621–5630 (2021).

28. Yu, C. J., Krzyaniak, M. D., Fataftah, M. S., Wasielewski, M. R. & Freedman, D. E. A concentrated array of copper porphyrin candidate qubits. *Chem Sci* **10**, 1702–1708 (2019).

29. Kopp, S. M. *et al.* Charge and Spin Transfer Dynamics in a Weakly Coupled Porphyrin Dimer. *J Am Chem Soc* **146**, 21476–21489 (2024).

30. Seeman, N. C. Nucleic acid junctions and lattices. *J Theor Biol* **99**, 237–247 (1982).

31. Chen, J. & Seeman, N. C. Synthesis from DNA of a molecule with the connectivity of a cube. *Nature* **350**, 631–633 (1991).

32. Goodman, R. P. *et al.* Rapid chiral assembly of rigid DNA building blocks for molecular nanofabrication. *Science* **310**, 1661–1665 (2005).

33. Winfree, E., Liu, F., Wenzler, L. A. & Seeman, N. C. Design and self-assembly of two-dimensional DNA crystals. *Nature* **394**, 539–544 (1998).

34. Zheng, J. *et al.* From molecular to macroscopic via the rational design of a self-assembled 3D DNA crystal. *Nature* **461**, 74–77 (2009).

35. Ke, Y., Ong, L. L., Shih, W. M. & Yin, P. Three-dimensional structures self-assembled from DNA bricks. *Science* **338**, 1177–1183 (2012).

36. Rothemund, P. W. K. Folding DNA to create nanoscale shapes and patterns. *Nature* **440**, 297–302 (2006).

37. Douglas, S. M. *et al.* Self-assembly of DNA into nanoscale three-dimensional shapes. *Nature* **459**, 414–418 (2009).

38. Wagenbauer, K. F., Sigl, C. & Dietz, H. Gigadalton-scale shape-programmable DNA assemblies. *Nature* **552**, 78–83 (2017).

39. Zhang, F. *et al.* Complex wireframe DNA origami nanostructures with multi-arm junction vertices. *Nature Nanotechnology* **10**, 779–784 (2015).

40. Tikhomirov, G., Petersen, P. & Qian, L. Fractal assembly of micrometre-scale DNA origami arrays with arbitrary patterns. *Nature* **552**, 67–71 (2017).

41. Wintersinger, C. M. *et al.* Multi-micron crisscross structures grown from DNA-origami slats. *Nature Nanotechnology* **18**, 281–289 (2022).

42. Madsen, M. & Gothelf, K. V. Chemistries for DNA Nanotechnology. *Chem Rev* **119**, 6384–6458 (2019).

43. Gopinath, A., Miyazono, E., Faraon, A. & Rothemund, P. W. K. Engineering and mapping nanocavity emission via precision placement of DNA origami. *Nature* **535**, 401–405 (2016).

44. Martynenko, I. V. *et al.* Site-directed placement of three-dimensional DNA origami. *Nature Nanotechnology* **18**, 1456–1462 (2023).

45. Helmi, S. & Turberfield, A. J. Template-directed conjugation of heterogeneous oligonucleotides to a homobifunctional molecule for programmable supramolecular assembly. *Nanoscale* **14**, 4463–4468 (2022).

46. Dadosh, T. *et al.* Measurement of the conductance of single conjugated molecules. **436**, 677–680 (2005).

47. Simoncelli, S. *et al.* Quantitative Single-Molecule Surface-Enhanced Raman Scattering by Optothermal Tuning of DNA Origami-Assembled Plasmonic Nanoantennas. *ACS Nano* **10**, 9809–9815 (2016).

48. Genereux, J. C. & Barton, J. K. Mechanisms for DNA charge transport. *Chem Rev* **110**, 1642–1662 (2010).

49. Simmons, J. G. Generalized Formula for the Electric Tunnel Effect between Similar Electrodes Separated by a Thin Insulating Film. *J Appl Phys* **34**, 1793–1803 (1963).

50. Schimelman, J. B. *et al.* Optical properties and electronic transitions of DNA oligonucleotides as a function of composition and stacking sequence. *Physical Chemistry Chemical Physics* **17**, 4589–4599 (2015).

51. Sedghi, G. *et al.* Single molecule conductance of porphyrin wires with ultralow attenuation. *J Am Chem Soc* **130**, 8582–8583 (2008).

52. Hanson, R., Kouwenhoven, L. P., Petta, J. R., Tarucha, S. & Vandersypen, L. M. K. Spins in few-electron quantum dots. *Rev Mod Phys* **79**, 1217–1265 (2007).

53. Li, X. & Liu, D. R. DNA-Templated Organic Synthesis: Nature's Strategy for Controlling Chemical Reactivity Applied to Synthetic Molecules. *Angewandte Chemie International Edition* **43**, 4848–4870 (2004).

54. Lutz, J. F., Ouchi, M., Liu, D. R. & Sawamoto, M. Sequence-controlled polymers. *Science* **341**, (2013).

55. McKee, M. L. *et al.* Programmable one-pot multistep organic synthesis using DNA junctions. *J Am Chem Soc* **134**, 1446–1449 (2012).

56. O'Reilly, R. K., Turberfield, A. J. & Wilks, T. R. The Evolution of DNA-Templated Synthesis as a Tool for Materials Discovery. *Acc Chem Res* **50**, 2496–2509 (2017).

Supporting Information: Programmable self-assembly for scalable molecular electronic architectures

Seham Helmi* [1,2], Junjie Liu* [3,4], Keith G Andrews [5], Robert Schreiber [3], Jonathan Bath [1,3], Harry L Anderson [2], Andrew J Turberfield [1,3], Arzhang Ardavan [3]

1 The Kavli Institute for Nanoscience Discovery, University of Oxford, Oxford OX1 3QU, UK

2 Department of Chemistry, University of Oxford, Mansfield Road, Oxford OX1 3TA, UK

3 The Clarendon Laboratory, Department of Physics, University of Oxford, Oxford OX1 3PU, UK

4 School of Physical and Chemical Sciences, Queen Mary University of London, London E1 4NS, UK

5 Department of Chemistry, Durham University, Lower Mount Joy, South Rd, Durham, DH1 3LE, UK

## 1. General information on porphyrin synthesis

Dry toluene, $CHCl_3$, $CH_2Cl_2$, THF, $Et_3N$, *N,N*-diisopropylamine and pyridine were obtained from the solvent drying system MBraun MBSPS-5-BenchTop under nitrogen atmosphere ($H_2O$ content < 20 ppm as determined by Karl-Fischer titration). All other solvents were used as commercially supplied. Flash chromatography was carried out using $SiO_2$ (60 Å, 230–400 mesh) under positive pressure. Analytical thin-layer chromatography was carried out on aluminium-backed silica gel 60 F254 plates. Petroleum ether (PE) 40−60 °C was used unless specified otherwise. All UV-vis-NIR spectra were recorded in solution using a Perkin-Lambda 20 spectrometer (1 cm path length fused silica cell). Chloroform (containing ca. 0.5% ethanol as stabilizer) was used for all UV measurements without any further purification (HPLC grade). $^{1}H$ and $^{13}C$ NMR spectra were recorded at 298 K using a Bruker AVIII HD 400, a Bruker AVII 500. $^{1}H$ and $^{13}C$ NMR spectra are reported in ppm; coupling constants are given in Hertz, to the nearest 0.1 Hz. $CDCl_3$ was calibrated to residual $CHCl_3$ at 7.26 ppm. MALDI-ToF spectra were measured using a Brucker Autoflex spectrometer utilizing dithranol or *trans*-2-[3-(4-*tert*-butylphenyl)-2-methyl-2-propenylidene]malononitrile (DCTB) as a matrix. Size exclusion chromatography (SEC) was carried out using Bio-Rad Bio-Beads S-X1 (40–80 μm bead size). Semi-preparative GPC was carried out on a Shimadzu Recycling GPC system equipped with a LC-20 AD pump, SPD-20A UV detector and a set of JAIGEL 3H (20 × 600 mm) and JAIGEL 4H (20 × 600 mm) columns in toluene + 1% pyridine as the eluent at a flow rate of 3.5 mL/min.

## 2. The synthesis of zinc porphyrin

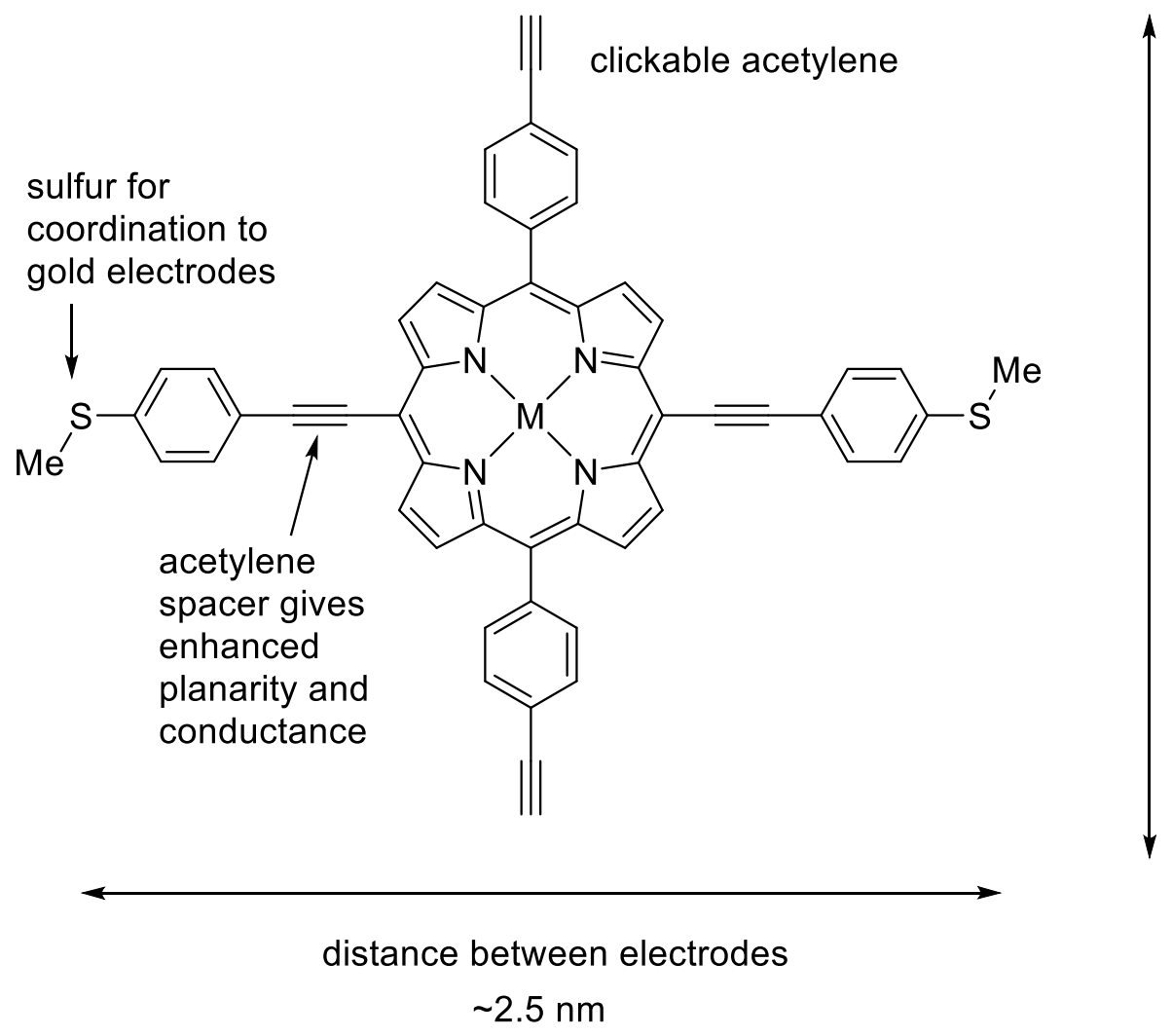

**Fig 1.** Porphyrin geometric design notes (distances measured from geometry calculated using MOPAC/PM7).

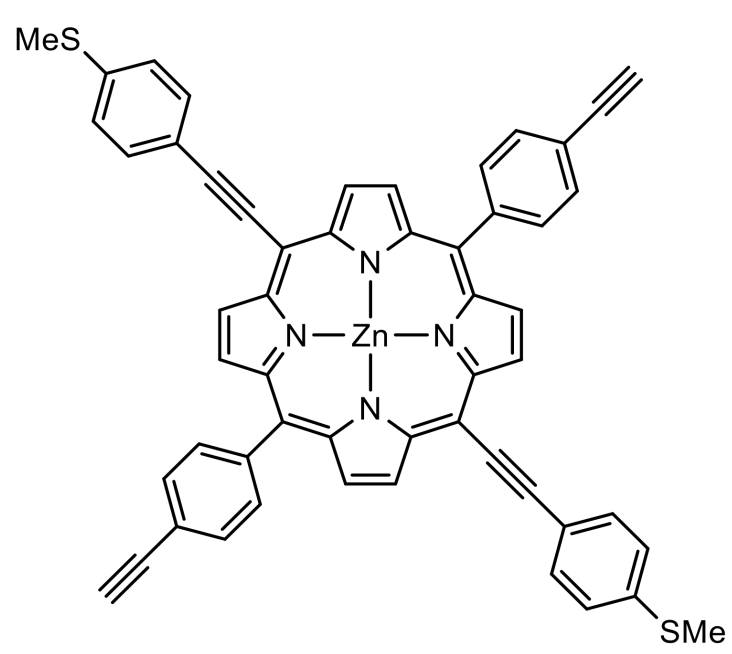

**Fig 2.** The studied zinc porphyrin.

## 2.1 Triisopropyl((4-(methylthio)phenyl)ethynyl)silane (3)

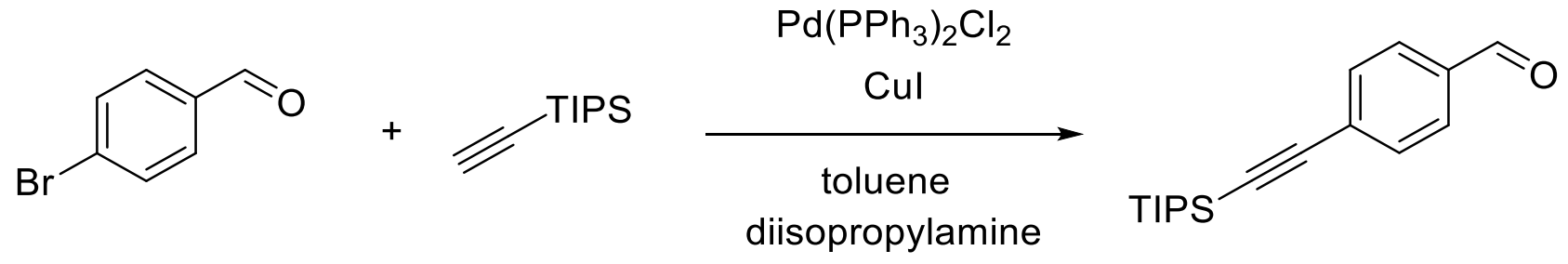

3

4-Bromobenzaldehyde (4.00 g, 21.6 mmol), bis(triphenylphosphine)palladium(II) dichloride (445 mg, 0.648 mmol) and copper(I) iodide (206 mg, 1.08 mmol) were dissolved in dry toluene (40 mL) and dry diisopropylamine (10 mL) and the solution was degassed with three argon/vacuum cycles. TIPS-acetylene (7.27 mL, 32.4 mmol) was added. The reaction mixture was stirred at 20 °C overnight, then filtered through a silica pad, eluting with petrol/dichloromethane 1:1, and purified by further flash column chromatography, eluting with 1:2 petrol/dichloromethane to give a pale yellow oil (5.78 g, 93%) with data consistent with the literature:[1] **$^1$H NMR** (400 MHz, chloroform-*d*) δ 10.00 (s, 1H), 7.84 – 7.80 (m, 2H), 7.64 – 7.59 (m, 2H), 1.21 – 1.07 (m, 21H); **$^{13}$C NMR** (101 MHz, chloroform-*d*) δ 191.4, 135.5, 132.5, 129.7, 129.4, 105.9, 95.8, 18.6, 11.3.

## 2.2 Triisopropyl((4-(methylthio)phenyl)ethynyl)silane (4)

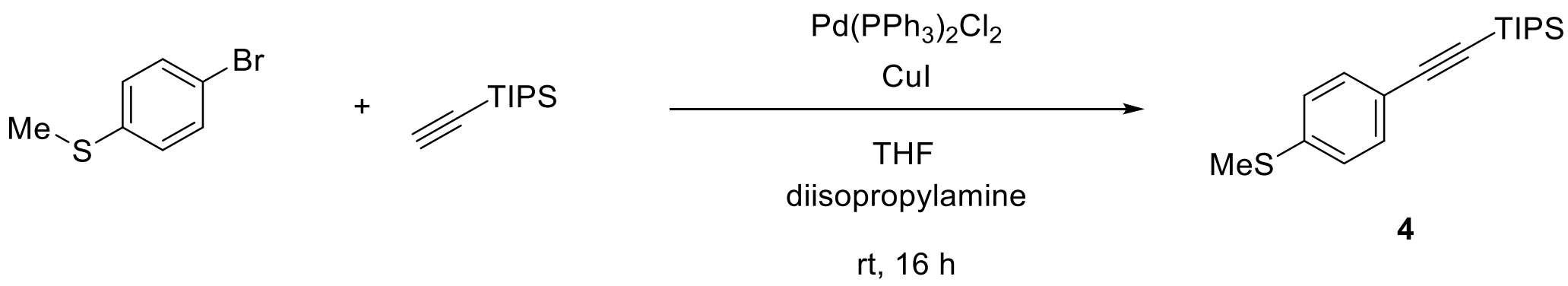

4

(4-Bromophenyl)(methyl)sulfane (813 mg, 4.00 mmol), bis(triphenylphosphine)palladium(II) dichloride (84.3 mg mg, 0.120 mmol) and copper(I) iodide (38.1 mg, 0.200 mmol) were dissolved in dry toluene (7.5 mL) and dry diisopropylamine (1.8 mL) and the solution was degassed with three argon/vacuum cycles. TIPS-acetylene (1.35 mL, 6.00 mmol) was added. The reaction mixture was stirred at 20 °C overnight, then filtered through a silica pad, eluting with dichloromethane, and resubmitted with fresh catalyst in dry THF (7.5 mL) and dry diisopropylamine (1.8 mL) at 50 °C for 14 h. The reaction was deemed complete by $^1$H-NMR, filtered through a silica pad, eluting with dichloromethane, and purified by further flash column chromatography, eluting with 9:1 petrol/dichloromethane to give a pale yellow oil (1.20 g, 98%) with data consistent with the literature:[2] **$^1$H NMR** (400 MHz, chloroform-*d*) δ 7.40 – 7.36 (m, 2H), 7.18 – 7.13 (m, 2H), 2.48 (s, 3H), 1.12 (s, 21H).

## 2.3 [5,15-Bis(4-((triisopropylsilyl)ethynyl)phenyl)porphyrin]zinc(II) (6)

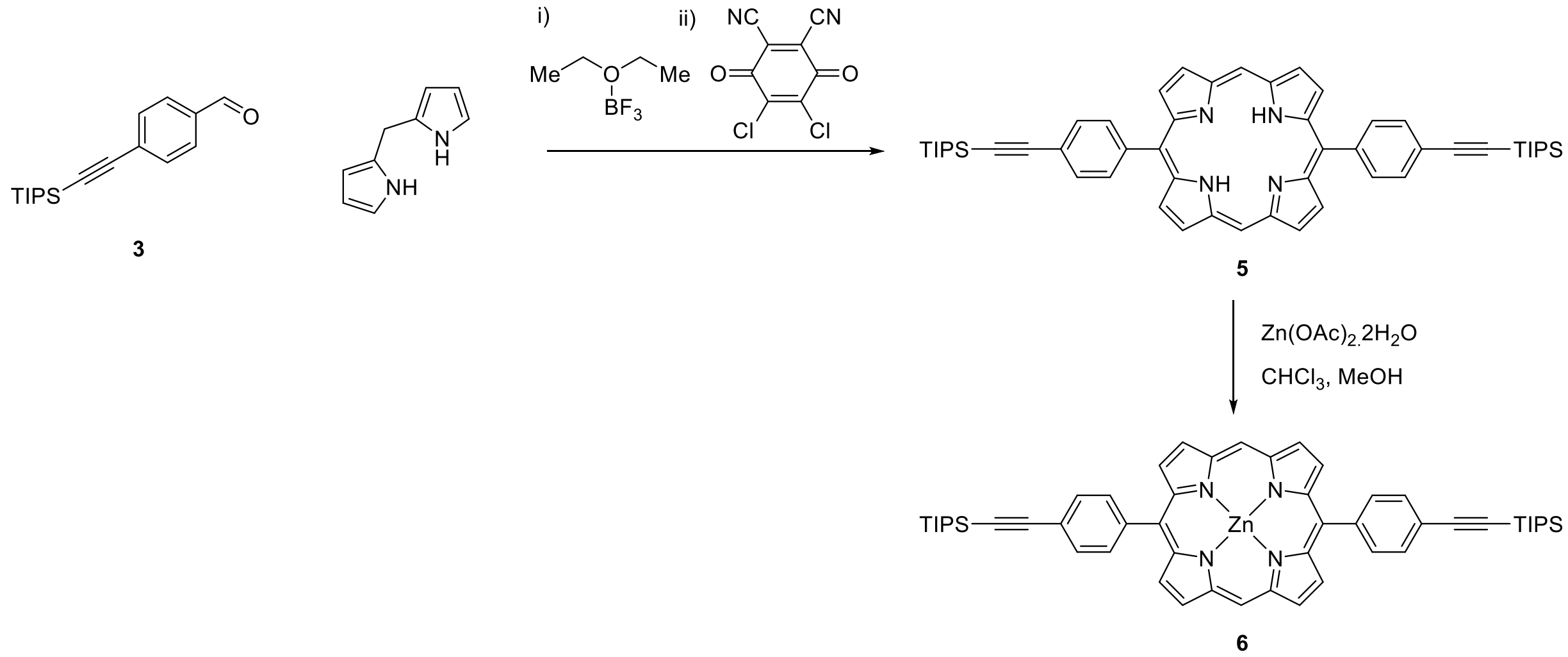

4-((Triisopropylsilyl)ethynyl)benzaldehyde **3** (1.84 g, 6.43 mmol) and dipyrromethane (0.940 g, 6.43 mmol) were added to a dry flask (1 L). Dry dichloromethane (360 mL) was added and the flask wrapped in foil. The stirring solution was degassed by bubbling argon through it for 30 min, and then boron trifluoride etherate (262 µL, 2.12 mmol) was added. The mixture was stirred in the dark for 45 min before the addition of 2,3-dichloro-5,6-dicyano-1,4-benzoquinone (2.19 g, 9.64 mmol). The reaction mixture was stirred for a further 10 min under air, and then poured directly through a silica plug (ø = 6 cm, height = 10 cm) collecting the red solution. The eluent was concentrated to ~ 100 mL and the silica filtration repeated 4 further times with fresh silica, eluting with dichloromethane, until no further black material stained the silica. The final deep red solution was concentrated to dryness yielding the freebase porphyrin **5** as a red solid (770 mg, 29%); **$^1$H NMR** (400 MHz, $CDCl_3$) δ 10.32 (s, 2H), 9.40 (d, *J* = 4.6 Hz, 4H), 9.07 (d, *J* = 4.6 Hz, 4H), 8.22 (d, *J* = 8.3 Hz, 4H), 7.93 (d, *J* = 8.3 Hz, 4H), 1.30 – 1.23 (m, 42H). **MALDI-TOF (R+)** 822.396 ($C_{54}H_{62}N_4Si_2$ (M)$^+$ requires 822.45).

The crude freebase porphyrin **5** (50.0 mg, 0.0607 mmol) was dissolved in chloroform (7.0 mL) and a solution of zinc(II) diacetate dihydrate (28.0 mg, 0.152 mmol) in methanol solution (1 mL) was added. The mixture was stirred at ambient temperature (~20 ° C) for 1 h, and passed through a short silica plug, eluting with $CH_2Cl_2$, collecting the pink/red liquid. After concentration, a dark red/burgundy solid remained, and was recrystallised by layer addition of methanol (~30 ml) onto a minimum dichloromethane solution of crude porphyrin (16 h, 4 °C). The pink/purple solid was collected by filtration (50.0 mg, 82%) to give the desired porphyrin with data consistent with the literature:[3] **$^1$H NMR** (400 MHz, chloroform-d) δ 10.20 (s, 2H), 9.36 (d, *J* = 4.4 Hz, 4H), 9.04 (d, *J* = 4.4 Hz, 4H), 8.20 – 8.16 (m, 4H), 7.90 – 7.85 (m, 4H), 1.29 – 1.23 (m, 42H). **MALDI-TOF (R+)** 884.24 ($C_{54}H_{60}N_4Si_2Zn$ (M)$^+$ requires 884.36).

## 2.4 [5,15-Dibromo-10,20-bis(4-((triisopropylsilyl)ethynyl)phenyl)porphyrin]zinc(II) (7)

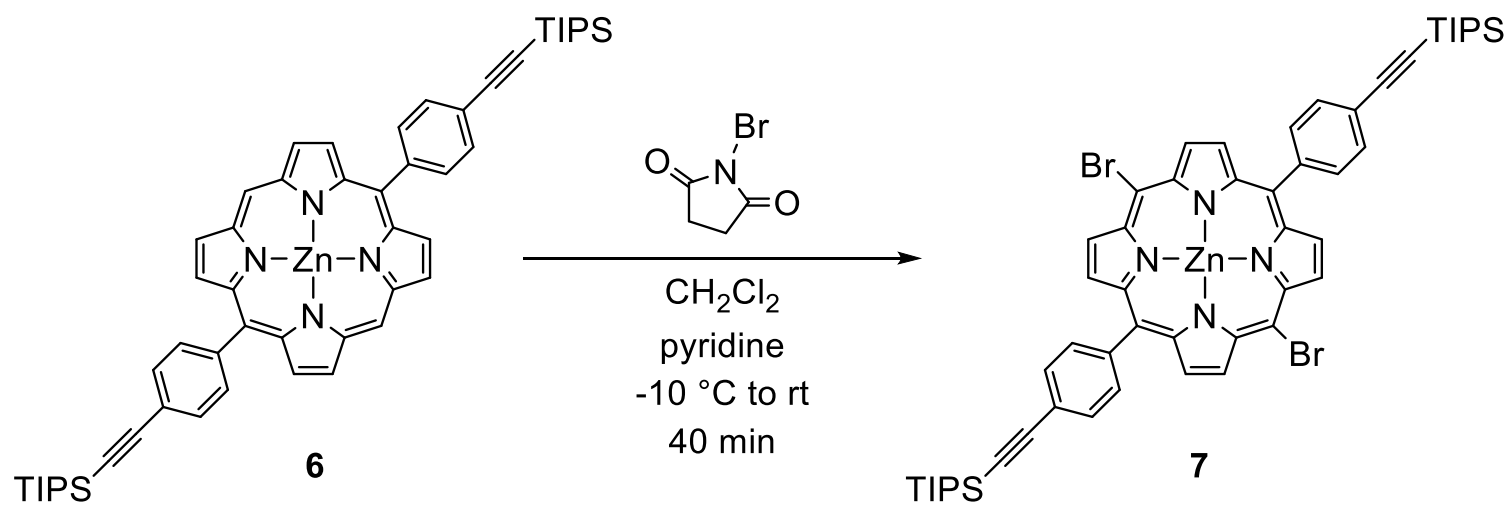

Zinc-porphyrin **6** (50.0 mg, 0.0564 mmol) was dissolved in dry dichloromethane (7.0 mL) and pyridine (70 μL) under argon and cooled to –40 °C with a dry ice acetone bath in the dark. Solid *N*-bromosuccinide (recrystallised from water, 22.1 mg, 0.124 mmol) was added and the reaction mixture was stirred while warming to 20 °C over 20 min, and for 20 min further at 20 °C. After this time, the blue/green reaction was quenched with acetone (0.5 mL) and stirred for a further minute before filtering through a short silica pad, eluting with dichloromethane. The solution was concentrated and dried to give the desired dibromide (50.5 mg, 86%). **$^1$H NMR** (400 MHz, chloroform-*d*) δ 9.65 (d, *J* = 4.6 Hz, 4H), 8.84 (d, *J* = 4.6 Hz, 4H), 8.09 – 8.05 (m, 4H), 7.88 – 7.83 (m, 4H), 1.28 – 1.23 (m, 42H). **MALDI-TOF (R+)** 1039.95 ($C_{54}H_{58}N_4Si_2Zn$ $(M)^+$ requires 1040.19).

## 2.5 (4-Ethynylphenyl)(methyl)sulfane (8)

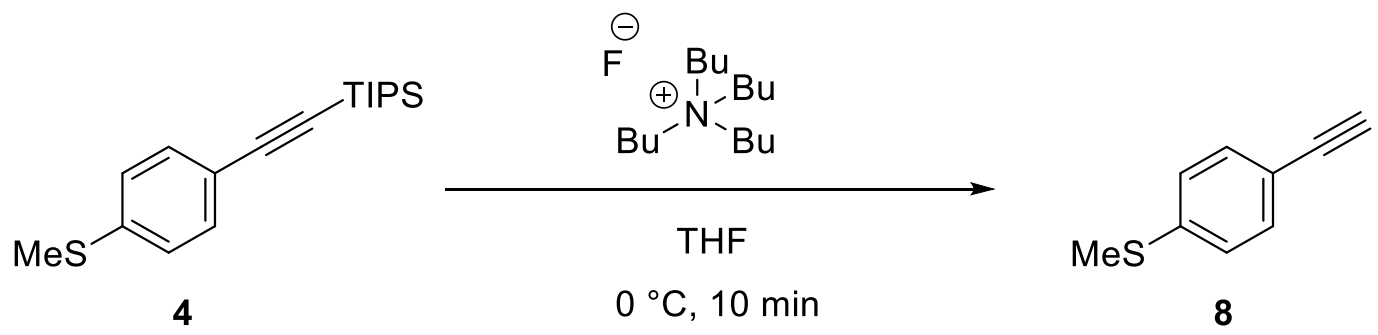

TIPS-acetylene **4** (400 mg, 1.31 mmol) was dissolved in dry THF (20 mL) and added was tetrabutylammonium fluoride (TBAF) solution (1.0 M in THF, 4.0 mL, 4.0 mmol) with stirring at –10 °C. The reaction mixture became instantly dark red, and after 10 min was diluted with diethyl ether (60 mL) and quenched with brine, becoming yellow. The organic layer was washed with further brine and dried over magnesium sulfate, filtered and concentrated. The crude diyne (still containing the silyl-containing byproduct(s) was used directly in the Sonogashira reaction (assuming an approximate mass fraction of 50%). Data from the crude spectrum were consistent with the previously reported[4] deprotected acetylene: **$^1$H NMR** (400 MHz, chloroform-*d*) δ 7.44 – 7.36 (m, 2H), 7.21 – 7.14 (m, 2H), 3.06 (s, 1H), 2.48 (s, 3H).

## [5,15-bis((4-(methylthio)phenyl)ethynyl)-10,20-bis(4-((triisopropylsilyl)ethynyl)phenyl)porphyrin]zinc(II) (9)

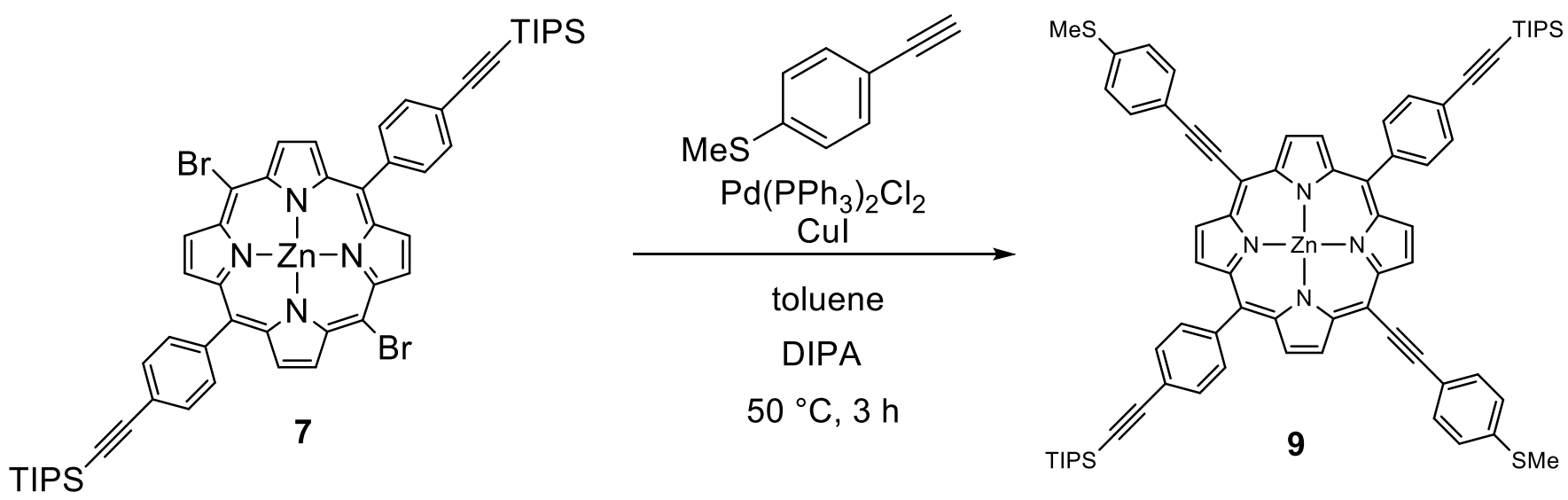

Dibromo-porphyrin **7** (50.0 mg, 0.0479 mmol), bis(triphenylphosphine)palladium(II) dichloride (1.7 mg, 0.0024 mmol) and the crude, freshly deprotected (4-ethynylphenyl)(methyl)sulfane **8** (17.7 mg, 0.120 mmol) were dissolved in a mixture of dry toluene (3.5 mL), dry diisopropylamine (0.9 mL) and dry pyridine (70 μL). The mixture was freeze-pump-thaw-degassed (3 cycles) and then heated to 50 °C. Copper(I) iodide (0.9 mg, 0.005 mmol) was added under a stream of argon, and the reaction stirred for 3 h, at which point the reaction was complete by TLC (1:4 EtOAc/petrol, conversion to single slower running greener spot). The reaction was cooled, and passed through a silica pad, eluting with dichloromethane (1% v/v pyridine). The desired was purified by passing through a size-exclusion column (toluene, 1% pyridine) to give a green solid (51.0 mg, 90%); **$^1$H NMR** (400 MHz, chloroform-d) δ 9.68 (d, *J* = 4.6 Hz, 4H), 8.82 (d, *J* = 4.6 Hz, 4H), 8.12 – 8.08 (m, 4H), 7.94 – 7.89 (m, 4H), 7.89 – 7.85 (m, 4H), 7.44 – 7.38 (m, 4H), 2.59 (s, 6H), 1.29 – 1.24 (m, 42H); **MALDI-TOF (R+)** 1176.97 ($C_{54}H_{32}N_4Si_2Zn$ $(M)^+$ requires 1176.40).

## 2.6 Zinc-porphyrin

**[5,15-bis(4-ethynylphenyl)-10,20-bis((4-(methylthio)phenyl)ethynyl)porphyrin]zinc(II) (1)**

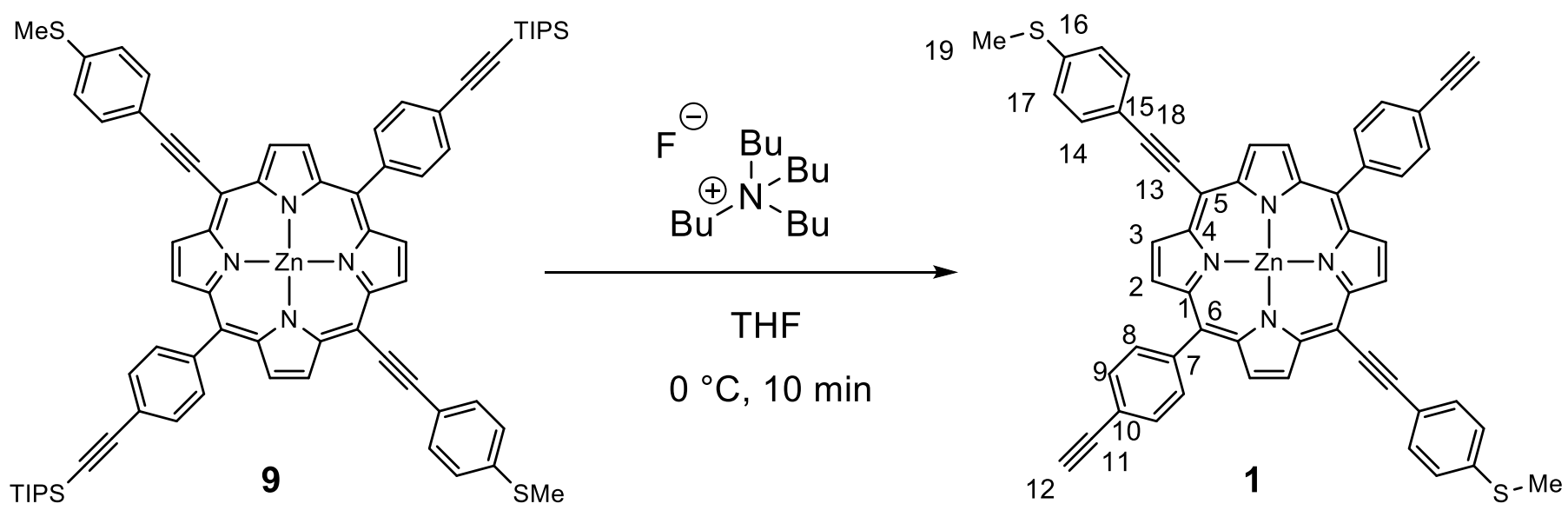

Di-TIPS protected porphyrin **9** (11.1 mg, 0.00941 mmol) was dissolved in dry THF (1 mL) and cooled to 0 °C. An excess of TBAF (1.0 M in THF, 110 μL, 0.11 mmol) was added and the reaction stirred for 10 min, warming to 20 °C. The reaction was judged complete by TLC (20% EtOAc/petrol) and diluted with ethyl acetate and washed twice with brine. The organics were dried over magnesium sulfate, filtered and concentrated. The crude porphyrin was purified further by size-exclusion chromatography (toluene, 1% v/v pyridine) to yield the desired as a green solid (8.1 mg, 99%) with data: **$^1$H NMR** (400 MHz, chloroform-*d*, 1% pyridine-*d*6) δ 9.69 (d, *J* = 4.6 Hz, 4H$^3$), 8.81 (d, *J* = 4.6 Hz, 4H$^2$), 8.18 – 8.10 (m, 4H$^8$), 7.91 (d, *J* = 8.3 Hz, 4H$^{14}$), 7.88 (d, *J* = 8.1 Hz, 4H$^9$), 7.45 – 7.37 (m, 4H$^{17}$), 3.33 (s, 2H$^{12}$), 2.59 (s, 6H$^{19}$); **$^{13}$C NMR** (126 MHz, chloroform-*d*, 1% pyridine-*d*6) δ 152.0 (C4), 149.6 (C1), 143.4 (C7), 139.4 (C16), 134.4 (C8), 132.2 (C2), 131.8 (C14), 130.9 (C3), 130.3 (C9), 126.2 (C17), 121.6 (C5), 121.2 (C10), 120.5 (C15), 101.3 (C6), 96.2 (C18), 93.3 (C13), 83.8 (C11), 78.1 (C12), 15.5 (C19); **MALDI-TOF (R+)** 864.5 ($C_{54}H_{32}N_4S_2Zn$ (M)$^+$ requires 864.1); **$\lambda_{max}$** ($CHCl_3$, 1% pyr) / nm (log ε) 459 (5.53), 614 (4.01), 668 (4.77).

3. **Test reaction to demonstrate click reactivity of the Zn-porphyrin under conditions approximating those to be used for the DNA conjugation experiments**

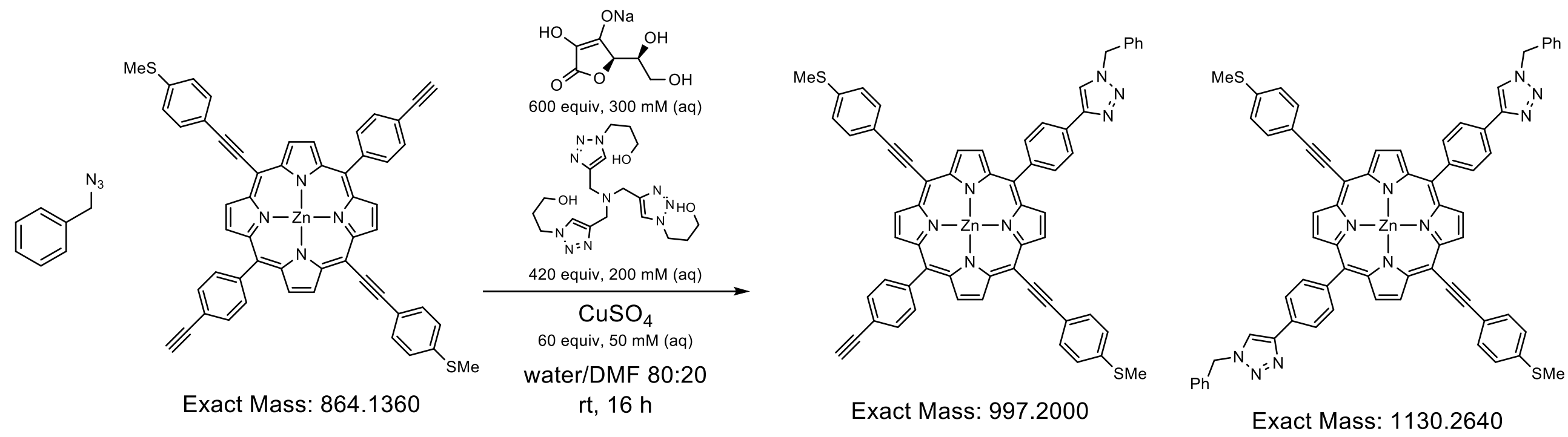

Two vials were prepared: vial A was designated as aqueous, and vial B as organic.

VIAL A: Sodium L-ascorbate (11.9 mg, 60.0 μmol) was dissolved in water (200 μL), and an aliquot (20 μL, 6.0 μmol) added to vial A. THPTA (tris(3-hydroxypropyltriazolylmethyl)amine) (21 μL of a 200 mM aqueous solution) was added to vial A. An aqueous solution of copper sulfate (5 μL, 2.5 M stock, 1270 equiv.) was added to vial A. Vial A was diluted with water (800 μL).

VIAL B: Benzyl azide (5.0 μL, 40 μmol) was dissolved in DMF (1 mL) and an aliquot (2.7 μL, 10 equiv.) added to vial B. Zn-porphyrin **1** (1 μL of a 10 mM solution in DMF, 1 equiv.) was added to vial B. Vial B was diluted with DMF (200 μL).

Finally, the solution in vial B was added to vial A with stirring. No degassing or oxygen exclusion precautions were taken. The solution was stirred slowly for 16 h at 20 °C. An aliquot was removed, diluted with dichloromethane, filtered through magnesium sulfate to remove the water, and blown down with nitrogen to remove the organic solvent. The residue was analysed by MALDI-ToF (mass spectrometry) using a DCTB matrix.

**Summary of the click test reaction conditions:** Sodium L-ascorbate (600 equiv.), THPTA (420 equiv.), benzyl azide (20 equiv., 10 fold excess wrt acetylene, final concentration is 0.2 μM), porphyrin **1** (1 equiv., final concentration is 0.01 μM), copper sulfate (13 mM, 1270 equivs) in a total volume of ~1 mL.

**3.1 Result:** MALDI-TOF analysis shows full conversion to the bis(triazole)porphyrin (MW = 1130) (the repeating minor peaks are likely salt clusters).

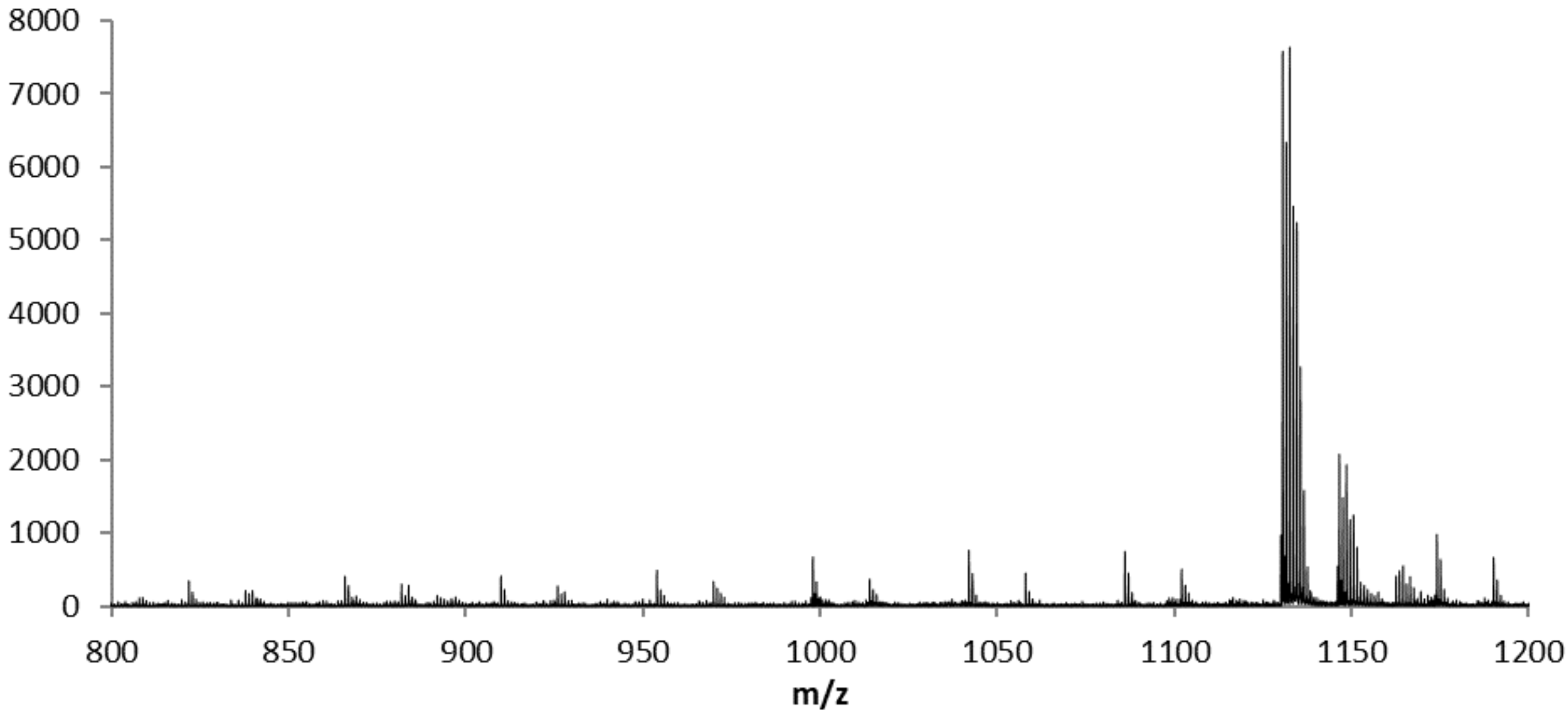

**Conclusion:** The reaction conditions used are very efficient for the click chemistry between porphyrin **1** and alkynes under the DMF/water conditions.

### 3.2 Supporting Spectra

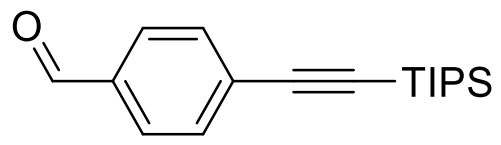

**3**

$^1$H-NMR ($CDCl_3$, 400 MHz) **triisopropyl((4-(methylthio)phenyl)ethynyl)silane (3)**

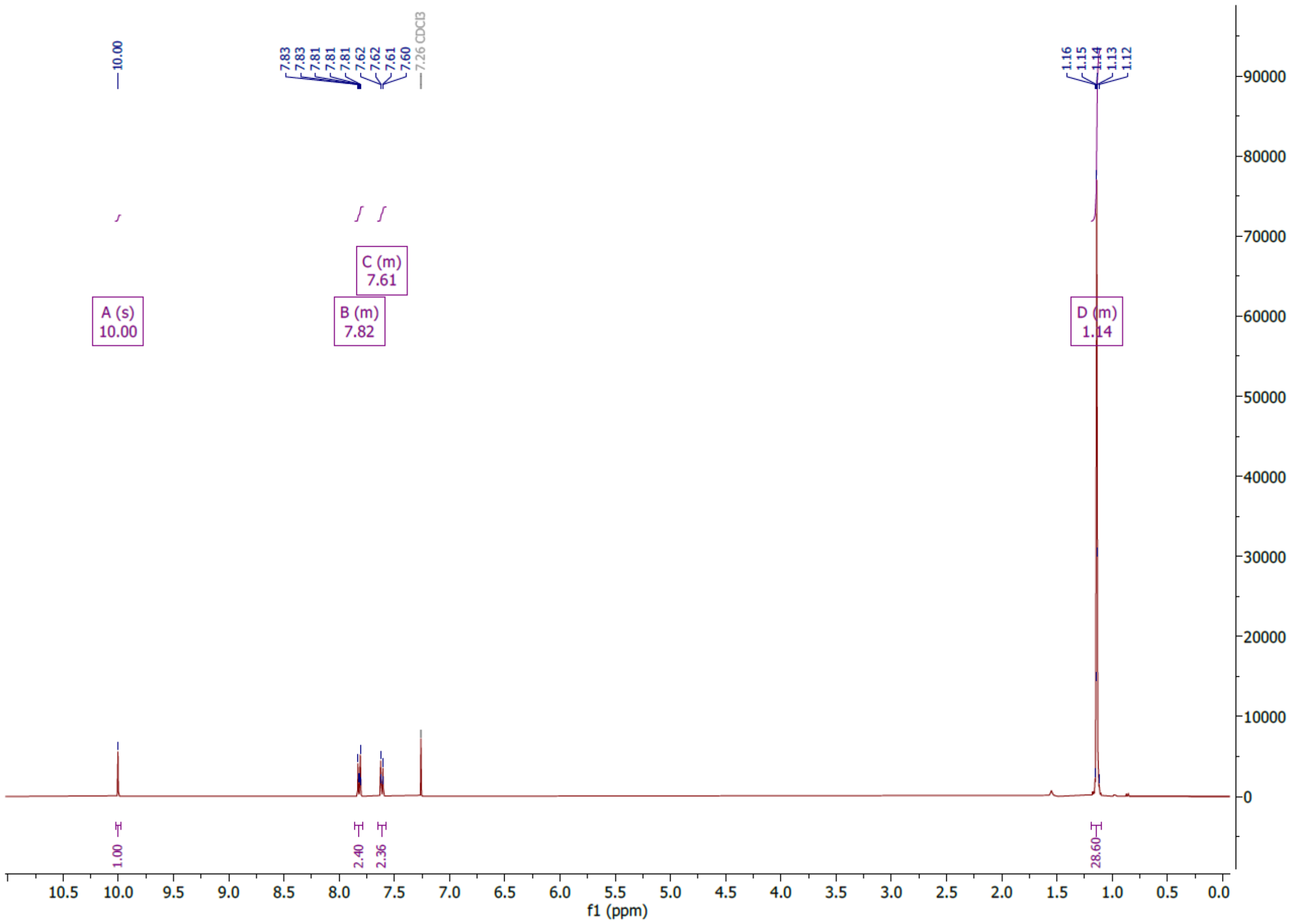

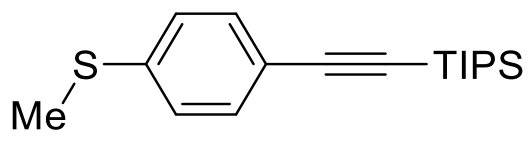

**4**

[1]H-NMR ($CDCl_3$, 400 MHz)

### 3.2.1 triisopropyl((4-(methylthio)phenyl)ethynyl)silane (4)

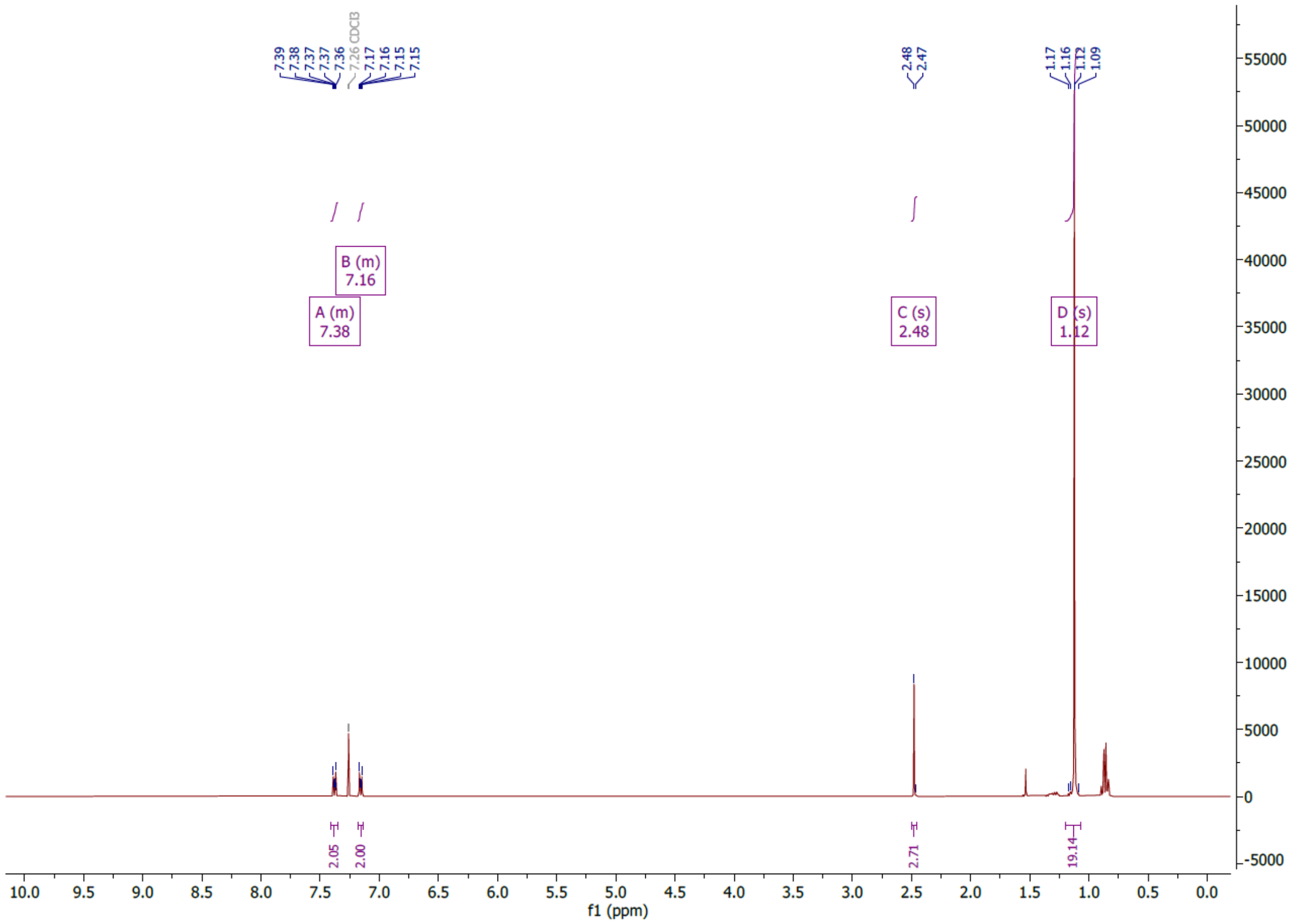

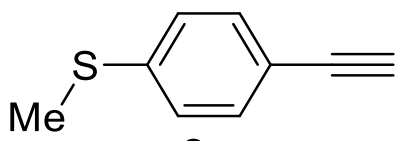

**8**

(crude material after TBAF deprotection)

[1]H-NMR ($CDCl_3$, 400 MHz)

### 3.2.2 (4-ethynylphenyl)(methyl)sulfane (8)

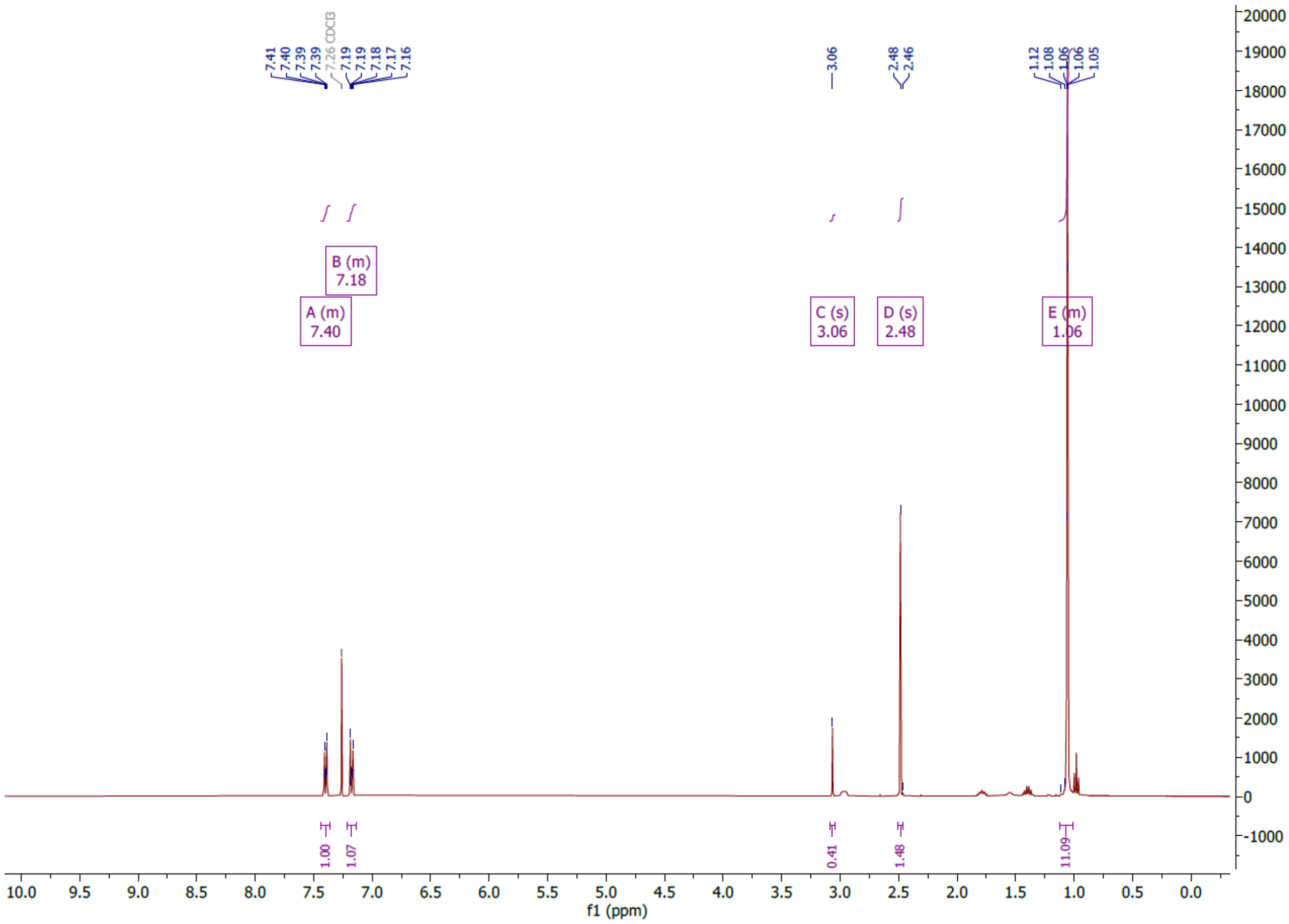

$^1$H-NMR ($CDCl_3$, 400 MHz)

### 3.2.3 [5,15-dibromo-10,20-bis(4-((triisopropylsilyl)ethynyl)phenyl)porphyrin]zinc(II) (7)

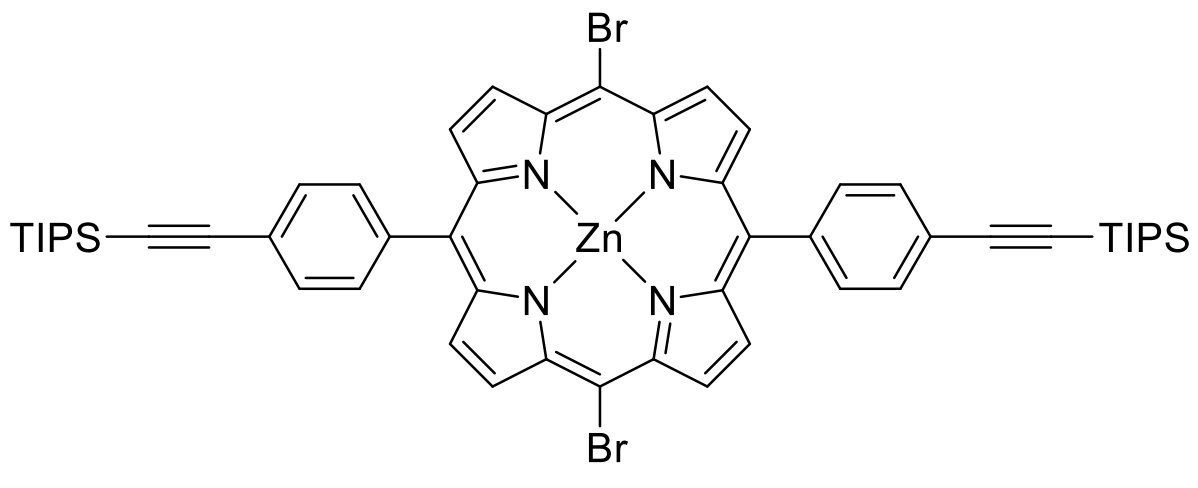

**7**

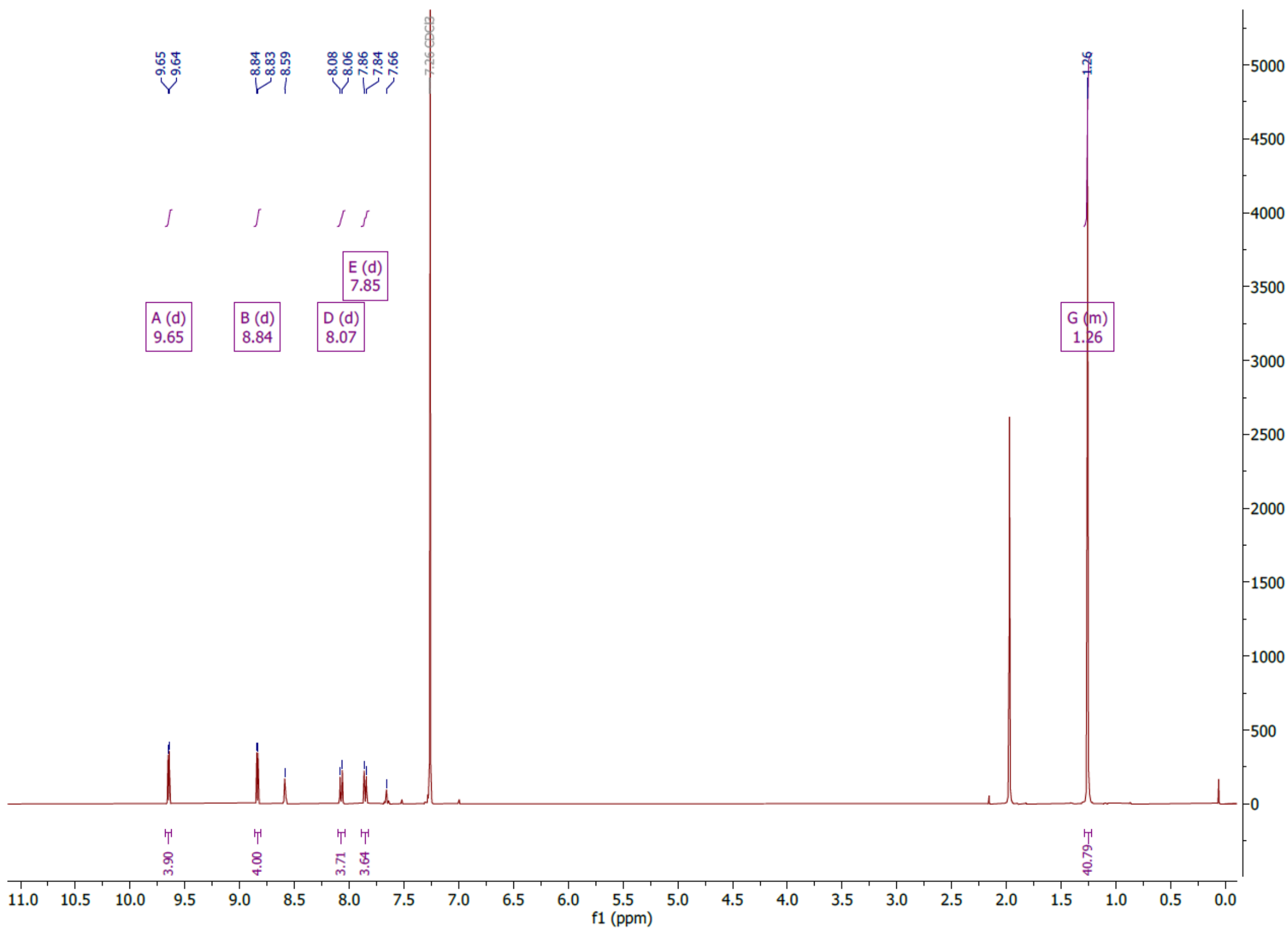

Expanded and full MALDI-TOF spectra for **7**

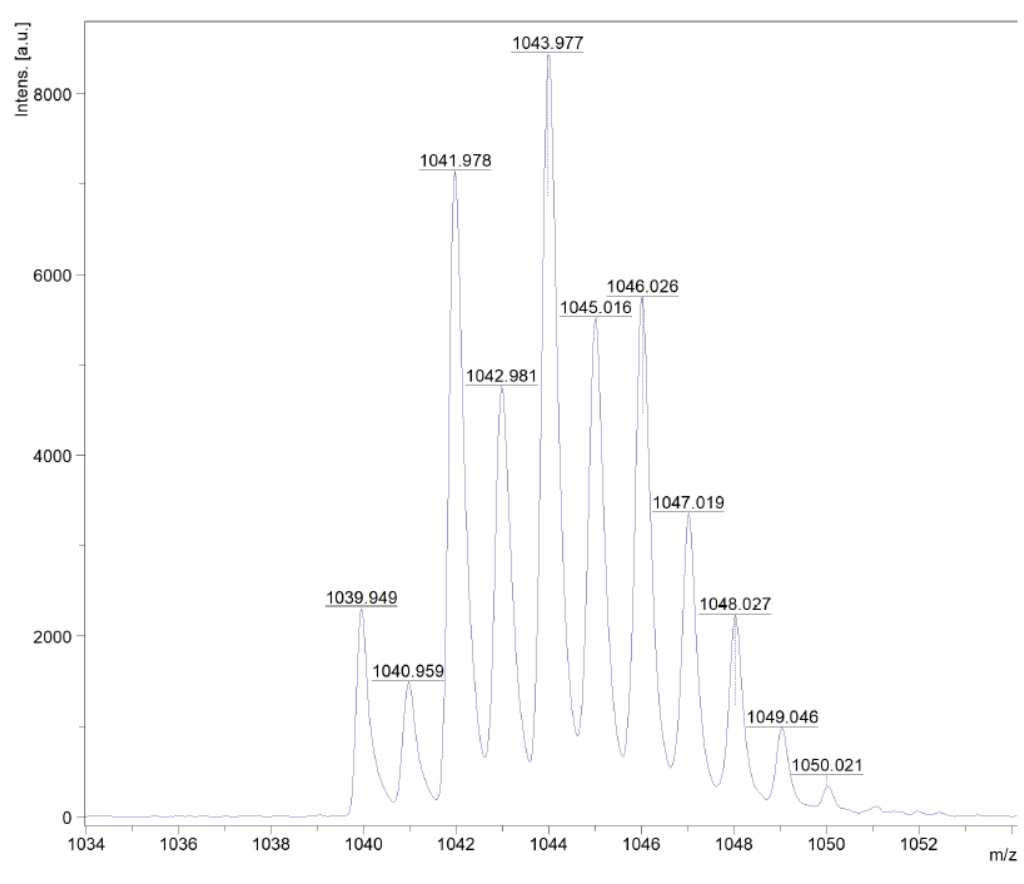

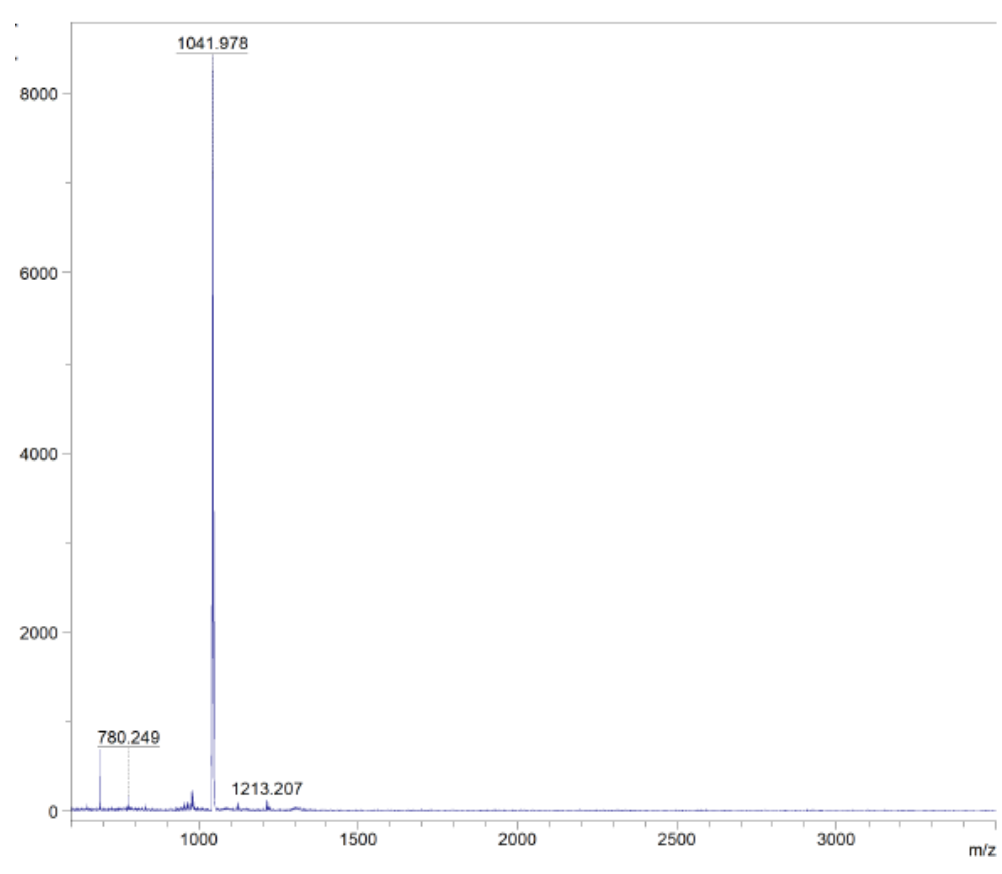

$^{1}H$-NMR ($CDCl_3$, 400 MHz)

### 3.2.4 [5,15-bis((4-(methylthio)phenyl)ethynyl)-10,20-bis(4 ((triisopropylsilyl)ethynyl)phenyl)porphyrin]zinc(II) (9)

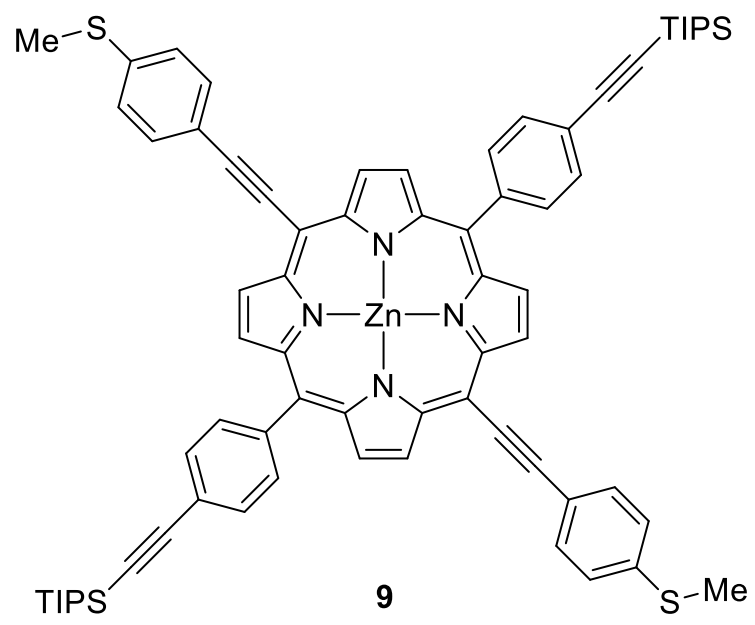

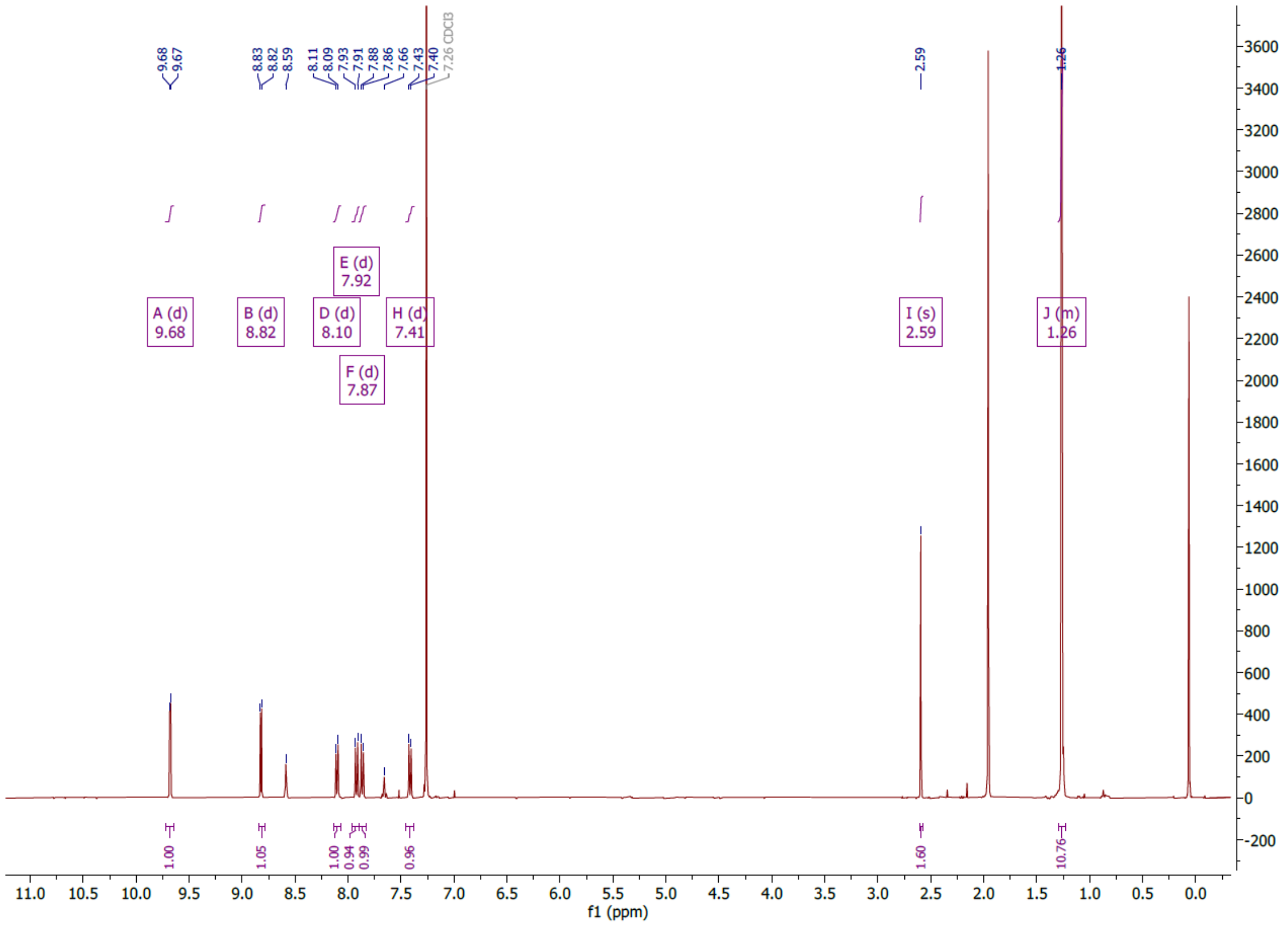

Expanded and full MALDI-TOF spectra for **9**

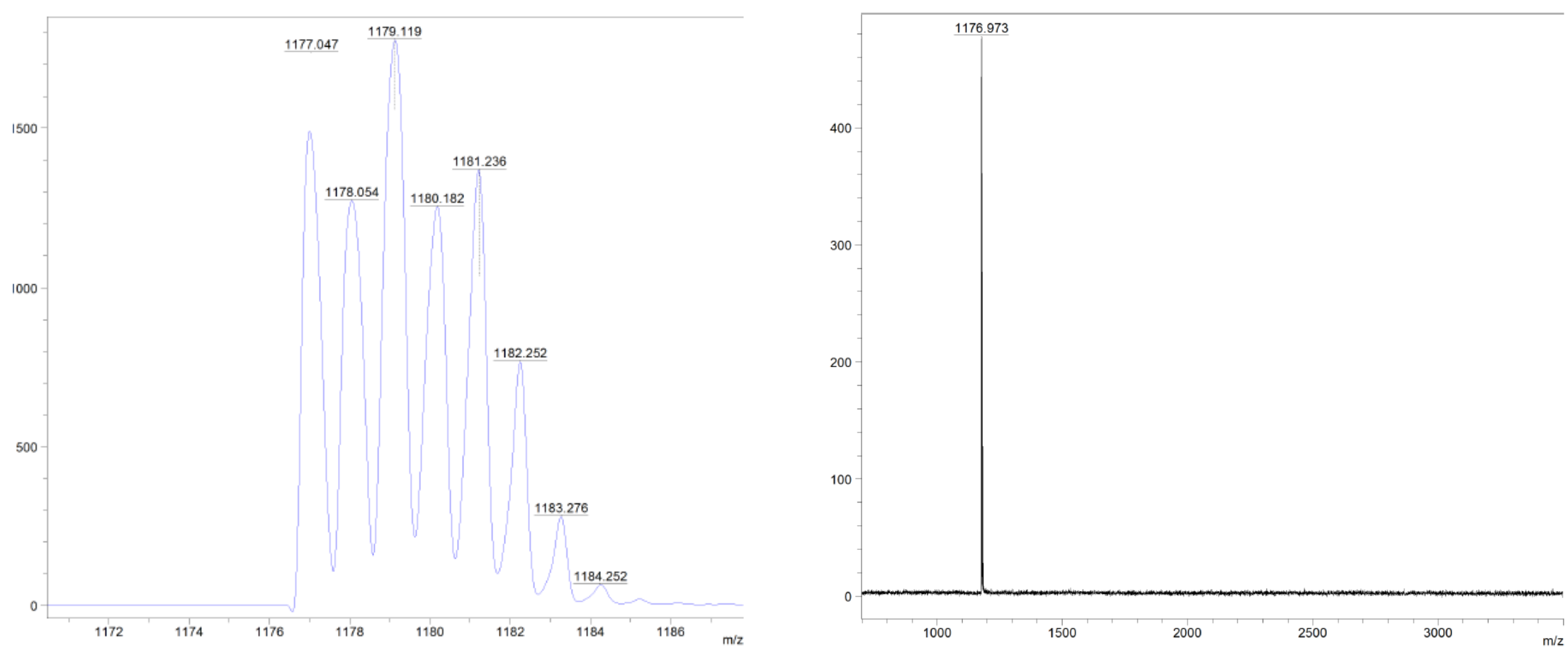

$^1$H-NMR ($CDCl_3$, 1% d6-pyridine, 400 MHz)

### 3.2.5 [5,15-bis(4-ethynylphenyl)-10,20-bis((4-(methylthio)phenyl)ethynyl)porphyrin]zinc(II) (1)

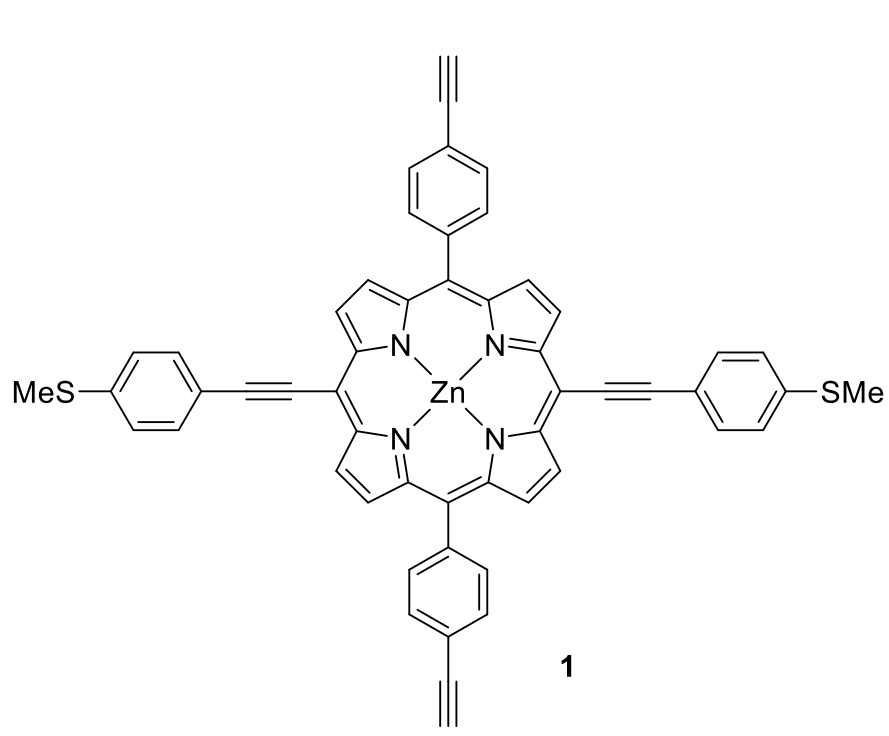

**1**

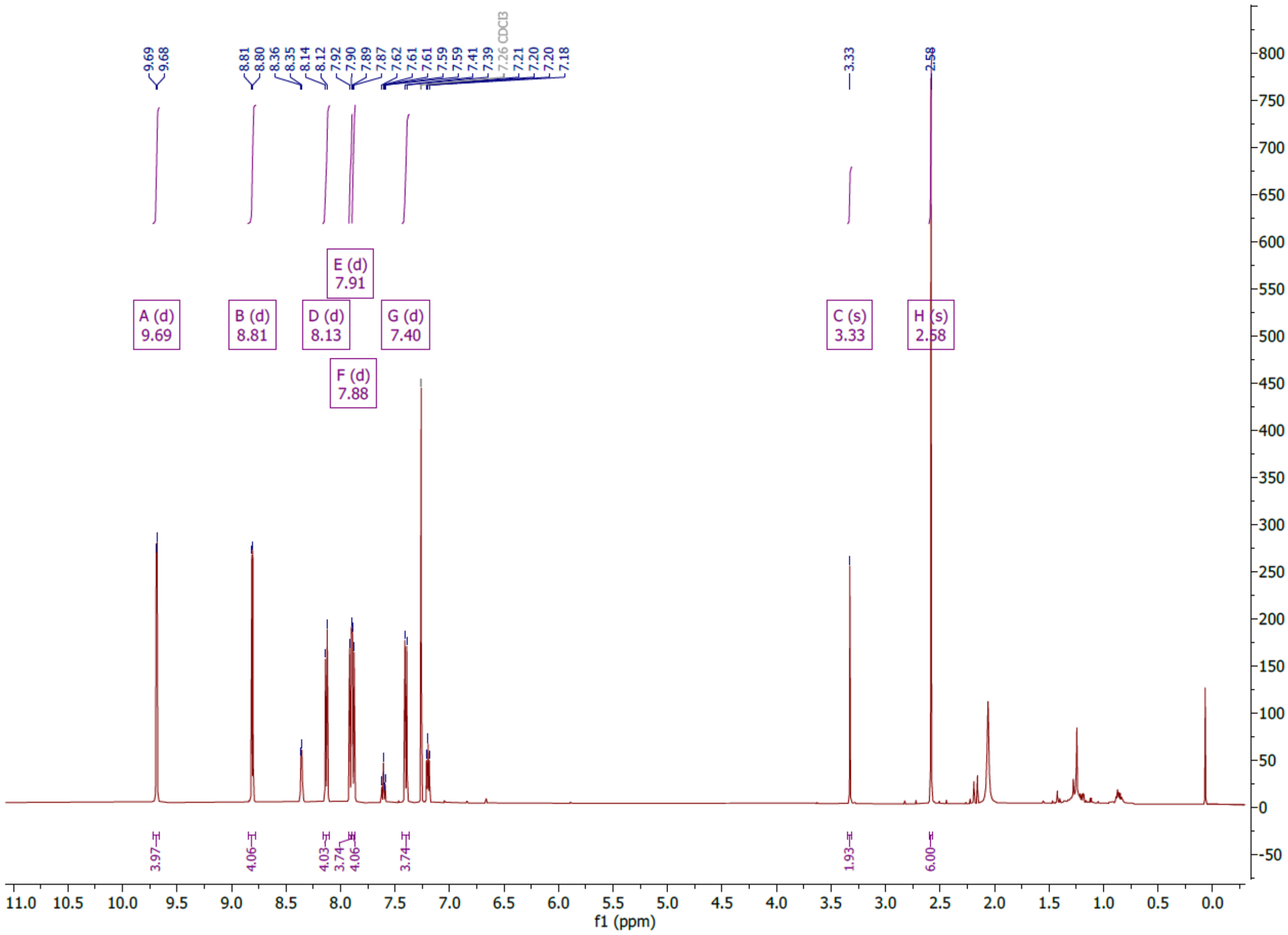

[13]C-NMR (CDCl$_3$, 1% d6-pyridine, 400 MHz)

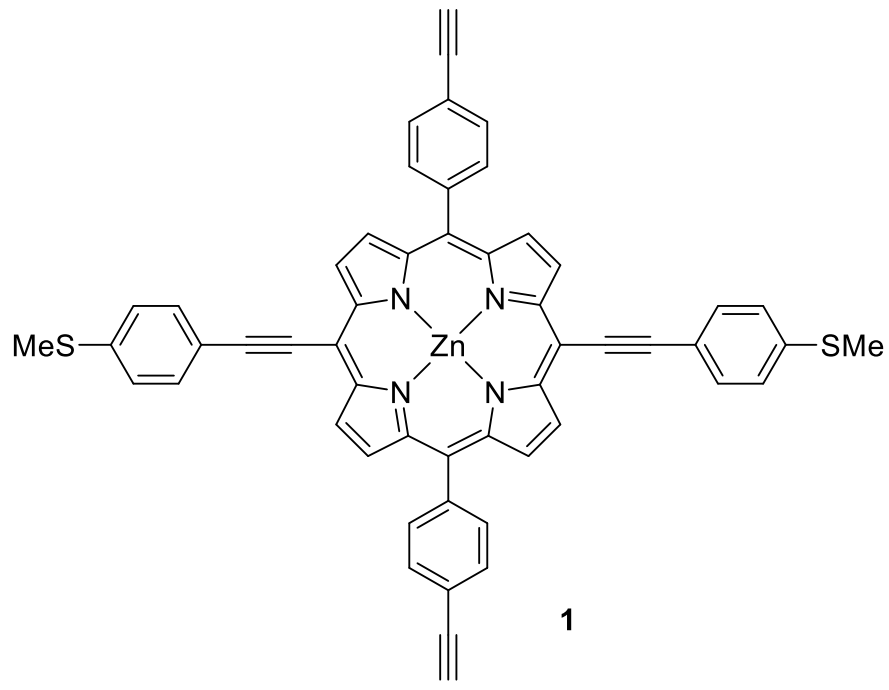

1

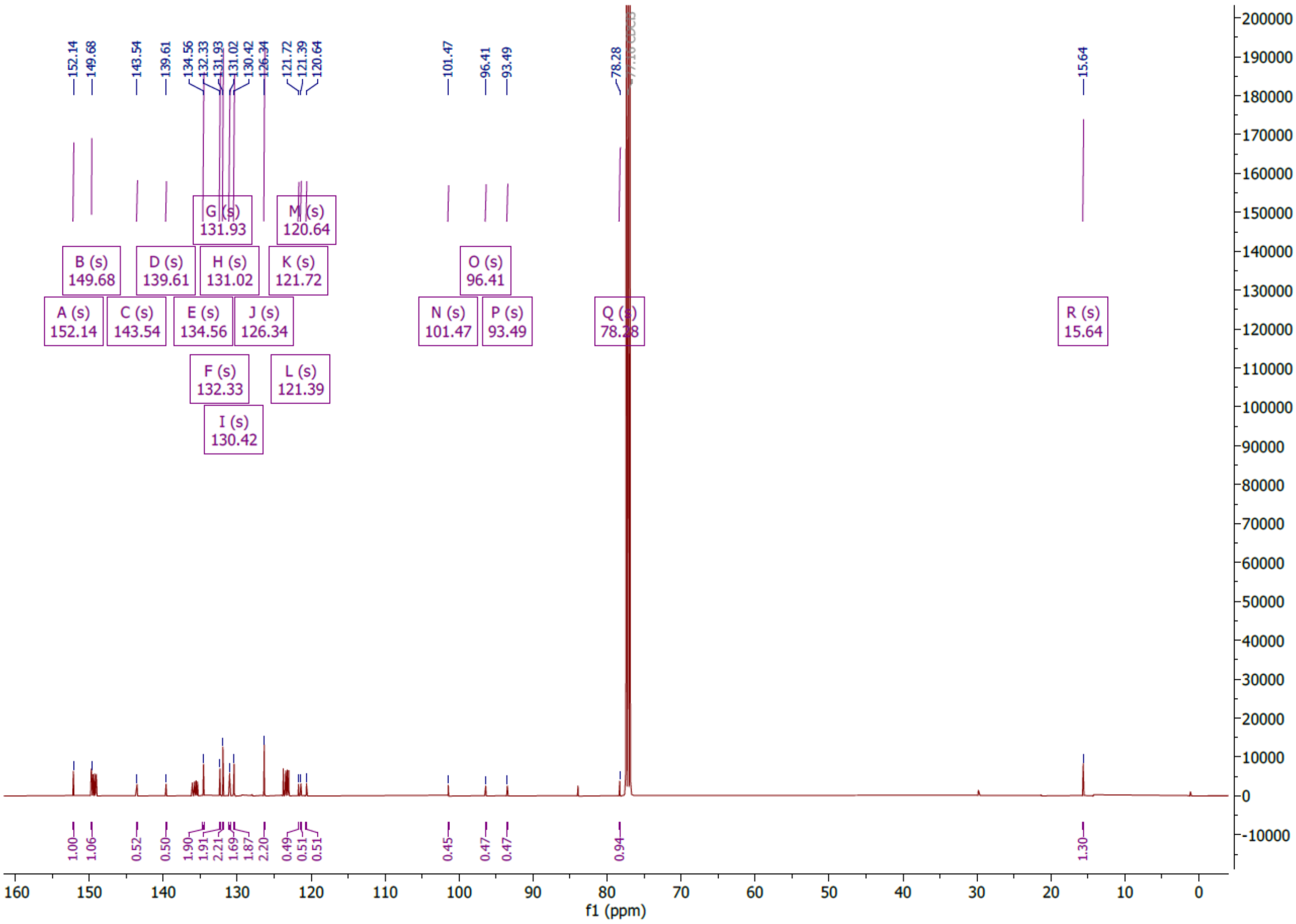

**3.2.6** UV-VIS spectrum **($CHCl_3$/1% pyridine) c=1.09x10$^{-6}$ M$^{-1}$**

**[5,15-bis(4-ethynylphenyl)-10,20-bis((4-(methylthio)phenyl)ethynyl)porphyrin]zinc(II) (2)**

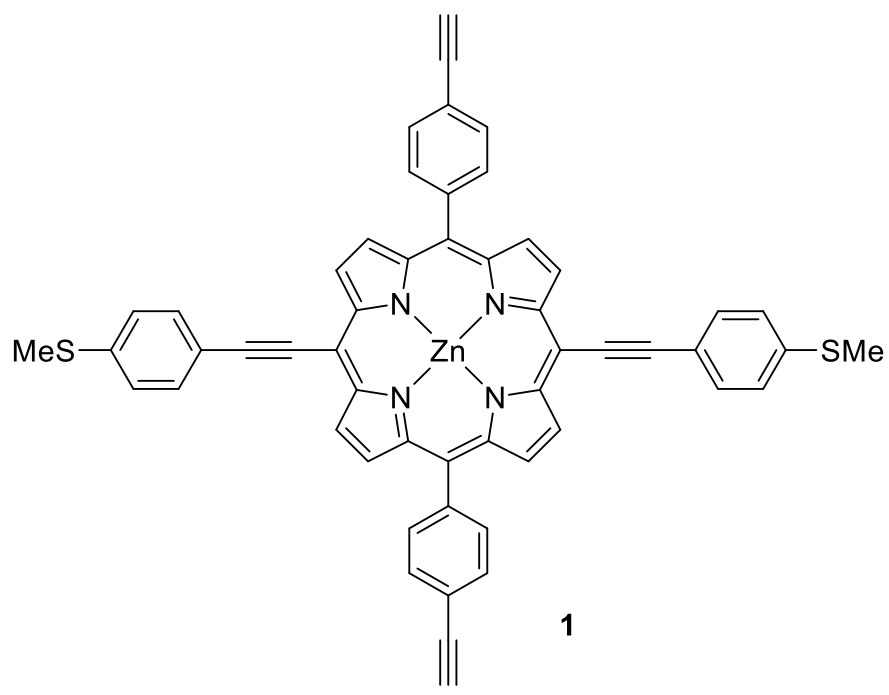

1

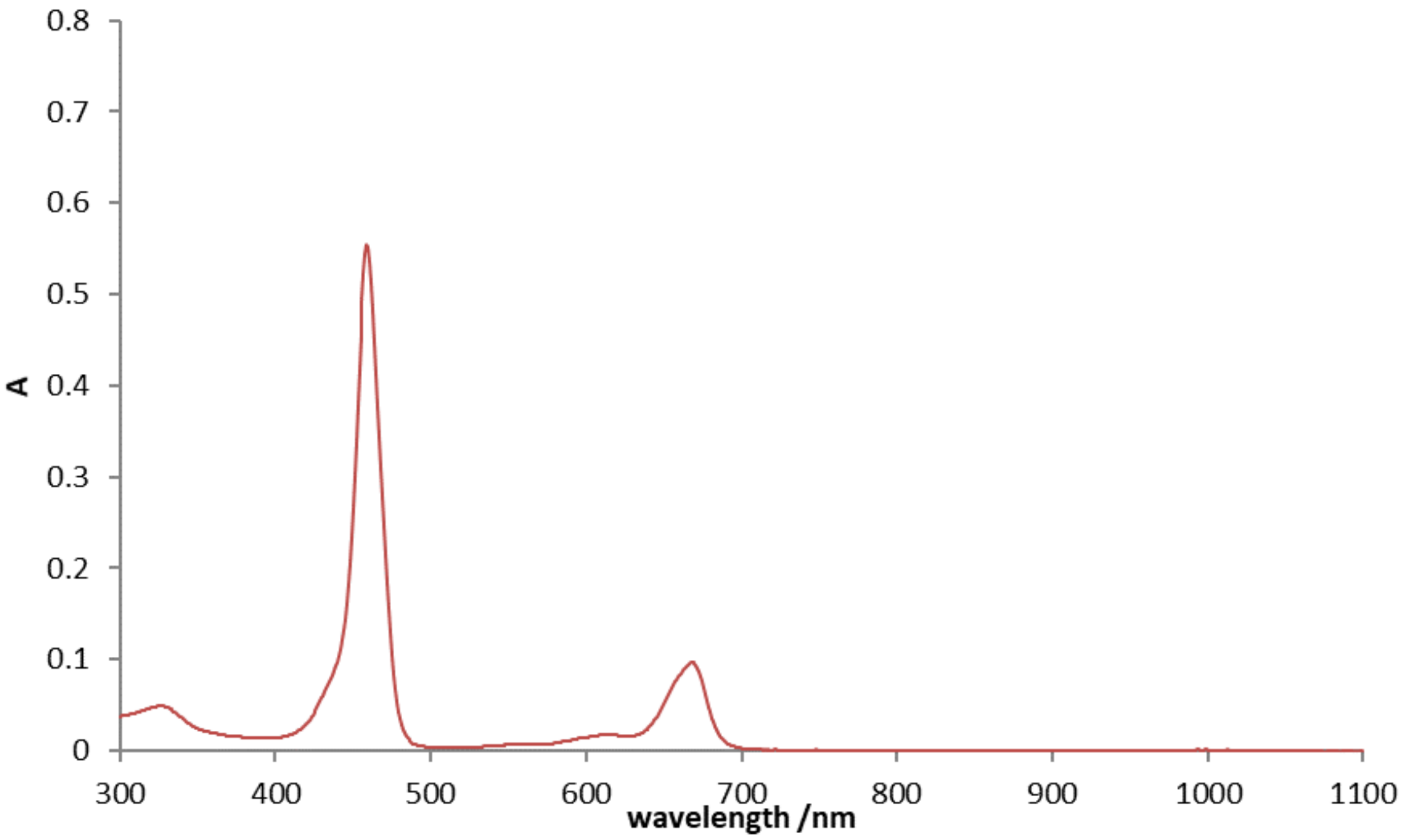

**Fig 2.** UV-VIS spectrum of Zn-porphyrin in chloroform, 1% pyridine.

### 3.2.7 Expanded MALDI-TOF spectrum for freebase porphyrin **10**

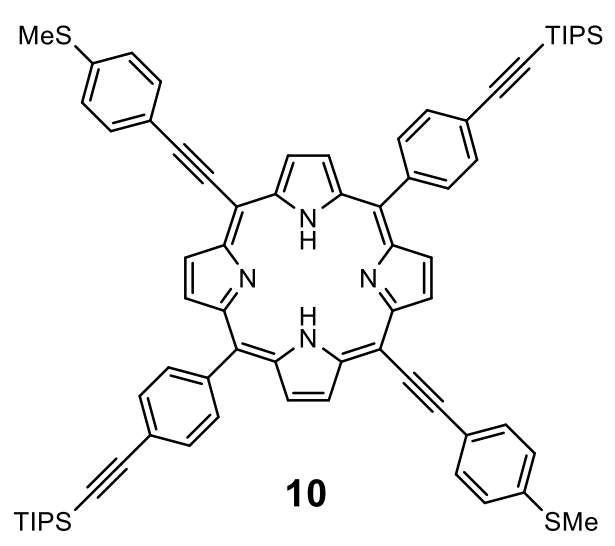

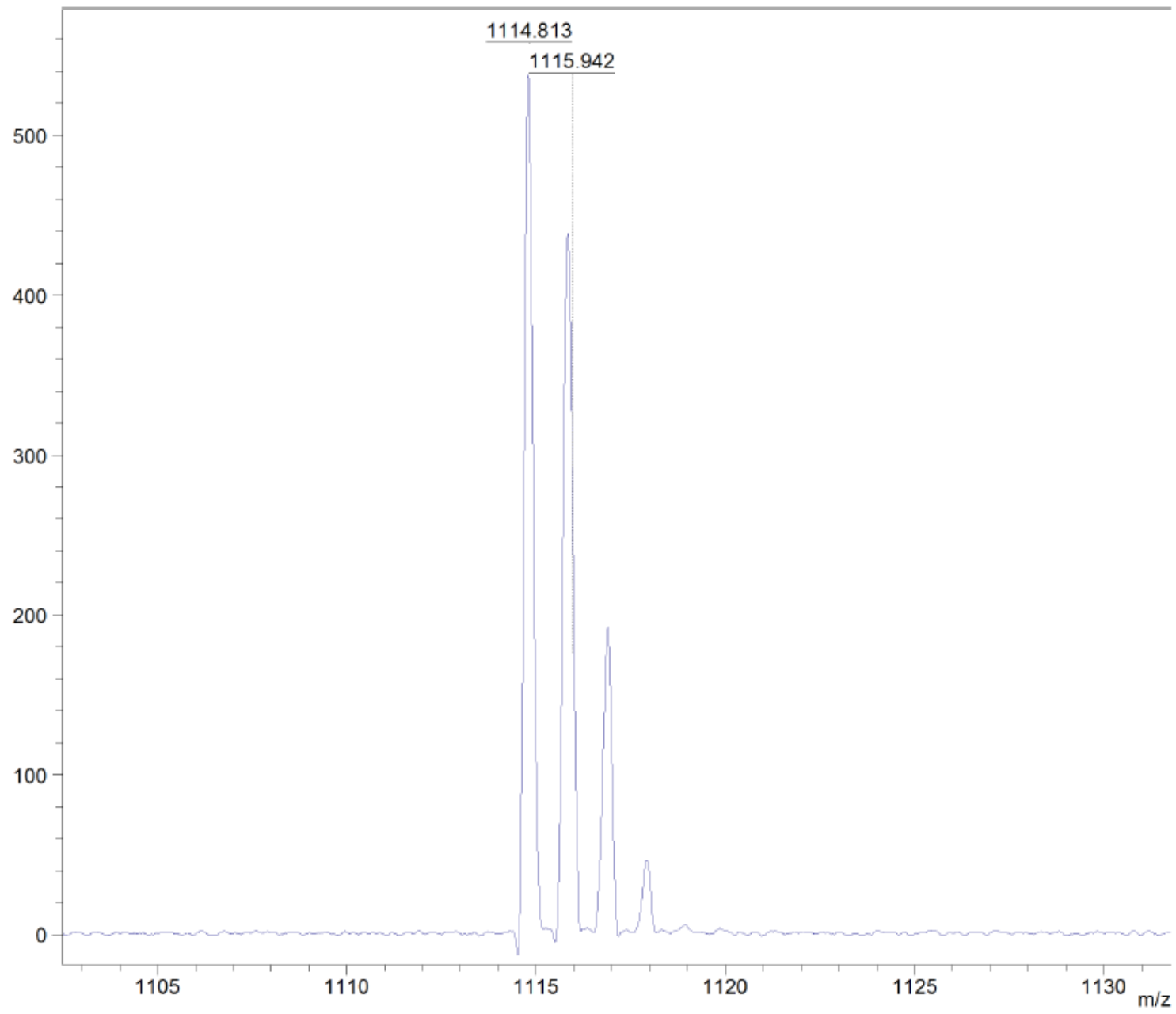

Full MALDI-TOF spectrum for freebase porphyrin **10**

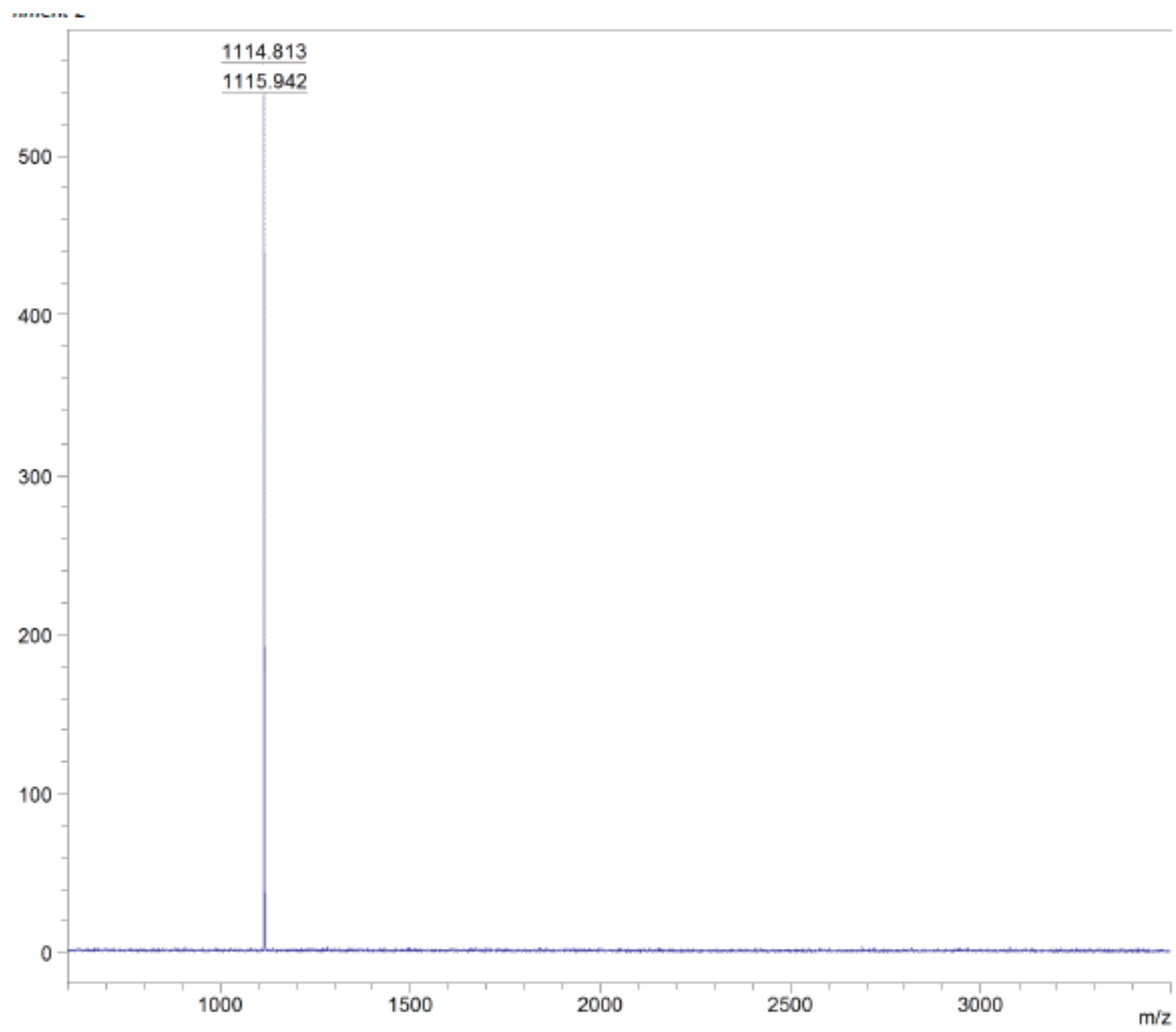

## 4. Conjugation Reactions of Zn porphyrin to DNA adapters

We conjugated the Zn porphyrin molecule (**1**), featuring dual terminal alkynes, to two distinct DNA adapters. These adapters were each modified at one end with a single azide group, allowing for copper-catalyzed azide-alkyne cycloaddition (CuAAC) reactions, following our recently published protocol [1].

### 4.1 Templated CuAAc reaction:

The DNA adapters were labelled with 3'-Cy3 or 5'-Cy5 fluorophores at the opposite ends of the reactive azide modification, enabling subsequent tracking and analysis.

During the annealing process, DNA adapters were combined with a DNA template to achieve a final concentration of 10 μM for both the DNA adapters and the template in a solution of 1× TBE buffer at pH 8, supplemented with 11 mM $MgCl_2$. This mixture underwent a short annealing programme in which the sample was heated to 96°C, followed by a gradual temperature reduction at a rate of 1°C every 3 seconds, ending by a temperature hold at 4°C.

Subsequently the zinc (Zn) porphyrin was introduced to the mixture containing the annealed DNA adapters. This was done at a 7× molar ratio of porphyrin to DNA, with the inclusion of 15% dimethylformamide (DMF) to facilitate the reaction.

### 4.2 Click Reaction:

To the total DNA adapters, sodium ascorbate added at a 600× molar equivalent, THPTA ligand at a 420× molar equivalent, and $CuSO_4$ at a 60× molar equivalent, were added. The reaction was degassed and incubated overnight on a shaker at room temperature. The templated-setup increases the local concentration of the reactive groups on the second adapter once the reaction to the first adapter has occurred, thereby promoting the selective attachment of the target molecule (porphyrin) to two adapters with different sequences and reducing side reactions in which two adapters of the same sequence, or only one adapter, are attached.

### 4.3 Non-templated CuAAc reaction:

Additionally, we optimized a non-templated CuAAc reaction for attaching porphyrin molecules to DNA adapters by adjusting the molar ratio of porphyrin to DNA, aiming to enhance the probability of a porphyrin molecule conjugating to two different DNA adapters. In the non-templated reaction, Zn porphyrin was mixed with DNA adapters at a reduced molar ratio (0.5×) in the presence of 85% DMF, with sodium ascorbate, THPTA ligand, and $CuSO_4$ added in the same molar equivalents as in the templated reaction. Following degassing, the mixture was incubated overnight on a shaker at room temperature.

This lower molar ratio of porphyrin to DNA adapters in the non-templated reaction helps to ensure that one porphyrin molecule is conjugated to two distinct DNA adapters, mitigating the risk of obtaining a product linked to only one DNA adapter. The non-templated approach was particularly useful for cases where polyacrylamide gel electrophoresis (PAGE) gel purification of the hetero-DNA-porphyrin was employed, as the template band and product band overlapped in the templated reactions, complicating the purification process.

Upon the CuAAc reaction, the reaction products were analysed by denaturing PAGE. The gels were then scanned for fluorescence in the Cy3, Cy5, and SYBR® gold (a DNA stain) channels. In the merged-colour image of the gel (Figure 4), a yellow band was observed, signifying the colocalization of Cy3 and Cy5 fluorescence. This band indicated the formation of the desired hetero-DNA-porphyrin product, wherein a porphyrin molecule is conjugated to one Cy3- and one Cy5-labelled DNA adapter. The hetero-DNA-porphyrin hybrids were subsequently purified either through High-Performance Liquid Chromatography (HPLC) (SI 10, Figure 11) or via PAGE purification. Analysis of these purified products was conducted using PAGE (Figure 4) prior to their incorporation into the DNA origami assembly reaction.

### 4.4 Denaturing PAGE gel:

The denaturing PAGE gel process involved the use of a 20% PAGE gel with a 29:1 acrylamide: bis-acrylamide ratio, which also contained 25% formamide and 30% W/V urea in 1× TBE buffer. Initially, the gels were subjected to a pre-run at 300 V for 30 minutes.
Prior to loading the reaction mixture onto the gel, samples were heated to 95 °C for 10 minutes and then promptly cooled on ice.
Following sample loading, the gel was run at 300 V for 50 minutes. The loading buffer used for the samples consisted of 80% formamide, 10% Tris-glycine, and 10% urea, ensuring the samples remained denatured for optimal separation during the electrophoresis process.

### 4.5 Gel scanning and analysis:

All gels were scanned in the fluorescence channels specific to each fluorophore, followed by staining with SYBR® Gold and then scanning in the SYBR® Gold channel. A Typhoon FLA 9500 gel scanner was utilized for all gel scans. ImageJ software was used for the analysis of the gel images. After scanning, merged colour images were created, displaying the Cy5 signal in red, the Cy3 signal in green, and the SYBR signal in blue. In the agarose gels: the yellow band in these merged images indicates the co-localization of Cy3 and Cy5 fluorescence, confirming the successful formation of the desired hetero-DNA-porphyrin product, i.e. single porphyrin molecule is conjugated to one of each of the Cy3- and Cy5-labelled DNA adapters.
The gel sections that contained the properly assembled structures were excised, and the origami structures were extracted by compressing these excised pieces of gel.

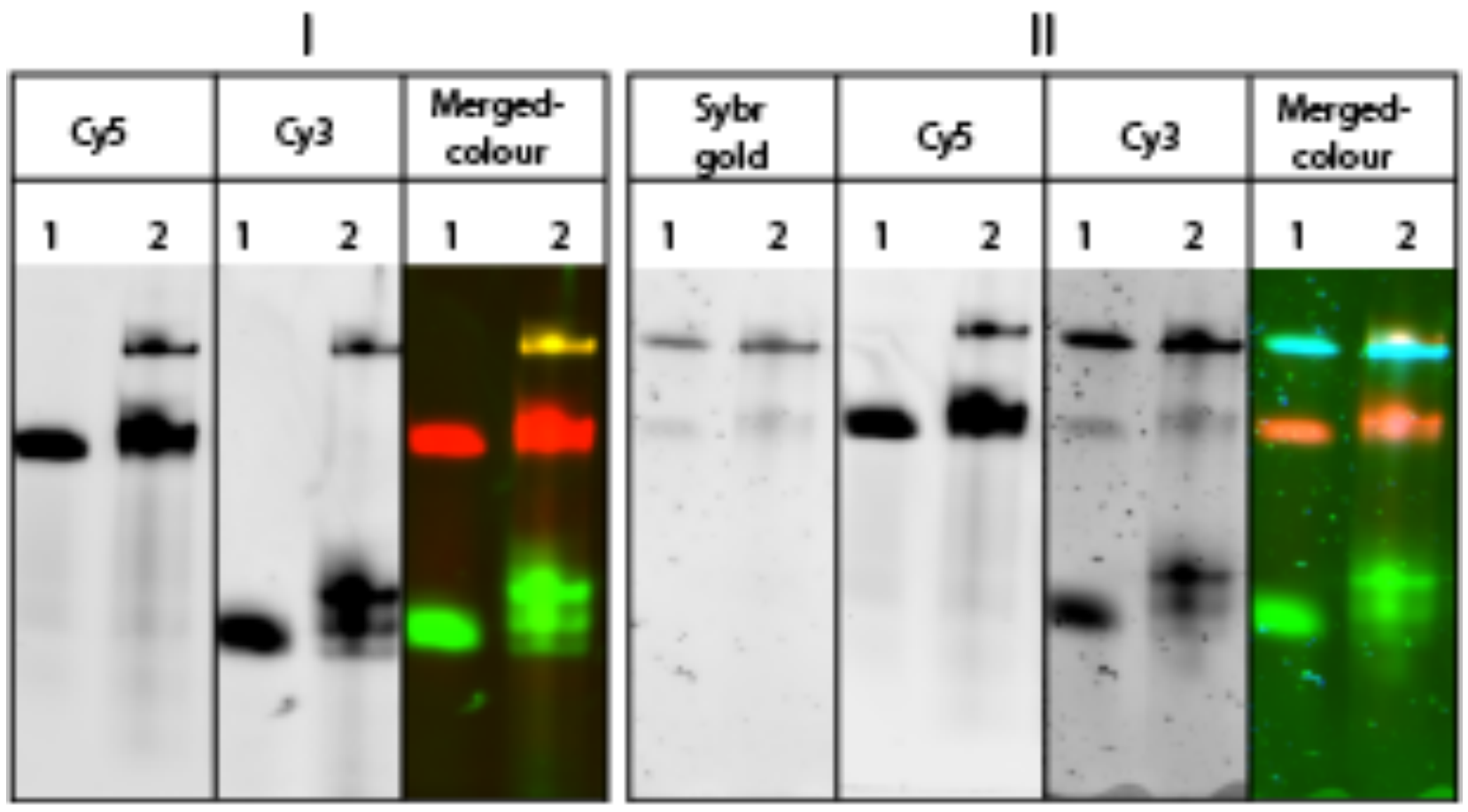

**Fig 3.** shows the analysis of the templated CuAAC reaction of zinc (Zn) porphyrin molecule, performed using 20% denaturing PAGE. To the left (I), the initial scan results for the Cy5 and Cy3 fluorophore channels are displayed prior to SYBR® Gold staining. These are accompanied by a merged-colour image of the signals from both channels. To the right (II), the figure showcases the scan results following SYBR® Gold staining, including individual scans for the SYBR® Gold , Cy5, and Cy5 channels. These are accompanied by a merged-colour image combining all three channels. Notably, the merged-colour image post-SYBR® Gold staining illustrates the overlap between the hetero-porphyrin product band and the DNA template band. This overlap makes it challenging to distinguish the product band, which was clearly visible before the staining, as seen in the left side of the figure, where the yellow band in the merged colour image of the Cy3 and Cy5 channels indicated the hetero-porphyrin product.

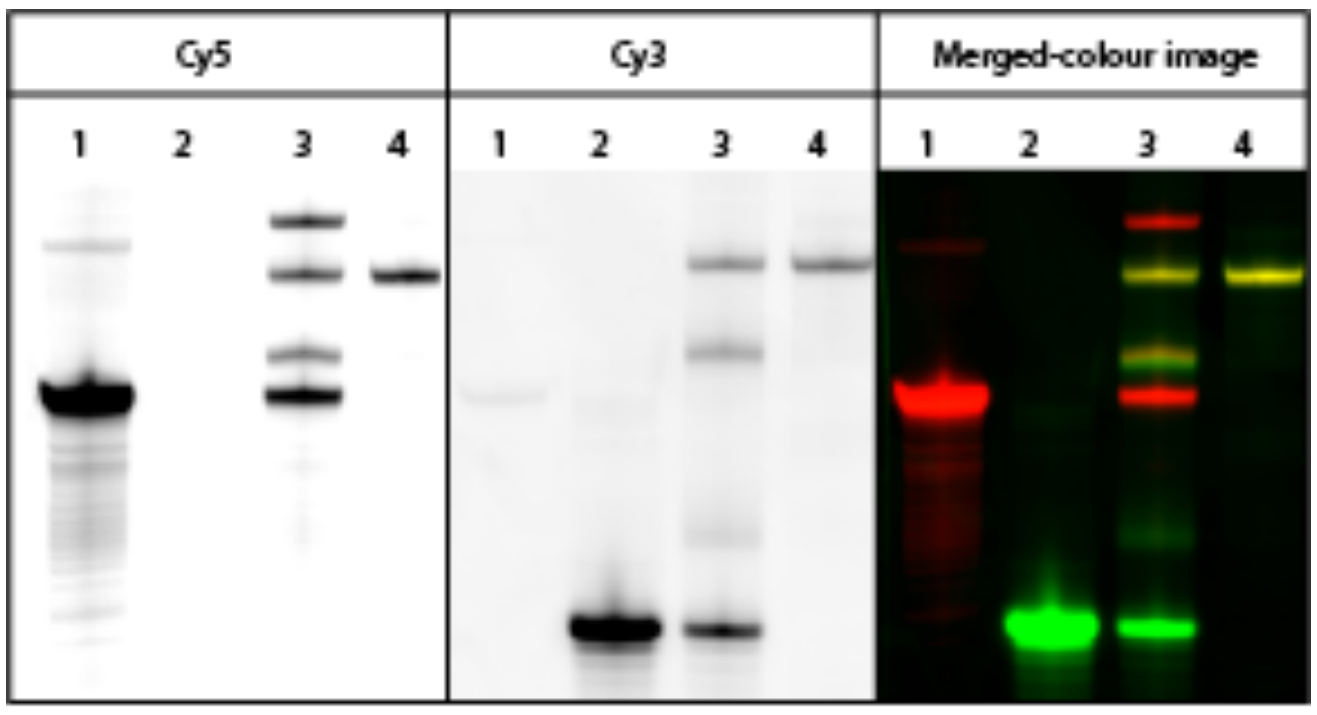

**Fig 4.** shows the analysis of the non-templated CuAAC reaction of zinc (Zn) porphyrin molecule, performed using 20% denaturing PAGE. From the left, the initial scan results for the Cy5 and Cy3 fluorophore channels are displayed, followed by a merged-colour image of the signals from both channels. Lane 1: shows the Cy5 DNA adapter, lane 2: shows the Cy3 DNA adapter, lane 3: shows the results of the non-templated reaction, and lane 4: shows the HPLC-purified hetero-bi-functionalized Zn-porphyrin (A-porphyrin-B) of the templated reaction for refence.

## 5. DNA origami Design

In this study, two origami designs were evaluated: the first, a five-layer DNA origami frame, inspired by a previously published design[5], measuring 45 nm x 36 nm x 10 nm. The central layer of this structure forms a complete scaffold sheet, while the top and bottom layers each encases a funnel-shaped groove.

The second design features a monolithic block-like shape, with dimensions of 60 nm x 40 nm x 6 nm. This structure is composed of three stacked layers of double-stranded DNA (dsDNA), each layer comprising 14 dsDNA helices. Within the core of this structure, the scaffold pathway is routed through a 5 nm pore from the top layer down to the bottom layer, facilitating the guidance of molecules along this predetermined route.

For both origami designs, two 60 nm gold nanoparticles (AuNPs) are designed to anneal in close proximity on either side of the origami structure. The separation distance between these AuNPs is determined by the thickness of the origami sheet itself—specifically, a single layer (2 nm) for the frame structure, and three layers (6 nm) for the monolithic design.

## 6. DNA origami assembly and purification:

The assembly of both origami designs involve combining 10 nM of the p7249 scaffold with each staple oligonucleotide at a concentration of 200 nM in a folding buffer solution of 10 mM Tris, 1 mM EDTA (pH of 8), and 16 mM $MgCl_2$. This mixture was then heated to 65 °C for 20 minutes to denature all DNA strands, followed by a gradual reduction in temperature to 20 °C over 16 hours, cooling at a rate of 1 °C every 45 minutes.

For assembling electronic devices using the frame-origami structures to include a single molecule porphyrin: hetero-DNA-Zn-porphyrin, were added to the assembly mix at a concentration of 280 nM. In contrast, for the control frame structures that do not include conductive molecules, an unmodified version of the two DNA adapters (without any fluorophores or azide modifications) was added to the assembly mixture at a concentration of 280 nM.

All the different origami structures were prepared in separate batch reactions.

Following the assembly of the different origami designs (monolith, and frame including Zn, and control versions), the structures were subjected to a series of purification steps using three cycles of PEG-purification to remove any unbound staple strands and to enhance the concentration of the structures. For each cycle, an equal volume of a PEG-precipitation solution, which contains 12.5 mM $MgCl_2$, 0.5 M NaCl, 15% PEG (w/v), and 1× TAE, was added to the samples. The samples were then spun at 20,000 rcf for 30 minutes at a temperature of 20 °C.

The assembled structure were analysed using gel electrophoresis, utilizing a 1.0% agarose gel in 1× TAE buffer (40 mM Tris, 40 mM acetic acid, 1 mM EDTA, with a pH of 8) and mM $MgCl_2$.

Assembly:

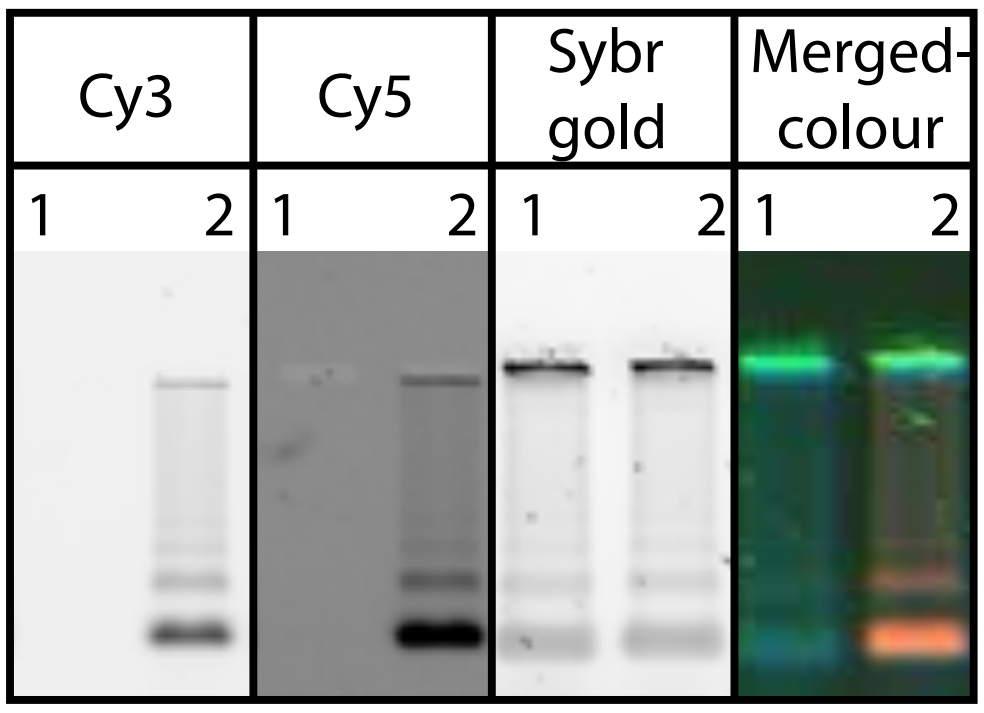

**Fig 5.** The figure presents an analysis using a 1% agarose gel to examine the assembly of the DNA origami frame structure, including control, and Zn, variations. Initially, the scan results for the Cy3 and Cy5 fluorophore channels are shown, taken before the application of SYBR® Gold staining. This initial scanning allows for the visualization of the structures assembled with the hetero-DNA porphyrin products.
Subsequently, the gel was stained with SYBR® Gold and a merged-colour image combining signals from all three channels is then provided, with Cy3 fluorescence depicted in green, Cy5 in red, and SYBR® Gold staining rendered in blue. The convergence of the Cy3 and Cy5 signals, as observed in the merged-colour image, signifies the successful integration of the Zn hetero-DNA-porphyrin product into the DNA origami nanostructure. This co-localization is crucial evidence of the precise assembly and incorporation of the porphyrin molecules within the specific sites of the origami frames. This showcase the effectiveness of our synthetic pathway for constructing complex, functionalized nanoscale architectures.

### 7. DNA- Functionalization of AuNPs:

Six ml of citrate-stabilized 60 nm gold nanoparticles (AuNPs), OD 1 (BBI solutions), were centrifuged twice at 8000 rcf for 5 minutes to both concentrate the nanoparticles and remove the citrate buffer. The resulting pellet was then re-suspended in 1 mL of Milli-Q (MQ). Into the suspension, 4 μM of freshly cleaved 19 nucleotide poly-thymine oligonucleotides (Tx19, Biomers) featuring a 5'-thiol modification, were introduced. Prior to addition, the thiol groups on these oligonucleotides were activated by incubation in 20 mM tris(2-carboxyethyl)phosphine hydrochloride (TCEP) for 1 hour, a step crucial for reducing the disulphide bonds. The mixture was then incubated at room temperature for 30 minutes, facilitating the binding of the thiol-modified oligonucleotides to the suspension.
Subsequently, the concentrations of phosphate-buffered saline (PBS) and sodium dodecyl sulfate (SDS) were adjusted to 0.01 M and 0.01%, respectively, before the mixture underwent a further incubation at room temperature for an additional 30 minutes. The concentration of NaCl was increased to 0.05 M using 2 M NaCl, while maintaining 0.01 M PBS, and 0.01%. SDS, concentrations. The solution was sonicated for approximately 10 seconds and then incubated for 20 minutes at room temperature. This procedure was repeated with an additional increment of 0.05 M NaCl and for every subsequent 0.1 M increase in NaCl concentration up to a final concentration of 1.0 M NaCl. Following each NaCl addition, the solution was sonicated for approximately 10 seconds and then incubated for 20 minutes at room temperature. Following the salting process, the mixture was left to incubate overnight at room temperature.

To assess the success of DNA functionalization on the gold nanoparticles (AuNPs), a $MgCl_2$ stability test was conducted. This involved mixing 10 µL of the DNA-functionalized AuNPs solution with 10 µL of $MgCl_2$ buffer, which consists of 200 mM $MgCl_2$ in 0.5x TBE. The colour of the nanoparticles acted as an indicator for the functionalization efficacy: a retention of the red colour suggested that the functionalization was successful, allowing the solution to move forward to the purification phase. Conversely, a shift in colour to blue/purple signalled inadequate functionalization with the thiolated-Tx19 oligonucleotides. To address this, more cleaved Tx19 oligonucleotides were added, and the mixture was incubated for an additional hour. The $MgCl_2$ stability test was then repeated to verify the adequacy of the DNA-AuNPs functionalization before advancing to the purification step.

After successfully functionalizing the AuNPs with DNA, the solution was purified to eliminate any excess of unbound oligonucleotides. This was achieved by centrifuging the solution at 8000 rcf for 5 minutes. Subsequently, the supernatant was discarded, leaving behind a pellet of the gold nanoparticles. These particles were then re-suspended in 0.01% SDS. This washing procedure was performed four times to ensure the complete removal of the supernatant. In the final step, the pellet was re-suspended in a running buffer consisting of 0.5x TBE and 11 mM $MgCl_2$.

This purification step plays a crucial role in ensuring a high yield of AuNPs attached to the DNA origami structures. By removing unbound single-stranded DNA (ssDNA), it prevents these excess oligonucleotides from blocking the attachment points (handles) on the origami template structure, thereby facilitating a more efficient attachment of the AuNPs to the designated sites on the DNA origami.

The concentration of the DNA-functionalized AuNPs was determined using a UV-Visible (UV-VIS) spectrophotometer. This involved measuring the UV absorption of the solution at a wavelength of 520 nm, a characteristic peak for gold nanoparticles, allowing for the accurate quantification of the AuNPs concentration post-functionalization.

### 8. Functionalization of DNA origami structures with AuNPs:

The AuNPs attachment to DNA origami structures relies on the complementary interaction between the single-stranded DNA (ssDNA) on the AuNPs and the hybridization sites on the DNA origami. For this purpose, 19x poly-thymine (T) were used on the AuNPs, and poly-adenosine (A) sequences were selected for the DNA origami. Forty-eight sites, each with a mix of 4 and 6-bases extensions, were designated for attaching each AuNP. The chosen number and length of these sites are designed to ensure sufficient affinity for the AuNPs to attach securely to their intended positions on the DNA origami structures, while avoiding entropic barriers. Too short or too few hybridization sites lead to unstable connections between the AuNPs and the DNA origami. Conversely, longer sequences might result in the attachment of multiple AuNPs to a single site or the nanoparticles becoming stuck at unintended locations.

The purified DNA-functionalized AuNPs were mixed with the purified DNA origami structures, in separate batch reactions, in a ratio of 5 AuNPs to each DNA origami structure. This mixture was then heated to 40 °C followed by a gradual cool down to 23 °C over the course of 1 hour, cooling at a consistent rate of 1 °C every 3.5 minutes. Following the successful functionalization of the AuNPs with DNA, they were combined with the purified DNA origami structures in separate batch reactions, using a ratio of 5 AuNPs for each origami structure. This mixture was then heated to 40 °C and allowed to cool gradually to 23 °C over an hour, cooling steadily at a rate of 1 °C every 3.5 minutes, facilitating the optimal interaction and attachment between the AuNPs and the DNA origami.

To separate the DNA origami structures, now functionalized with two 60 nm AuNPs, from any excess AuNPs, a purification process was employed using a 1% agarose gel in 0.5x TBE buffer and 11 mM $MgCl_2$. Gel sections containing the correctly assembled structures were identified and excised. The DNA origami structures were then carefully extracted from these gel sections by squeezing the excised gel pieces.

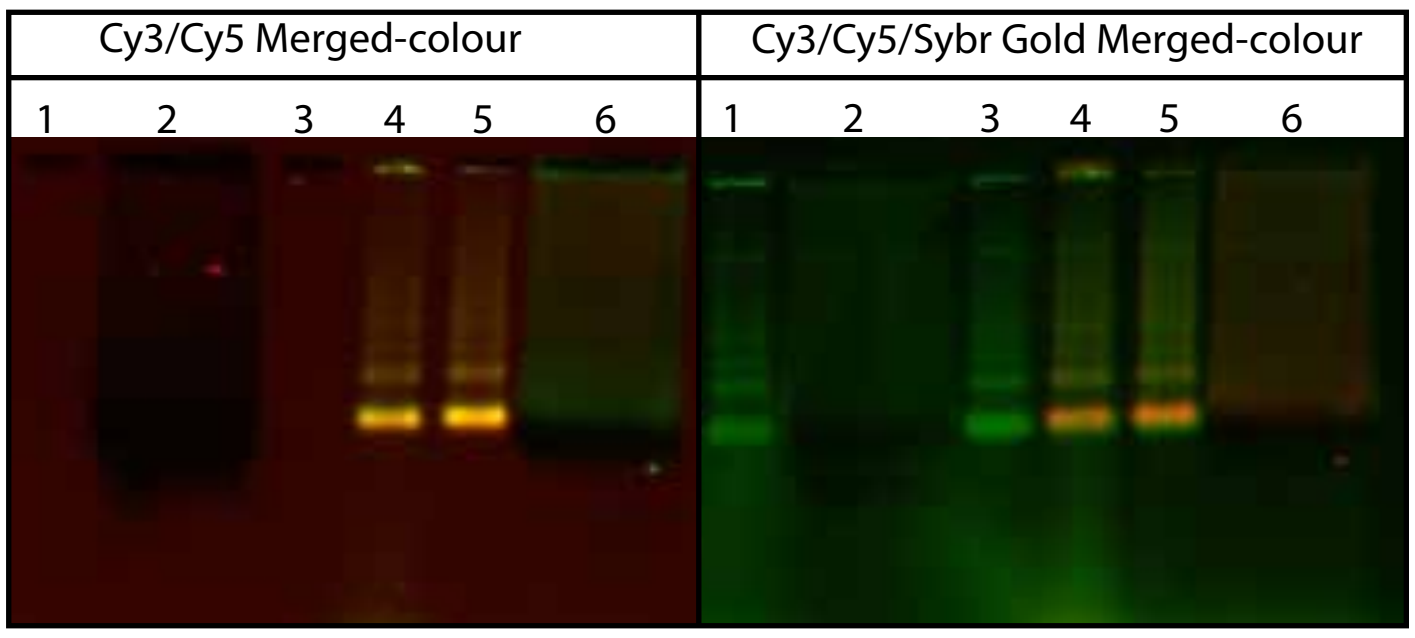

**Fig 6.** The figure provides a detailed analysis through a 1% agarose gel, focusing on the assembly of origami frame structures that include control and zinc (Zn) variants, subsequent to the annealing of two 60 nm gold nanoparticles (AuNPs). Initially depicted are the scan results for the Cy3 and Cy5 fluorophore channels, captured prior to the application of SYBR® Gold staining. These preliminary scans facilitate the observation of the structures that have been assembled with the hetero-DNA porphyrin products.
To the right, the gel after staining with SYBR® Gold, and merging signals from the Cy3, Cy5, and SYBR® Gold channels. Lane: 1, and 3: show the control origami, lanes 4, and 6 shows the Zn-porphyrin devices. Lanes 2, and 6: shows the full assembled (structures plus AuNPs) control and Zn-porphyrin structures, respectively.

## 9. Transmission electron microscopy (TEM) analysis:

For TEM analysis, a 5 μL of the diluted origami sample solution (1 nM) was applied to glow-discharged, carbon-coated grids. These samples were then stained with a 2% solution of filtered uranyl format in 5 mM NaOH, for a duration of 2 minutes. TEM imaging was conducted using a (JEOL 1400 Flash 120kV TEM) transmission electron microscope, operating at an acceleration voltage of 100 kV.

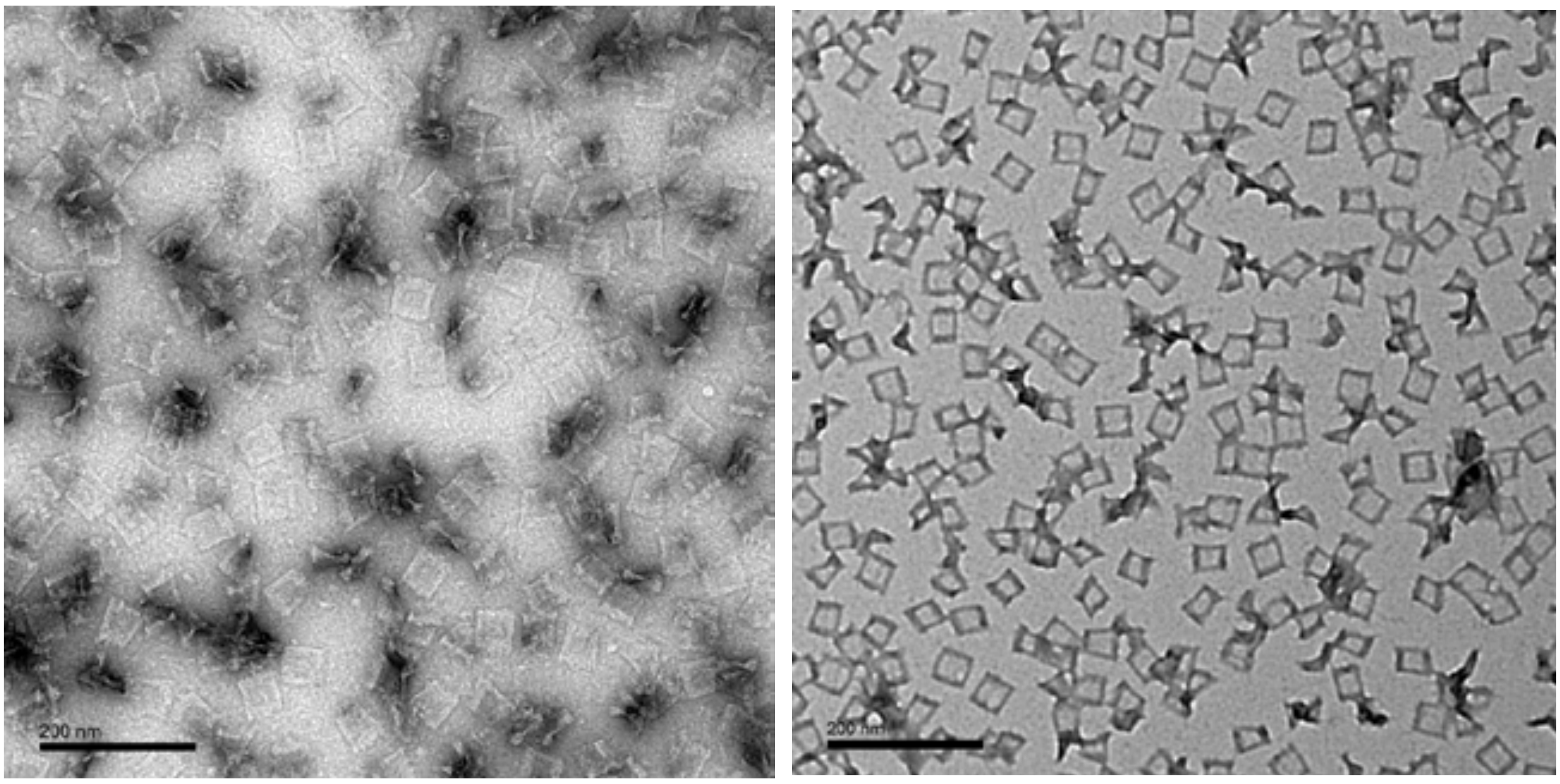

**Fig 7.** Overview TEM images of the five-layer DNA origami frame.

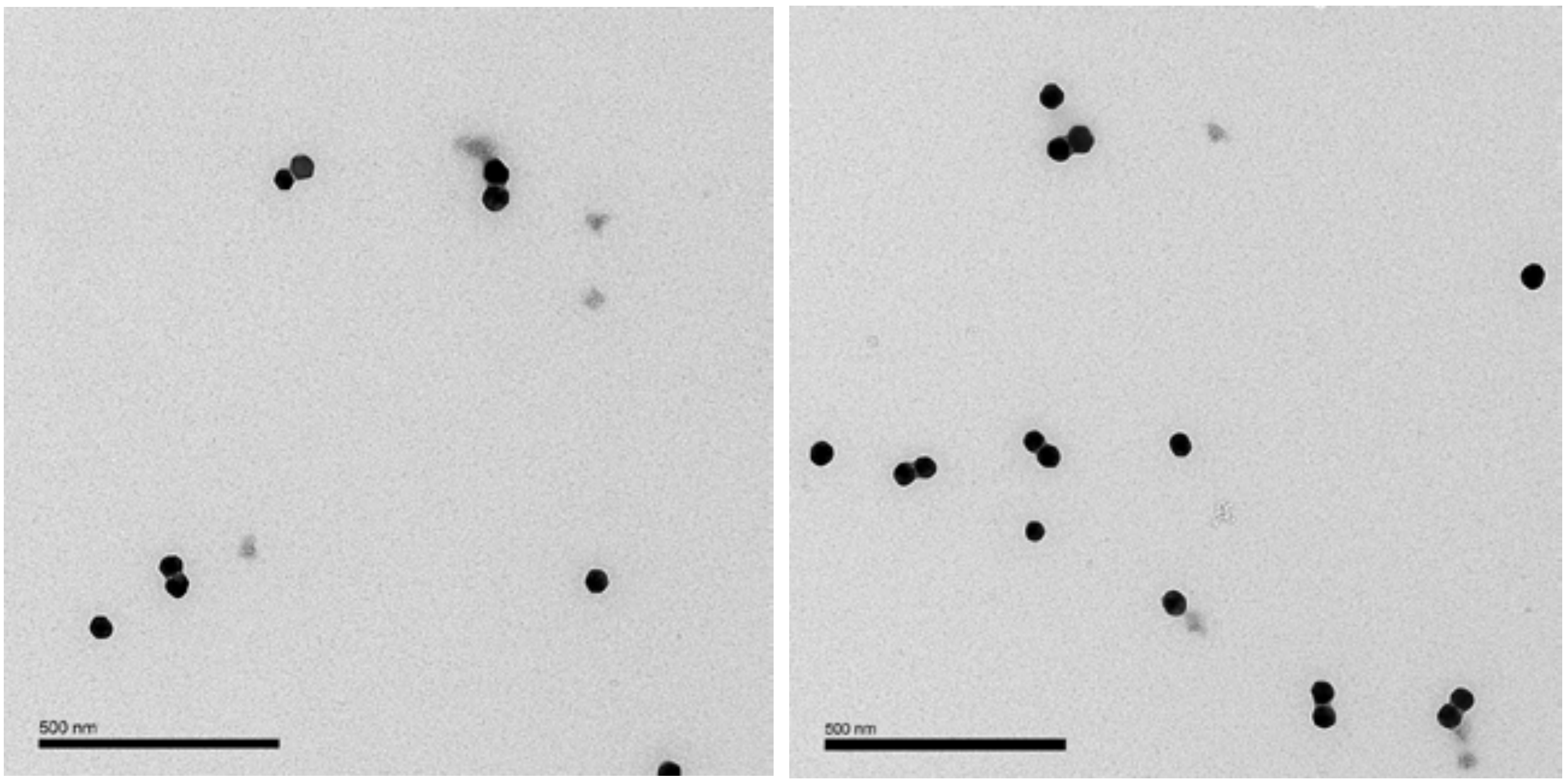

**Fig 8.** Overview TEM images of the five-layer DNA origami frame functionalized with 2x 60nm AuNPs.

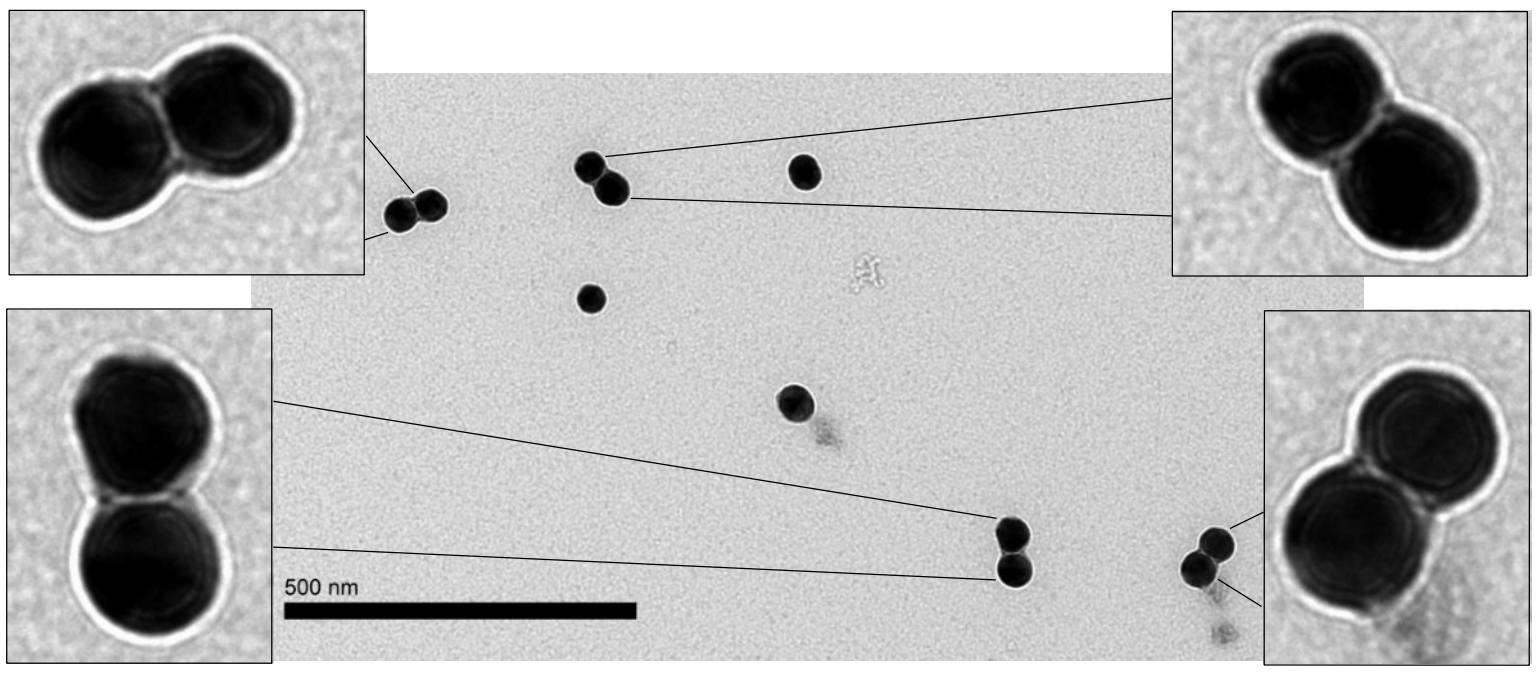

**Fig 9.** Overview TEM image of the DNA-templated molecular devices (four devices in the field of view) containing the target molecule and two 60 nm AuNPs introduced to create contacts.

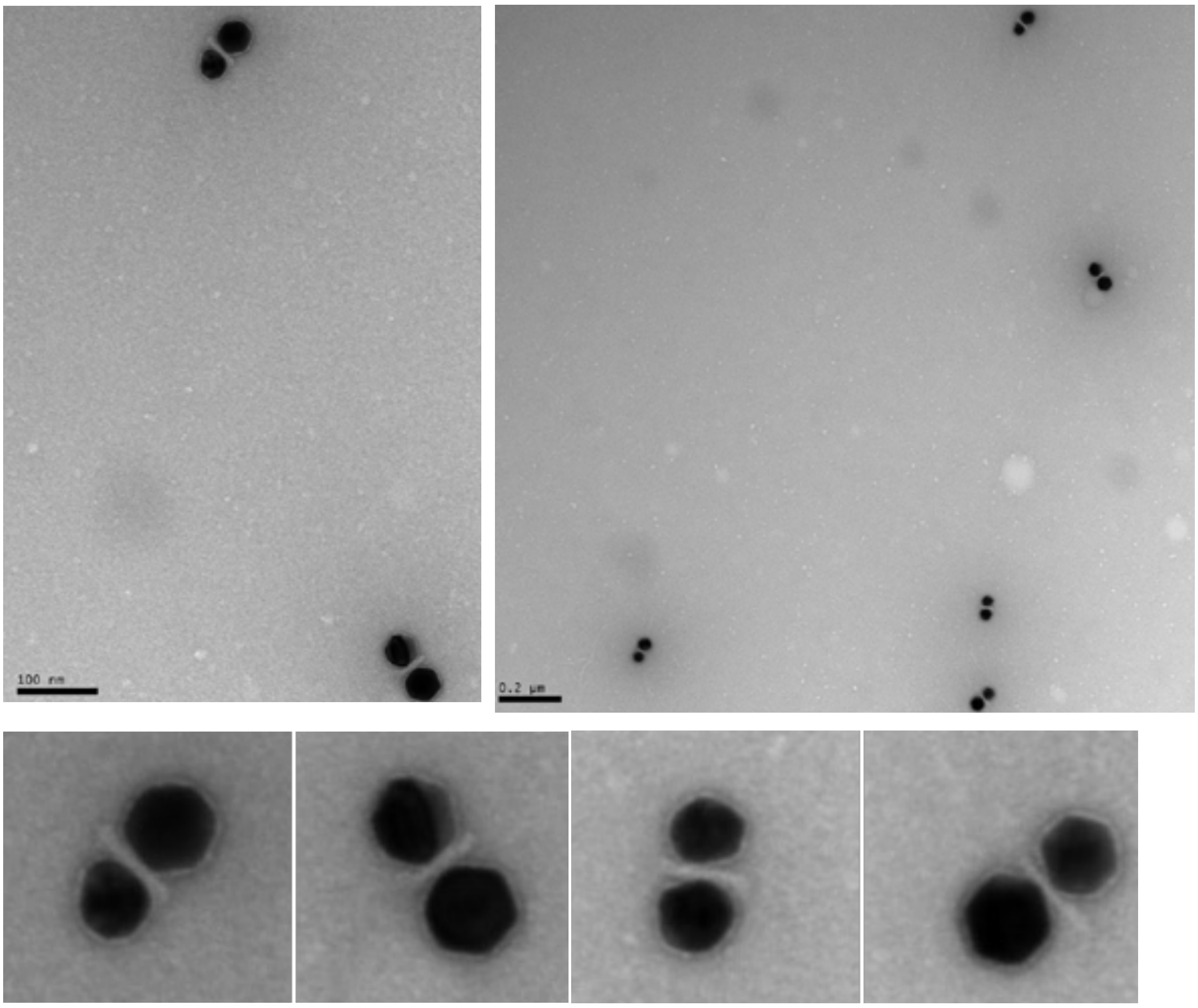

**Fig 10.** TEM images of the three-layered origami structure functionalized with 2x 60 nm AuNPs.

## 10. HPLC (Agilent 1200)

- Buffer A: 5% MeCN, 0.1 M TEAA
- Buffer B: 70% MeCN, 0.1 M TEAA
- Column: Waters XBridge® Oligonucleotide BEH C18 column, 130 Å, 2.5 µm 4.6×50 mm
- Temperature: 60°C

| Time | %B | Flow |
|---|---|---|
| 0.0 | 0.0 | 1 mL/min |
| 1.0 | 0.0 | 1 mL/min |
| 20.0 | 100.0 | 1 mL/min |
| 21.0 | 100.0 | 1 mL/min |
| 22.0 | 100.0 | 1 mL/min |
| 23.0 | 100.0 | 1 mL/min |
| 23.10 | 0.0 | 1 mL/min |

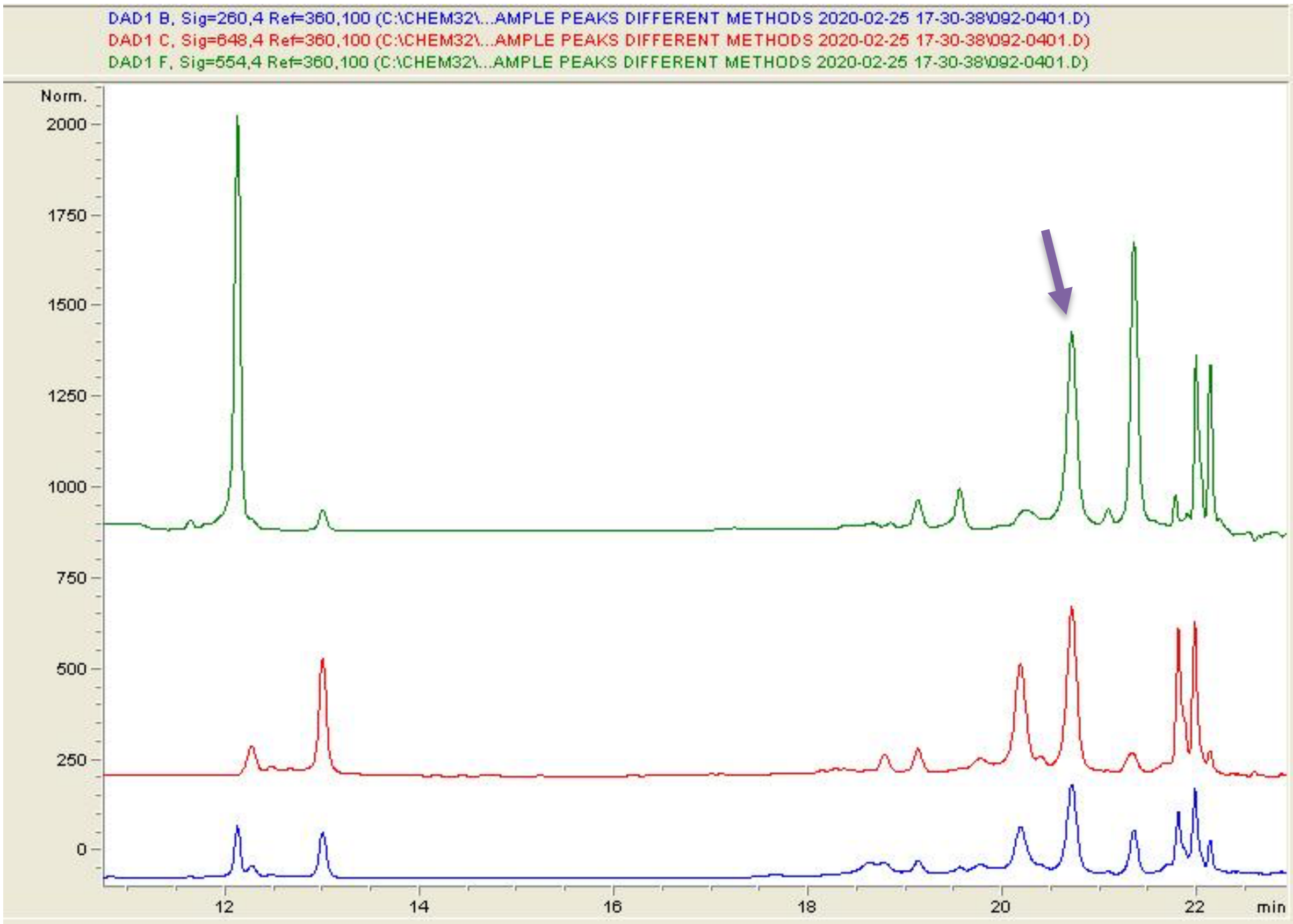

**Fig 11.** The HPLC trace of the CuAAC reaction of the Zn-porphyrin molecule shows the specific detection of various components at distinct wavelengths corresponding to their unique optical properties. In this analysis:
- The DNA content is monitored at 260 nm, represented in blue, pinpointing the nucleic acid components within the sample.
- The signal for Cy5 is detected at 648 nm, illustrated in red, identifying the presence of this specific fluorophore.
- The Cy3 fluorophore is observed at 554 nm, shown in green, marking its distinct contribution to the reaction mixture.

A notable peak, where the blue, green, and red signals converge, is highlighted with a yellow arrow. This coalescence of signals at a singular chromatographic location unequivocally signifies the hetero-DNA-Zn porphyrin product of the reaction. The simultaneous detection of DNA, Cy5, and Cy3 signals at this peak confirms the successful conjugation of the Zn-porphyrin molecule with DNA strands modified with Cy3 and Cy5 fluorophores. This demonstrates the specificity and efficiency of the CuAAC reaction in synthesizing the intended hybrid molecular structure, providing a clear, visual indication of successful product formation within the reaction mixture.

## 11. Agarose gel:

- 2% agarose.
- 11 mM $MgCl_2$.
- 1× TAE 40 mM (Tris, 40 mM acetic acid, 1 mM EDTA, pH 8)
- 60 V for 2 h in ice.

## 12. UV-VIS spectrum of Zn-DNA conjugate

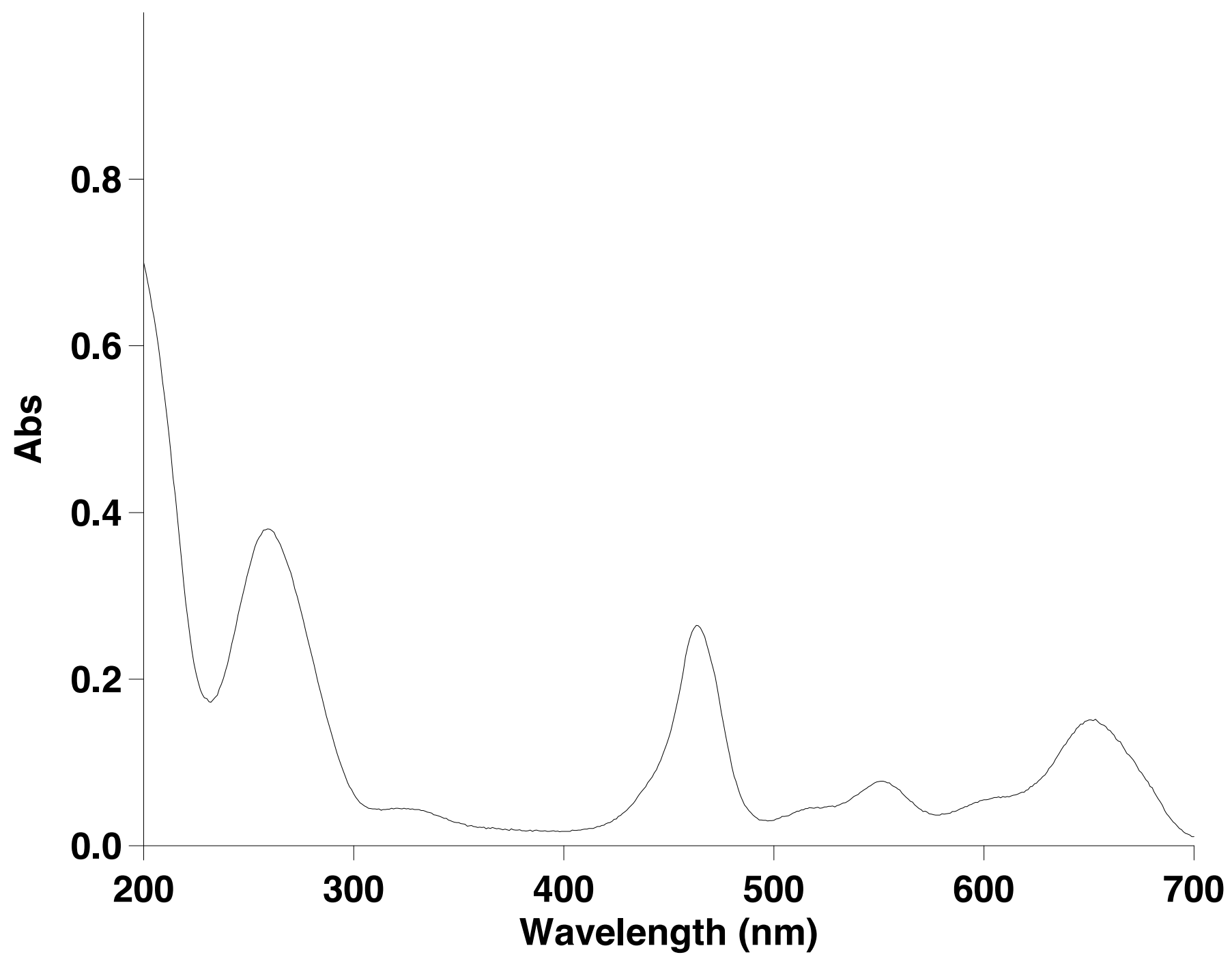

**Fig 12.** The UV-Visible (UV-VIS) spectrum of the purified Zn porphyrin-hetero DNA products measured in TE buffer (10 mM Tris, 1 mM DETA (pH of 8)), following HPLC purification, provides a detailed insight into their optical properties, aligning well with their respective control spectra obtained before the reaction. This spectrum prominently features the characteristic absorption peaks associated with the porphyrin molecules, alongside specific peaks attributed to the Cy3 and Cy5 fluorophores, and a distinct peak representing DNA.
- The Soret band: This strong absorption peak, evident in the UV region around 400-430 nm, confirms the presence of porphyrin structures. It is a hallmark of the porphyrin core, visible in the Zn porphyrin spectra.
- Q-bands: Observable in the visible range from 500 to 700 nm, these peaks are indicative of the metal centre (Zn) within the porphyrin, reflecting the influence of the metal on the electronic structure of the molecule.

- Cy3 and Cy5 Peaks: The Cy3 fluorophore demonstrates a peak around 550 nm, emitting in the green region, while the Cy5 fluorophore exhibits its peak closer to 650 nm, corresponding to red fluorescence. These peaks confirm the successful attachment of these fluorophores to the porphyrin-DNA conjugates.
- DNA Peak: The absorption peak at approximately 260 nm is characteristic of DNA, verifying its presence within the conjugated structure.

This spectrum, by combining the distinctive absorption features of the Soret band, Q-bands, and the fluorophore peaks of Cy3 and Cy5 with the DNA peak, not only confirms the molecular composition of the hetero-DNA-porphyrin constructs but also underscores the successful synthesis and conjugation of these components. The visual representation of these combined spectral features offers a clear confirmation of the hybrid nature of the products and serves as a pivotal validation of the conjugation process's efficacy and the purity of the final constructs. This detailed spectral profile serves not only to confirm the presence of porphyrin, fluorophores, and DNA within the sample but also highlights the successful conjugation of the porphyrin molecules to the two DNA adapters.

## 13. Device fabrication and electrical characterisation

The assembled Au-DNA-Au nanostructures are deposited onto a wafer surface and connected to macroscopic metal electrodes using a top-down electron-beam lithography (EBL) technique, forming single-molecule devices for electrical transport measurements. Since the Au-DNA-Au nanostructures are deposited from a buffer solution and land randomly on the wafer, it is essential to pre-pattern the wafer with predefined features to enable precise localisation for electrical contacting of the nanostructures.
The devices were fabricated on silicon wafers with a thermally grown $SiO_2$ layer of 100 nm. Fig. 13a shows the predefined features. The large blue crosses serve as the metal markers for the EBL alignment, ensuring accurate overlay between successive lithography steps. Additionally, small red crosses arranged in an 8×8 matrix, with a 2 μm separation between crosses, form the "local coordination grid" for locating the deposited Au-DNA-Au nanostructures. All these features are defined using standard EBL processes prior to nanostructure deposition. To prevent aggregation of Au-DNA-Au nanostructures on the pre-patterned metal features, we coat the entire wafer with a 300 nm layer of polymethyl methacrylate (PMMA). The PMMA in the areas between the red crosses (highlighted as the cyan-hatched regions in Fig. 13a) is selectively removed by EBL exposure followed by development and gentle oxygen plasma cleaning. This selective removal not only prevents unwanted aggregation but also allows subsequent removal of excess nanostructures in later fabrication steps. Each 1×1 $cm^2$ wafer includes 16 sets of the predefined features, allowing a maximum of 16 single-molecule transistors. The number of devices is limited by the control/measurement lines in our experimental apparatus. It can be scaled up easily using techniques such as multiplexers.

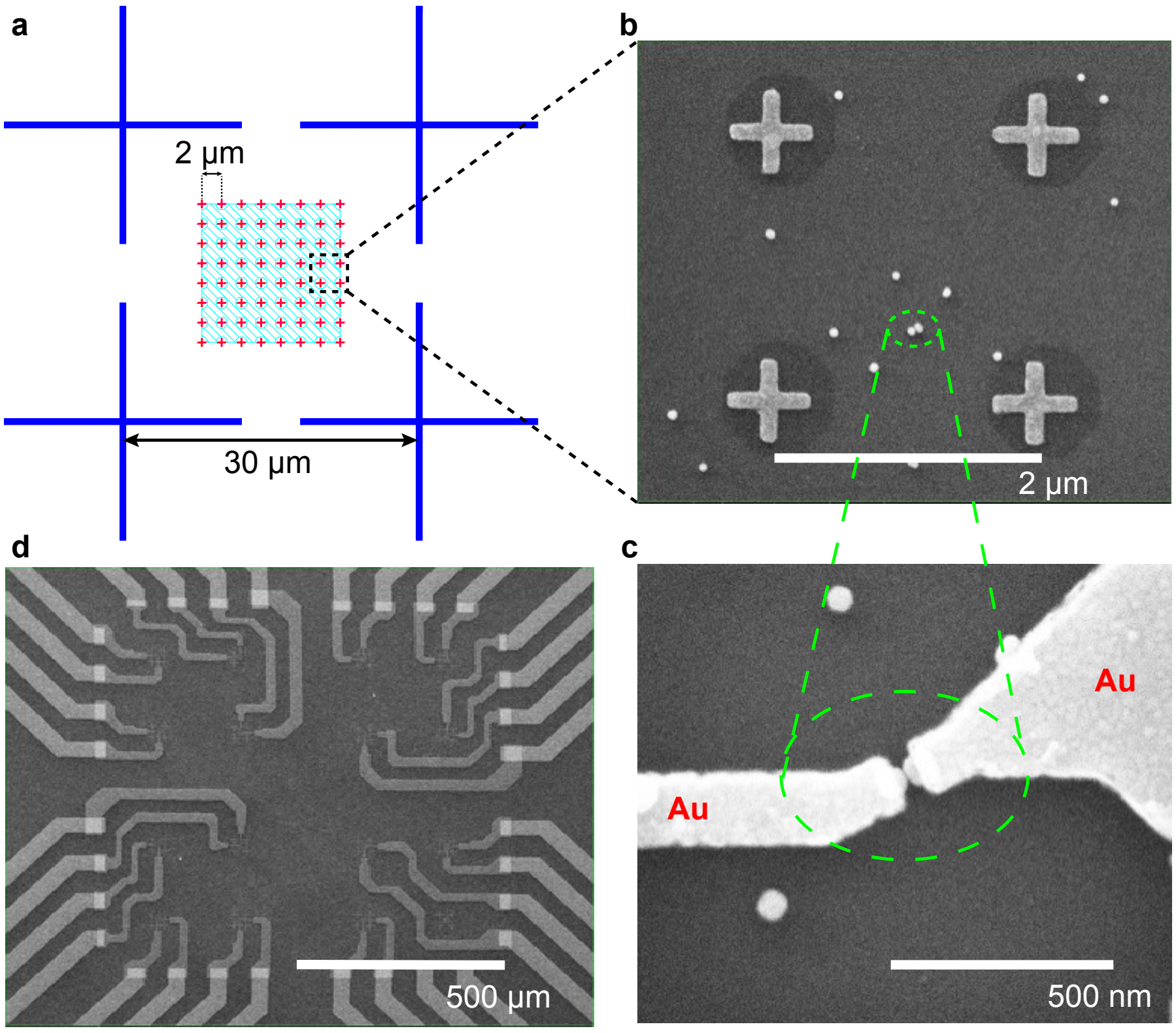

**Fig 13.** a. Schematic plot showing the prepatterned features used for wafer alignment and locating nanostructures. b. After the deposition and the removal of the PMMA layer, the wafer is imaged using SEM, where the location of the deposited nanostructure (indicated by the green oval box) can be defined using the small cross. c. the Au-DNA-Au nanostructure is then connected to metal electrodes using EBL technique. d. SEM image showing multiple fabricated devices (16 on this wafer) measured simultaneously on a single chip.

The Au-DNA-Au nanostructures are introduced onto the $SiO_2$ wafer surface by the deposition of buffer droplets containing Au-DNA-Au nanostructures. After deposition, the wafer is left to dry under ambient conditions and subsequently rinsed with deionized water to remove any excess material from the solution. The PMMA layer, along with any nanostructures adhered to it, is then removed using acetone. A scanning electron microscope (SEM) image of a nanostructure deposited within the predefined area is shown in Fig. S13b. The exact position and orientation of the nanostructure on the wafer can be determined from the SEM image, enabling precise alignment for connecting the structure to metal electrodes using the EBL technique.

Standard fabrication processes typically involve several high-temperature (> 120°C) baking steps to treat the PMMA EBL resist. However, to preserve the structural integrity of the DNA origami, we adopted a fabrication procedure that avoids heating the wafer above room temperature.

Following the SEM imaging step, a single layer of PMMA with a thickness of approximately 450 nm (PMMA 950K A6, spin-coated at 4500 rpm for 1 minute) is applied to the wafer. The thickness is chosen to ensure sufficient height for subsequent metal deposition and lift-off processes. The PMMA layer is dried at room temperature without any baking. Electrode pairs are then designed based on the SEM images and patterned using a 50 kV JEOL EBL system. After exposure, the wafer is developed in a methyl isobutyl ketone (MIBK) and isopropyl alcohol (IPA) solution with a 1:3 volume ratio (MIBK:IPA). This is followed by the deposition of 5 nm chromium and 120 nm gold layers via thermal evaporation. Finally, the unexposed PMMA and excess metal are removed through lift-off in acetone.

Fig. S13c shows the metal electrodes in contact with the Au-DNA-Au nanostructure in Fig. S13b. The fabrication process enables us to reliably connect electrodes to the nanostructures with an alignment error of less than 10 nm. This small alignment error, combined with the designed 30 nm gap between electrodes, allows for device fabrication with a high success rate. Moreover, any unsuccessful fabrication, typically caused by alignment errors, can be readily identified using optical microscopy or SEM imaging. Such defective devices are excluded from electrical transport measurements.

Notably, toward the later stages of this project, we identified that most alignment errors originated from inconsistencies in the EBL system's focus calibration. This issue can be mitigated by implementing on-chip focus calibration before each EBL exposure step, which could reduce the alignment error to below 5 nm.

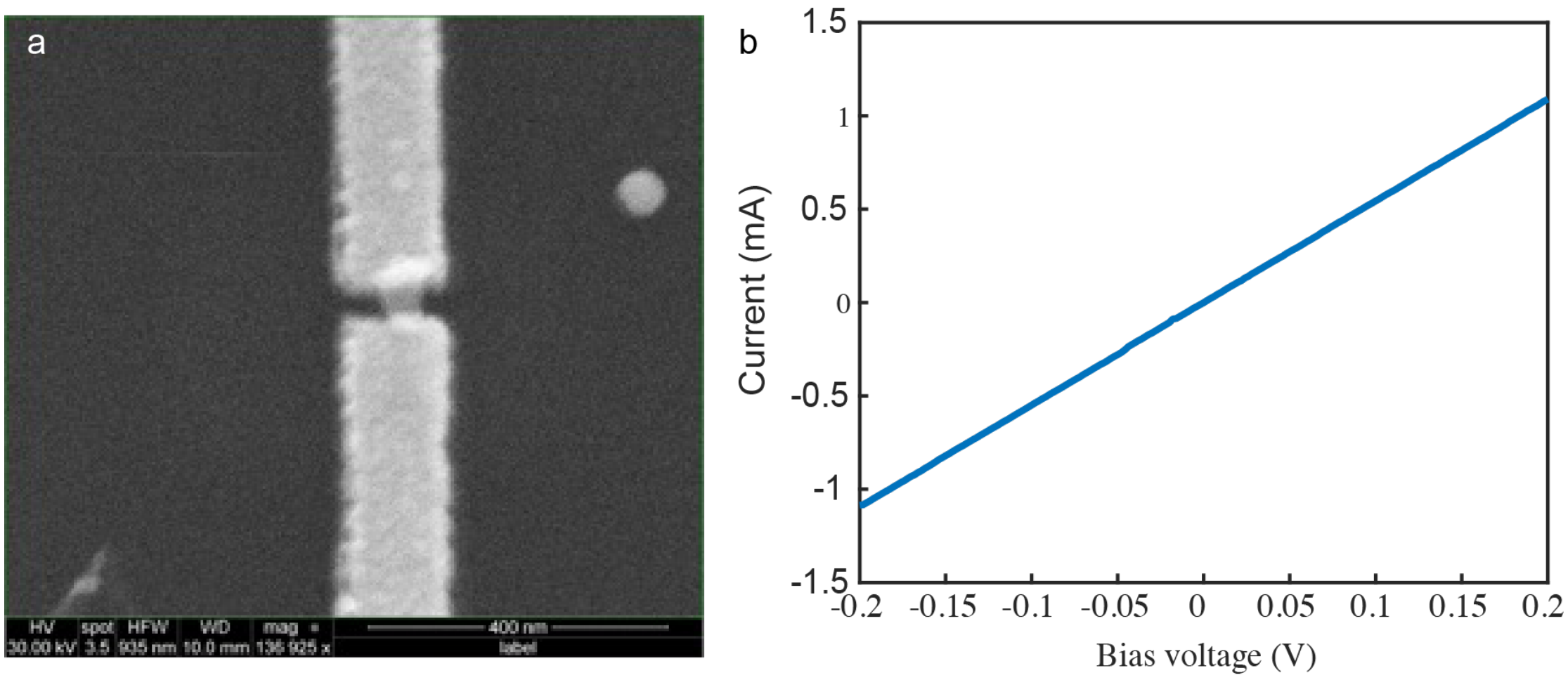

**Fig 14.** a. SEM image showing a single Au nanoparticle connected to two electrodes, using the fabrication procedure described above. b. The bias trace of the Au nanoparticle measured at 77 K. It shows a pure ohmic behaviour and is capable of passing a large current up to 1 mA.

As an initial test of the fabrication procedure, we created electrical contacts to a single Au nanoparticle, as shown in Fig. S14a, to evaluate the contact resistance. The typical measurement result is shown in Fig. S14b, where we observed purely ohmic behaviour with a contact resistance of approximately 200 Ω. This test was repeated on multiple wafers, yielding consistent and reproducible results. Given that the resistance of single-molecule devices typically ranges from several hundred kilo-ohms to much higher values, the measured contact resistance is negligible. Moreover, the contacts are capable of sustaining high electric currents up to the milliampere range at low temperatures, ensuring they do not limit device performance during measurements. Therefore, we conclude that our fabrication procedure is well suited for the reliable fabrication of single-molecule electronic devices.

**Electrical transport measurement for 6 nm nanogap devices**

Fig. S15 shows the representative data for two control devices with ~6 nm nanogaps. All 6 nm devices exhibit low conductance at low bias voltage, typically below $10^{-8}$ $G_0$. Additionally, linear current-voltage characteristics were observed over a wide voltage window of ±500 mV for all 6 nm devices. This behaviour significantly contrasts with that for the 2 nm devices, where tunnelling-like behaviour are clearly evident with the same voltage range.

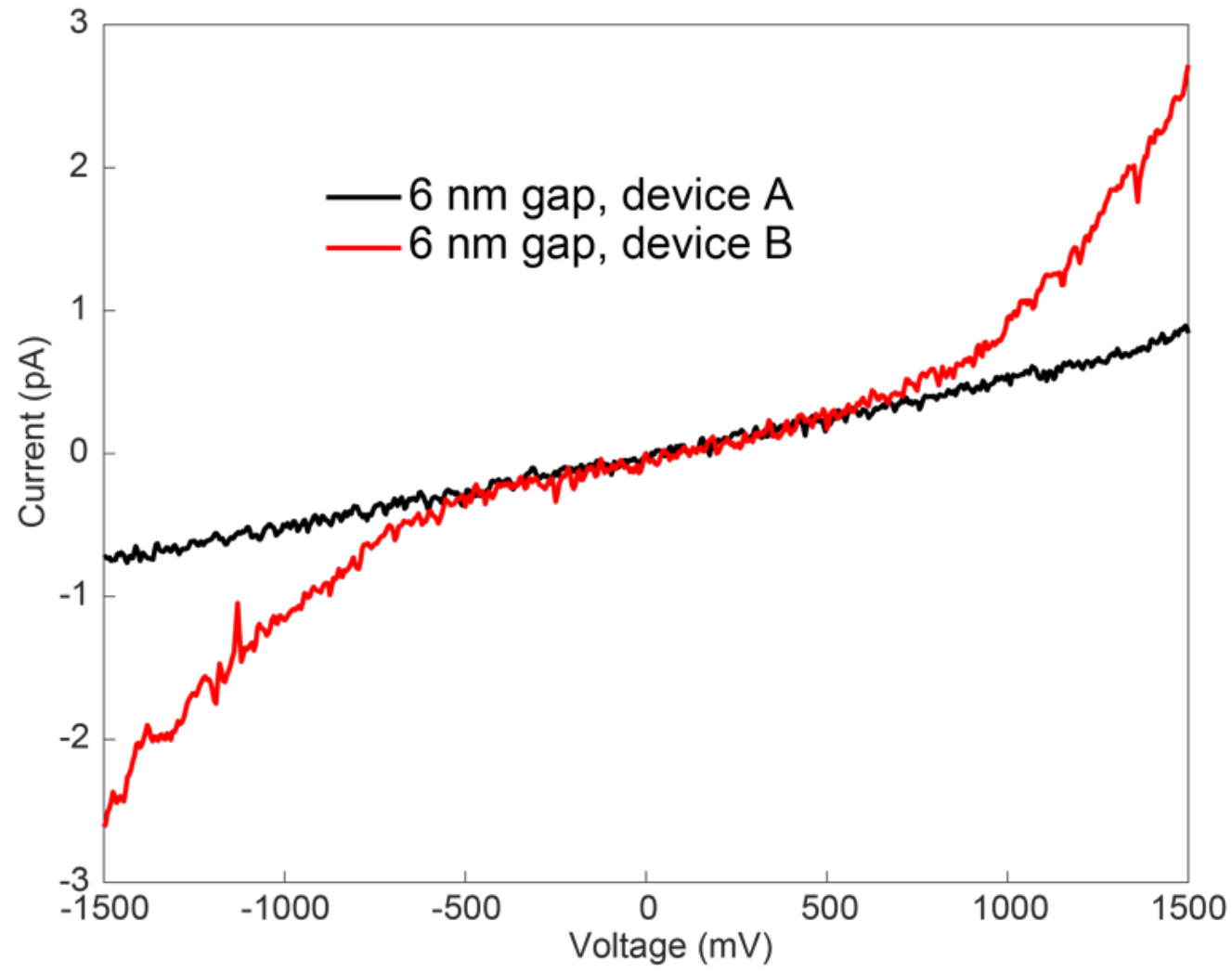

**Fig 15.** Representative IV characteristics for devices with ~6 nm nanogaps. All devices show low conductance and linear IV behaviour up to ±500 mV. Two distinct types of IV curves were observed at high voltages: device A remains a linear IV dependence, while device B displays tunnelling-like behaviour at high voltages.

Up further increasing the bias voltage, the 6 nm devices exhibit two distinct types of behaviour. Among the 35 devices we measured, 26 devices showed a linear IV characteristics up to ±1.5 V with the maximum current below or around 1 pA (Fig. S15, device A), confirming the highly resistive nature of the 6 nm nanogaps. On the other hand, 9 devices began to show identifiable non-linear IV dependences resembling tunnelling at high voltages (Fig. S15, device B). However, the observed tunnelling current remains significantly smaller than those measured for the 2 nm nanogaps, consistent with the increased tunnelling resistance expected for larger gap distances.

**Electrical transport data for all 2 nm gap control devices and Zn-porphyrin devices**

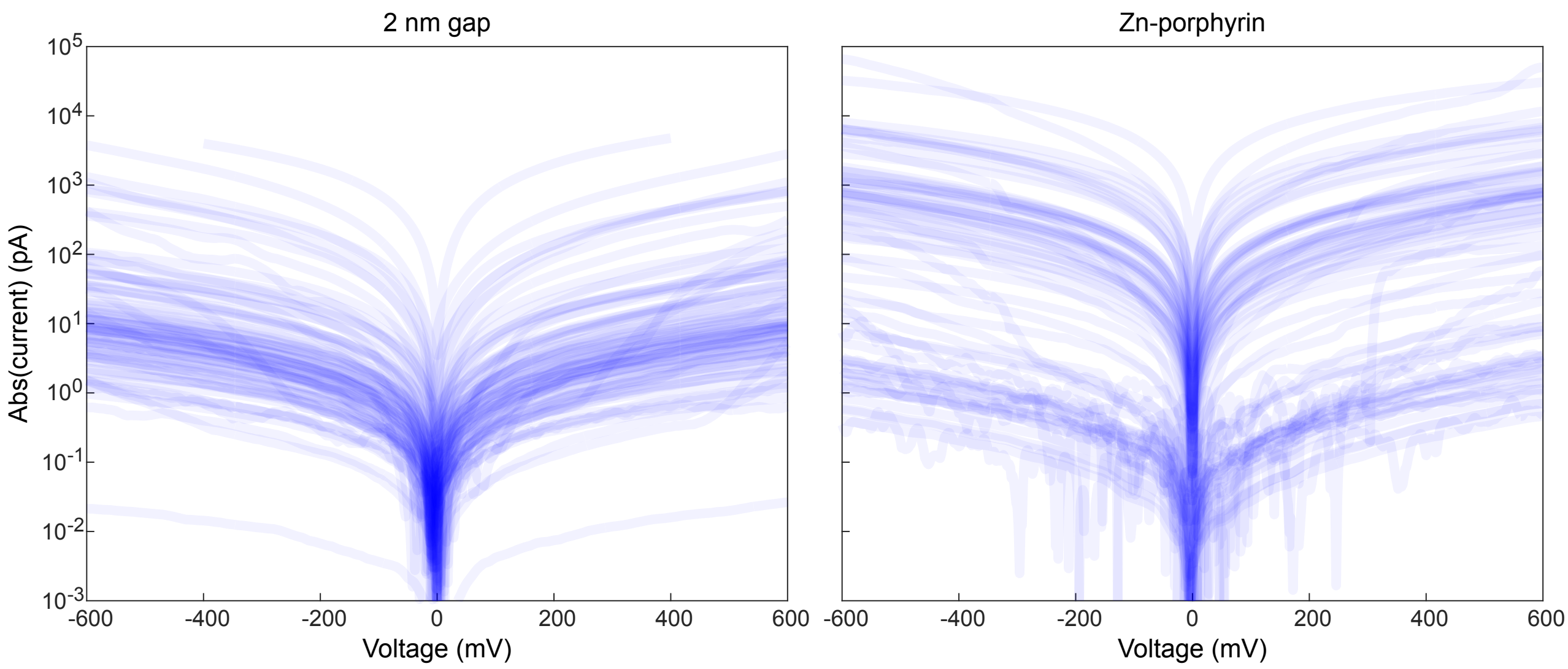

**Fig.** 16. Current-Voltage characteristics for all devices. IV for all 2 nm gap control devices (left) and Zn-porphyrin devices (right). A single band is observed for the 2 nm gap devices, while two distinct bands are found for the Zn-porphyrin devices.

**Representative temperature dependence of 2-layer control device electrical transport**

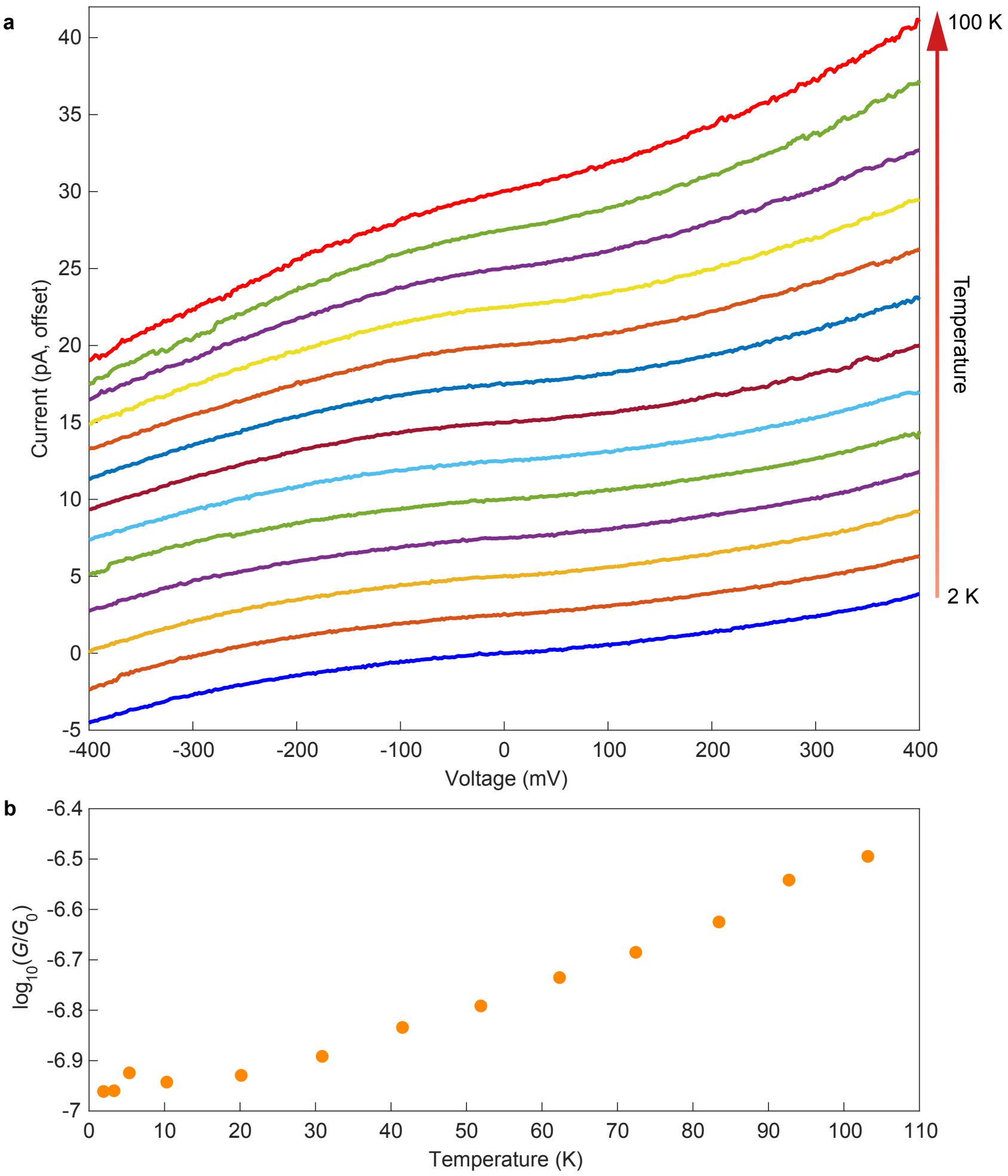

**Fig.** 17. Temperature-dependent transport in a 2 nm gap junction. a, IV characteristics recorded at different temperatures. b, Temperature dependence of the device conductance, extracted at a bias of 300 mV.

## 14. References

1. Kavthe, R. D., Ishikawa, Y., Kusuma, I. & Asao, N. Chemoselective Aerobic Cross-Dehydrogenative Coupling of Terminal Alkynes with Hydrosilanes by a Nanoporous Gold Catalyst. *Chem. – Eur. J.* **24**, 15777–15780 (2018).

2. Castagnolo, D. & Botta, M. Iron-Catalyzed Cross-Coupling between 1-Bromoalkynes and Grignard-Derived Organocuprate Reagents. *Eur. J. Org. Chem.* **2010**, 3224–3228 (2010).

3. Takanami, T., Matsumoto, J., Kumagai, Y., Sawaizumi, A. & Suda, K. A facile one-pot preparation of meso-hydroxymethylporphyrins via a sequential SNAr reaction with (2-pyridyldimethylsilyl)methyllithium followed by hydrolysis and aerobic oxidation. *Tet. Lett.* **50**, 68–70 (2009).

4. Pagano, M. et al. The Fight against the Influenza A Virus H1N1: Synthesis, Molecular Modeling, and Biological Evaluation of Benzofurazan Derivatives as Viral RNA Polymerase Inhibitors. *ChemMedChem* **9**, 129–150 (2014).

5. Simoncelli, S. et al. Quantitative Single-Molecule Surface-Enhanced Raman Scattering by Optothermal Tuning of DNA Origami-Assembled Plasmonic Nanoantennas. ACS Nano **10**, 9809–9815 (2016).